\documentclass[trackchanges,twocolumn]{aastex701}

\usepackage{amsmath} 
\usepackage{tabularx}
\usepackage{hyperref}
\usepackage{xspace}
\let\oldAA\AA
\renewcommand{\AA}{\text{\oldAA}\xspace}

\newcommand{\OIII}{\mbox{O\,\textsc{iii}}} 
 
\newcommand{\NII}{\mbox{N\,\textsc{ii}}} 
\newcommand{\SII}{\mbox{S\,\textsc{ii}}} 
\newcommand{\OI}{\mbox{O\,\textsc{i}}}

\newcommand{\hei}{\mbox{He\,\textsc{i}}}

\begin{document}

\title{A Census of Na D-traced neutral ISM and outflows at $0.6<z<4$}

\correspondingauthor{Yang Sun}
\email{sunyang@arizona.edu}

\author[0000-0001-6561-9443]{Yang Sun}
\affiliation{Steward Observatory, University of Arizona,
933 North Cherry Avenue, Tucson, AZ 85719, USA}
\email{sunyang@arizona.edu}

\author[0000-0001-7673-2257]{Zhiyuan Ji}
\affiliation{Steward Observatory, University of Arizona,
933 North Cherry Avenue, Tucson, AZ 85719, USA}
\email{}

\author[0000-0003-2388-8172]{Francesco D'Eugenio}
\affiliation{Kavli Institute for Cosmology, University of Cambridge, Madingley Road, Cambridge, CB3 0HA, UK}
\affiliation{Cavendish Laboratory, University of Cambridge, 19 JJ Thomson Avenue, Cambridge, CB3 0HE, UK}
\email{}

\author[0000-0003-3307-7525]{Yongda Zhu}
\affiliation{Steward Observatory, University of Arizona,
933 North Cherry Avenue, Tucson, AZ 85719, USA}
\email{}

\author[0000-0003-2303-6519]{George H. Rieke}
\affiliation{Steward Observatory, University of Arizona,
933 North Cherry Avenue, Tucson, AZ 85719, USA}
\email{}

\author[0000-0003-0215-1104]{William M.\ Baker}
\affiliation{DARK, Niels Bohr Institute, University of Copenhagen, Jagtvej 155A, DK-2200 Copenhagen, Denmark}
\email{william.baker@nbi.ku.dk}

\author[0000-0002-8651-9879]{Andrew J. Bunker}
\affiliation{Department of Physics, University of Oxford, Denys Wilkinson Building, Keble Road, Oxford OX1 3RH, UK}
\email{andy.bunker@physics.ox.ac.uk}

\author[0000-0002-6719-380X]{Stefano Carniani}
\affiliation{Scuola Normale Superiore, Piazza dei Cavalieri 7, I-56126 Pisa, Italy}
\email{stefano.carniani@sns.it}

\author[0000-0003-4337-6211]{Jakob M.\ Helton}
\affiliation{Department of Astronomy \& Astrophysics, The Pennsylvania State University, University Park, PA 16802, USA}
\email{jakobhelton@psu.edu}

\author[0000-0002-0362-5941]{Michele Perna}
\affiliation{Centro de Astrobiolog\'ia (CAB), CSIC–INTA, Cra. de Ajalvir Km.~4, 28850- Torrej\'on de Ardoz, Madrid, Spain}
\email{michele.perna@cab.inta-csic.es}

\author[0000-0003-4528-5639]{Pablo G. P\'erez-Gonz\'alez}
\affiliation{Centro de Astrobiolog\'ia (CAB), CSIC–INTA, Cra. de Ajalvir Km.~4, 28850- Torrej\'on de Ardoz, Madrid, Spain}
\email{pgperez@cab.inta-csic.es}

\author[0000-0002-5104-8245]{Pierluigi Rinaldi}
\affiliation{Space Telescope Science Institute, 3700 San Martin Drive, Baltimore, Maryland 21218, USA}
\email{prinaldi@stsci.edu}

\author[0000-0003-4891-0794]{Hannah \"Ubler}
\affiliation{Max-Planck-Institut f\"ur extraterrestrische Physik (MPE), Gie{\ss}enbachstra{\ss}e 1, 85748 Garching, Germany}
\email{hannah@mpe.mpg.de}

\author[0000-0001-9262-9997]{Christopher N. A. Willmer}
\affiliation{Steward Observatory, University of Arizona, 933 North Cherry Avenue, Tucson, AZ 85719, USA}
\email{cnaw@as.arizona.edu}

\begin{abstract}
We present a statistical census of the Na D-traced neutral interstellar medium (ISM) and outflows in 309 galaxies at $0.6<z<4$ using JWST/NIRSpec medium-resolution spectroscopy from the SMILES, JADES, Blue Jay, and Aurora surveys. After subtracting the stellar continuum, we model the Na D $\lambda\lambda 5890, 5896$ \AA and identify 73 Na D ISM absorptions, including 63 robust (4 in broad-line AGNs) and 10 tentative detections. Of the robust detections, 88\% are found in massive galaxies ($\log(M_*/M_\odot)>10$), and 12\% in lower-mass systems. At high mass, ISM absorption is seen in both star-forming and quiescent galaxies, whereas in lower-mass systems it is observed only in star-forming galaxies. In massive quiescent galaxies, Na D detectability appears linked to star formation history: it is preferentially detected in older systems with larger 4000 \AA breaks, and younger, rapidly quenching galaxies with strong Balmer absorption H$\delta_A$. We identify Na D {\it outflows} in 25 galaxies, revealing a possible dichotomy in driving mechanisms between star-forming and quiescent galaxies. In star-forming galaxies, outflow properties correlate with star-formation properties, consistent with a star-formation-driven origin. In quiescent galaxies, however, outflows are not associated with residual star formation and often require more energy than such star formation can provide. Together with the high AGN fraction among outflow-detected quiescent galaxies, this suggests that AGN dominate Na D-traced neutral outflows in cosmic noon quiescent systems. We further identify four quiescent galaxies with possible AGN fossil outflows, suggesting that AGN-driven outflows can persist beyond the active accretion phase and may help maintain quiescence.

\end{abstract}

\keywords{
\href{http://astrothesaurus.org/uat/572}{Galactic Winds (572)},
\href{http://astrothesaurus.org/uat/594}{Galaxy Evolution (594)},
\href{http://astrothesaurus.org/uat/1569}{Star Formation (1569)},
\href{http://astrothesaurus.org/uat/2040}{Galaxy Quenching (2040)}
}


\section{Introduction} 
\label{sec:intro}

Observations in both the local Universe and at high redshift have revealed a clear bimodality in galaxy color, morphology, and star formation activity, with galaxies broadly divided into star-forming and quiescent populations \citep[e.g.,][]{Strateva2001,Kauffmann2003,Brinchmann2004,Noeske2007,Speagle2014,Tacchella2016,Ji2023}. Understanding quenching, i.e. the processes that causes star formation to cease and transforms galaxies from star-forming systems into quiescent ones, remains a central problem in galaxy evolution.

Feedback is widely thought to be essential for quenching \citep[e.g.,][]{Keres2009,Hopkins2014,Somerville2015,Naab2017,Silk1998,Hopkins2008,Man2018}, but its detailed physics remain poorly understood. By injecting energy and momentum into galaxies, feedback shapes the gas within and around them, but this is observationally difficult to characterize because galaxies host multiphase gas spanning a wide range of temperatures and densities. To understand how feedback drives galaxy quenching, we therefore need a census of outflows across gas phases and galaxy evolutionary stages. So far, the ionized phase has been studied most extensively, thanks to bright emission lines such as [O~III] and H$\alpha$ \citep[e.g.,][]{Woo2016,Concas2017,Forster2019}. In contrast, cooler gas, especially the neutral and molecular phases ($T \lesssim 10^4$~K), remains much less constrained, despite being more directly tied to star-forming fuel. This is particularly true in quiescent galaxies, where emission lines are often faint or even entirely absent. Observations of these phases remain limited to small samples, despite their importance for tracing gas removal and redistribution \citep[][and references therein]{Veilleux2020,French2015,Spilker2018,Williams2021,Bezanson2022}.

The Na~D doublet is particularly powerful in this context because it probes the cool neutral gas phase that is otherwise difficult to access, especially in quiescent galaxies. It is a resonant transition of neutral sodium at rest-frame 5890~\AA\ and 5896~\AA, commonly observed in absorption against the stellar continuum and also capable of exhibiting resonant emission through scattering and re-emission in outflowing gas \citep[e.g.,][]{Prochaska2011}. Its absorption profile provides a sensitive probe of the column density and kinematics of neutral gas along the line of sight, thereby offering a direct way to connect feedback signatures to the cool ISM and its associated dust during galaxy evolution.

In the local Universe, Na~D has been extensively studied as a tracer of the neutral ISM and its connection to feedback. Na~D-traced ISM is commonly detected in star-forming and starburst galaxies \citep[e.g.,][]{Rupke2005b,Chen2010,Cazzoli2016,Concas2019} and in AGN hosts \citep[e.g.,][]{Sarzi2016,Fluetsch2021,Perna2019}. It is also frequently detected in local post-starburst galaxies (PSBs) that were quenched within the past $\sim$1~Gyr, and which in many cases show signatures of outflowing neutral gas \citep[e.g.,][]{Tremonti2007,Coil2011,Luo2022,Baron2022,Sun2024}. By contrast, Na~D-traced ISM is rare in local quiescent galaxies that have remained quenched for longer periods \citep{Sun2024}, although we note that there exists a population of quenched galaxies with low-level AGN activity and ioinzied outflow \citep[``Red Geysers"; e.g.,][]{Cheung2016,Moghni2026}. These results suggest that substantial cool gas reservoirs can persist into the quenching phase and are often kinematically disturbed, but that this gas is largely depleted in galaxies long after quenching.

At $z>1$, Na~D studies were long hampered by the wavelength coverage and sensitivity limits of pre-JWST facilities. Yet constraining the neutral ISM at cosmic noon is particularly important, because this is the epoch when star formation and black hole growth peak \citep{Madau2014} and galaxies transition rapidly across evolutionary stages \citep[e.g.,][]{Muzzin2013}. With the launch of JWST \citep{Gardner2006}, Na~D-traced neutral ISM detections, especially outflows, have increasingly been reported in galaxies at cosmic noon and beyond, initially in individual systems \citep{Belli2024,D'Eugenio2024,Perez-Gonzalez2025,Wu2025,Sun2026} and more recently in a few survey-level studies, though still with limited sample sizes (5--30 detections; \citealt{Davies2024,Taylor2026,ZhuP2026}).

Even within these limited samples, Na~D studies have already revealed several intriguing trends at cosmic noon. Na~D outflows have been identified in quiescent galaxies with ongoing AGN activity \citep[e.g.,][]{Belli2024,Davies2024,D'Eugenio2024,Wu2025}, but also in a few $z>1$ quiescent galaxies without clear evidence for such activity \citep{Perez-Gonzalez2025,Sun2026,Taylor2026,ZhuP2026}. This suggests that the physical drivers of Na~D-traced neutral outflows are not straightforward: in some cases they may be linked to ongoing AGN feedback, while in others they may reflect fossil AGN (i.e., previously active but that have faded below detection limits) or other mechanisms. Taken together, these results imply that neutral outflows may arise in quenching systems under a range of physical conditions, underscoring the need for a statistical census of the neutral-phase ISM and outflows across galaxy evolutionary stages at cosmic noon to determine how common they are and what role they play in early massive-galaxy quenching.

To this end, we present a census of Na~D-traced ISM and outflows at $0.6<z<4$ based on a compilation of JWST/NIRSpec medium-grating ($R\sim1000$) spectroscopy from Cycles 1--3 programs, including SMILES \citep{Rieke2024,Alberts2024}, JADES \citep{Bunker2024,Eisenstein2026,Eisenstein2023b}, Blue Jay \citep{Belli2025}, and Aurora \citep{Shapley2025}. This yields the largest dataset to date for this purpose, comprising 309 galaxies with spectra of sufficient quality for detailed Na~D analysis and spanning evolutionary stages from star-forming to quiescent. Using this sample, we measure the incidence of Na~D-traced ISM absorption, the prevalence and kinematics of neutral outflows, and their dependence on host-galaxy star formation and AGN properties. This dataset provides a statistical view of the cool neutral phase across the quenching sequence at cosmic noon.

This paper is organized as follows:
We present the data used in this study in Section~\ref{sec:data}, and the sample drawn from it in Section~\ref{sec:sample}. We then describe the spectral analysis and the SED modelling in Section~\ref{sec:analy}. The census results for Na D-traced ISM are in Section~\ref{sec:nad_ism_census} and outflows in Section~\ref{sec:nad_outf}. We discuss their implications on the outflow driving mechanisms, multiphase natures, and correlation with different galaxy quenching pathways in Section~\ref{sec:discuss}. Finally, we summarize our work in Section~\ref{sec:concl}. Throughout this paper, we assume a standard $\Lambda$CDM universe whose cosmological parameters are $\mathrm{H_0} =
70~\mathrm{km~s^{-1}~Mpc^{-1}}$, $\Omega_{\Lambda} = 0.7$, and $\Omega_{\mathrm{m}} = 0.3$.

\section{Data} \label{sec:data}


\begin{table*}
\centering
\caption{Survey sample summary}
\label{tab:survey_sample}
\begin{tabular}{lcccc}
\hline
Survey & Spectroscopy & Photometry & $N_{\rm initial}^{a}$ & $N_{\rm final}^{b}$ \\
\hline
SMILES  & SMILES DR2$^c$  & JADES DR5$^d$ & 166  & 63  \\
JADES   &JADES DR4$^e$ & JADES DR5 & 4752 & 159 \\
Blue Jay & DJA$^f$ & Blue Jay DR$^g$ & 153  & 56  \\
Aurora  & DJA &JADES DR5 \& DJA & 97   & 31  \\
\hline
Total    &&& 5168 & 309 \\
\hline
\end{tabular}
\tablecomments{$^{a}$Total number of published spectra of each survey; $^{b}$Final sample defined by requiring $0.6<z<4$, and $\mathrm{SNR_{cont}}>5$ in the continuum adjacent to Na D, measured over rest-frame 5850--5870\,\AA\ and 5910--5930\,\AA, and with available HST and JWST photometry measurements.$^c$ \citet{Zhu2026}; $^d$ \citet{Johnson2026,Robertson2026}; $^e$ \citet{Curtis-Lake2025,Scholtz2025}; $^f$ Dawn JWST Archive; \citet{heintz2025,deGraaff2025}; $^g$ \url{https://zenodo.org/records/13292819}}
\end{table*}

We compiled JWST NIRSpec/Micro Shutter Assembly (MSA; \citealt{Jakobsen2022,Ferruit2022}) G140M and G235M medium-resolution grating (R$\sim$1000) spectra from four surveys that have good coverage of cosmic noon galaxies: SMILES in GOODS-S/UDF, JADES in GOODS-S and -N, Blue Jay in COSMOS, and Aurora in COSMOS and GOODS-N.
These setups cover the Na D feature for galaxies at $0.6<z<4$, enabling an analysis of neutral gas absorption across a broad range of redshift and galaxy types.

The spectroscopic data reductions adopted for each survey are listed in Table~\ref{tab:survey_sample}. We use spectra reduced with local background subtraction using the default DJA setup, namely three-nod subtraction for SMILES, JADES, and Aurora, and ``A$-$B'' subtraction for the two-shutter Blue Jay slitlets\footnote{Blue Jay spectra are observed with a two-shutter slitlet, so the DJA reduction is based on local background subtraction, i.e., ``A$-$B'' subtraction.}. Although this approach might lead to self-subtraction in extended galaxies, Appendix~\ref{ap:nod_comp} shows that the effect is negligible for our Na D ISM profile modeling and scientific conclusions. In addition, because our $z\sim2$ targets are often spatially extended, the MSA shutter usually does not cover the full galaxy. We therefore estimate slit-loss corrections by matching synthetic slit photometry to the full-galaxy NIRCam photometry. This procedure implicitly assumes spatially uniform line-to-continuum ratios and similar spatial distributions for nearby lines (e.g., [\NII] and H$\alpha$), which may not always hold. To minimize potential systematics, we apply the slit-loss correction only when deriving quantities that require absolute line luminosities, such as H$\alpha$-based SFRs (Appendix~\ref{ap:SFR_comp}) and [\OIII]-based AGN luminosities (Section~\ref{sec:Eout_pout}). For stellar-population fitting (Section~\ref{sec:ppxf}), Na D profile measurements (Section~\ref{sec:nad_profile}), and line-ratio diagnostics including Balmer decrements (Section~\ref{sec:ppxf}) and BPT diagrams (Section~\ref{sec:BPT}), we use the uncorrected spectra.

For photometry, we use the published catalogs in each survey field, also summarized in Table~\ref{tab:survey_sample}. We adopt all available 
HST/ACS (F435W, F606W, F775W, F814W and F850LP), and WFC3 (F105W, F125W, F140W, and F160W), and JWST/NIRCam (F090W, F115W, F150W, F182M, F200W, F210M, F309W, F356W, F410M, F430M, F444W, F460M and F480M)\footnote{Only a subset of the JADES and SMILES galaxies have medium-band observations available in the JADES photometric catalog.} measurements to sample the rest-frame optical-to-near-IR SEDs of the galaxies (about 0.3--3\micron~at z$\sim$0.6, and 0.1--1\micron~at z$\sim$4), enabling reliable constraints on their stellar properties.

\section{The Sample}\label{sec:sample}

Our initial sample is drawn from the full set of galaxies with spectroscopy in the surveys listed in Table~\ref{tab:survey_sample}, where the total number of spectra in each field is also reported. We first restrict the redshift range to $0.6<z<4$, such that the Na D doublet is covered by either the G140M or G235M grating.
Because detecting and modeling Na D absorption requires robust continuum measurements, we further refine the sample by requiring the average per-pixel continuum SNR adjacent to Na D, measured over rest-frame 5850--5870 \AA\ and 5910--5930 \AA, to satisfy $\mathrm{SNR_{cont}}>5$. 
This yields a spectral parent sample of 314 galaxies at $0.6<z<4$ (Table~\ref{tab:survey_sample}). After cross-matching with the photometric catalogs, two JADES galaxies and three Aurora galaxies are excluded from further analysis because of missing photometric counterparts or large photometric uncertainties.

\begin{figure}
\centering
\includegraphics[width=\columnwidth]{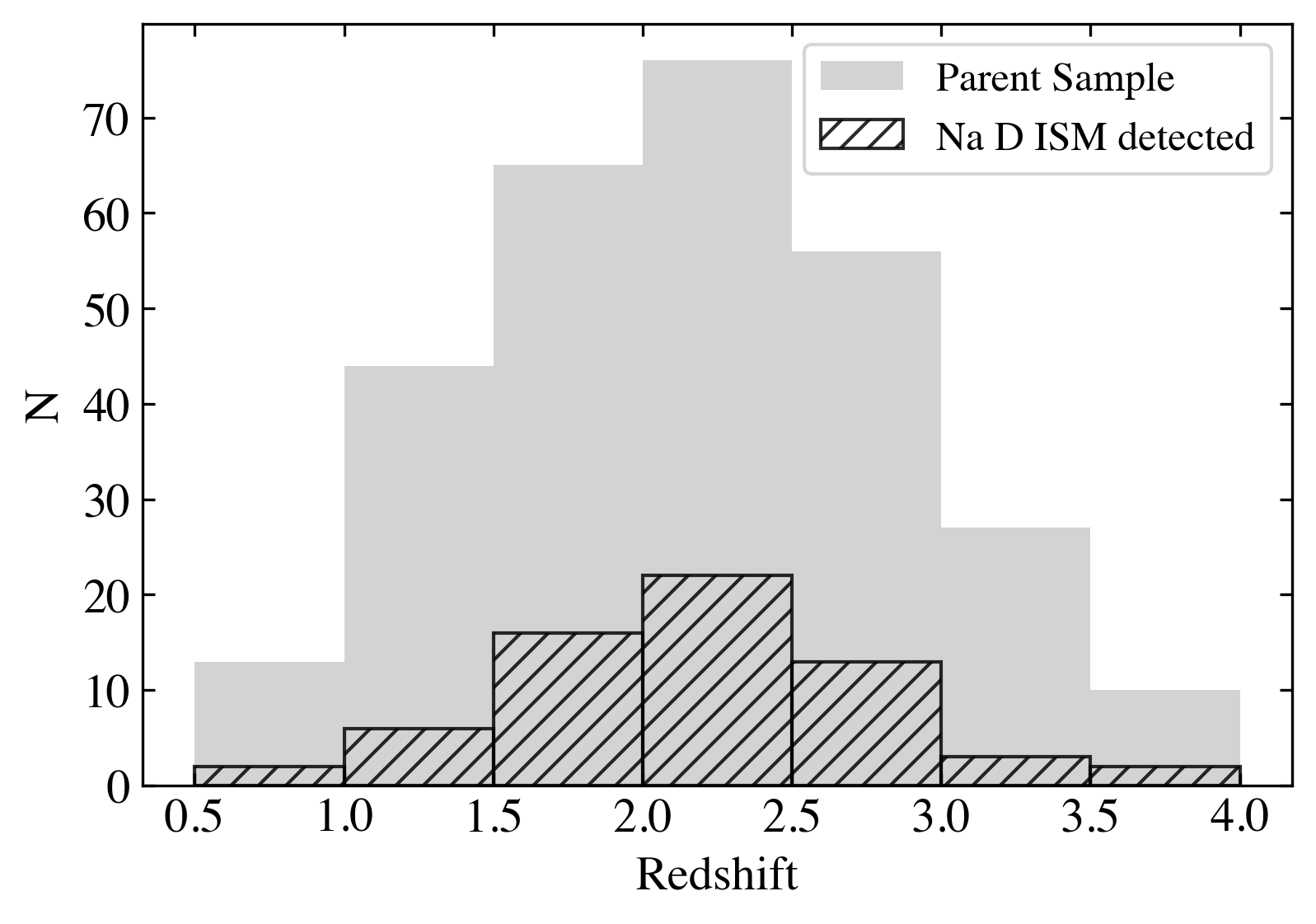}
\caption{Redshift distributions of the final sample of 309 galaxies and the subsample of 59 galaxies with robust detected Na D ISM absorption.}
\label{fig:z_dist}
\end{figure}

Our final sample comprises 309 galaxies at $0.6<z<4$, all with NIRSpec $R\sim1000$ spectra of sufficient quality for Na~D modeling and with UV-to-NIR photometry for stellar-population inference. The sample includes 63 SMILES galaxies, 159 JADES galaxies, 56 Blue Jay galaxies, and 31 Aurora galaxies. The redshift and $M_*$--sSFR distributions of the parent sample are shown in Figures~\ref{fig:z_dist} and \ref{fig:MS}, respectively.

\section{Analysis} \label{sec:analy}

In this section, we describe the steps used to isolate the Na D absorption arising from the ISM in the observed spectra. Specifically, we (1) correct the Na D absorption for the contribution due to stars rather than the ISM (Section~\ref{sec:ppxf}); (2)  fit possible Na D line profiles and measure their strength (Section~\ref{sec:nad_profile}); and (3) evaluate whether these results indicate a detection at a significant level and if so, whether the detection is an inflow, outflow, or at the systemic velocity (Section~\ref{sec:detect&class}). Then, in Section~\ref{sec:SED}, we describe the SED modeling we use to determine the host galaxy properties, particularly their current levels of star formation and their star formation histories. 

\subsection{Stellar continuum and nebular emission modeling using pPXF}
\label{sec:ppxf}
Both the stellar photosphere and the ISM can contribute to Na D spectral features; therefore, we model the stellar continuum of each galaxy 
spectrum over the wavelength range of 3500-7500 \AA~ 
in the rest-frame using penalized pixel fitting \citep[pPXF,][]{cappellari2004,cappellari2017}. We adopt the instrument line spread function from \citet{Jakobsen2022}. Our pPXF setup is similar to \citet{D'Eugenio2024}. We build stellar templates using the Flexible Stellar Population Synthesis (FSPS; \citealt{Conroy2009, Conroy2010}) model built by \citet{Cappellari2023}, with the C3K stellar spectral library \citep{park2025}, the MIST isochrones \citep{Choi2016} and the Salpeter IMF \citep{Salpeter1955}. The templates span a logarithmically spaced grid of ages and metallicities, covering the range of 1 Myr--15.85 Gyr with 0.1 dex sampling and [Z/H]$=$-1.75--0.25 with 0.25 dex sampling, respectively. The age of the input templates is required not to exceed the Universe age at the redshift of each galaxy. 

In pPXF fitting, we also simultaneously fit the nebular emission lines, including Balmer lines, [$\OIII$] $\lambda\lambda4959, 5007$, [$\OI$] $\lambda6300$, [$\NII$] $\lambda\lambda6548, 6584$, and [$\SII$] $\lambda\lambda6717, 6731$. A single Gaussian profile is assigned to each line, sharing the same velocity and velocity dispersion. The line velocity dispersions are limited within $300$ $\mathrm{km\,s^{-1}}$, and the velocity offset is restricted within $\pm 200 \,\mathrm{km\,s^{-1}}$ relative to the systemic redshift.

Finally, a multiplicative polynomial\footnote{In most cases, we adopt a fourth-degree polynomial. However, for a small subset of galaxies in which visual inspection reveals a significant mismatch between the G140M and G235M continuum levels, likely due to contamination, we increase the polynomial degree to 12. } is applied to avoid mismatches between galaxy spectra and stellar templates and ensure the accuracy of fitted kinematic parameters. The Na D line is masked during fitting to avoid the influence of Na D ISM absorption on the stellar Na D absorption modeling.

The systemic redshift is determined by pPXF. Ideally, we aim to anchor the systemic redshift using stellar absorption features, as the Na D outflow we seek is defined by Na D ISM absorption that is blueshifted relative to background stellar light. However, in reality, most galaxies in our sample do not exhibit clear stellar absorption lines but rather strong emission lines. In this case, we  measure the systemic redshift using the emission lines. 

We measure emission line fluxes from the best-fit nebular components in the pPXF model. To estimate flux uncertainties, we perform 100 Monte Carlo realizations of the pPXF fitting. We then derive the dust attenuation ($A_V$) primarily from the Balmer decrement H$\alpha$/H$\beta$. We assume  Case B recombination and  $T_e=10^4\,K$, and hence an intrinsic Balmer flux ratio $\rm (H\alpha/H\beta)_0 = 2.86$. If the SNR of either H$\alpha$ or H$\beta$ is less than three, we instead adopt the SED-based $A_V$ (SED modeling will be presented in Section~\ref{sec:SED})\footnote{In fact, about 60\% of $A_V$ of our parent galaxy sample is measured through the Balmer Decrement, and the other 40\% is derived from SED fitting. As a sanity check, we compare the two independent $A_V$ measurements for galaxies with both available and find that they are broadly consistent.}. Finally, we use the \citet{Calzetti2000} reddening curve to dust-correct the line fluxes.

Finally, we verify the robustness of our pPXF stellar continuum modeling by confirming that the Mg b (5167, 5173, and 5184 \AA) absorption feature, which is purely stellar in origin, is well reproduced by the pPXF best-fit model. Details of the Mg b residual test are presented in Appendix~\ref{ap: Mgbtest}.

\subsection{Na D excess profile modeling}
\label{sec:nad_profile}

After subtracting the stellar Na D contribution with pPXF, we model the residual Na D absorption from the ISM following \citet{Sun2026}:
\begin{align}
    F(\lambda) = F_{*} \times F_{\text{Na D, ISM}},
\label{eq1}
\end{align}
where $F_*$ is the pPXF stellar continuum and $F_{\text{Na D, ISM}}$ is described with the standard partial-covering formalism of \citet{Rupke2005a}:
\begin{align}
    F_{\text{Na D, ISM}}(\lambda) = 1 - C_f + C_f\,e^{(-\tau_{b}(\lambda) - \tau_{r}(\lambda))},
\end{align}
with covering fraction $C_f$ and Gaussian optical-depth profiles for the blue ($\tau_b$) and red ($\tau_r$) doublet components, assuming $\tau_b/\tau_r=2$. The model-free parameters, therefore, are the line centroid $\lambda_0$, optical depth $\tau_b$, Doppler width $b$ after considering the instrumental broadening, and covering fraction $C_f$. 

In addition, with our spectral resolution of $R\sim1000$, the $\hei$ $\lambda5877$ emission would be partially blended with Na D. Therefore, we add an additional Gaussian component to model $\hei$ $\lambda5877$, for which the line width and velocity are fixed to that of other lines determined by pPXF (Section~\ref{sec:ppxf}). We apply the Bayesian information criterion (BIC) method to assess whether $\hei$ $\lambda5877$ is present. If the difference of BIC between the model without and with $\hei$ emission exceeds 10 ($\Delta \text{BIC} = \text{BIC}_{\text{w/o\,HeI}} - \text{BIC}_{\text{w/\,HeI}}>10$), we consider   $\hei$ to be detected. 
To this end, for the model with $\hei$, one more parameter, $F_{\hei}$, is added.

Finally, we run Markov Chain Monte Carlo (MCMC) to fit the model with \texttt{emcee} to estimate the parameter uncertainties and account for the well-known degeneracy between optical depth and covering fraction in blended Na D absorption \citep{Rupke2005a,Davies2024}. Full details of the model setup, fitting procedure, and validation are given in \citet{Sun2026}.

\subsection{Na D ISM detection and flow classification}\label{sec:detect&class}

To quantify the amount of Na D ISM absorption and evaluate the confidence of the Na D ISM detection, we measure the equivalent width ($\mathrm{EW_{NaD, ISM}}$, where positive means absorption and negative means emission), and the SNR of the Na D ISM absorption minimum relative to the mean noise of the adjacent continuum ($\mathrm{SNR_{NaD, ISM}}$). We define the ``robust'' Na D ISM absorption detected galaxies as: (1) 
$\mathrm{EW_{NaD, ISM}} -2\sigma(\mathrm{EW_{NaD, ISM}})>0$ and (2) $\mathrm{SNR_{NaD, ISM}}$ $>$ 2. We also classify galaxies as ``tentative" Na D ISM detected if $\mathrm{SNR_{NaD,ISM}}$ or the $\mathrm{EW_{NaD,ISM}}$ significance levels are slightly lower then the robust threshold ( $1.5-2\sigma$). We then perform a false-positive test by fitting any Na D-like absorption in the continuum regions on both sides of the Na D wavelength, where no strong absorption features are expected. We exclude galaxies as possible false Na D detections if either continuum region contains a spurious Na D-like feature that also satisfies our selection thresholds. Finally we do visual inspection across all the selected Na D detections.

After this procedure, 63 galaxies have a robust Na D ISM absorption detection, including 12 from SMILES, 27 from JADES, 19 from Blue Jay, and 5 from Aurora. We additionally identify 10 galaxies with ``tentative'' Na D ISM detections. However, to be most conservative and ensure the result robustness, we only list those tentative detections in Table~\ref{Tab:tentative} and show their locations in Figure~\ref{fig:MS}, while \textit{exclude} them from the subsequent analysis.

We note that, \citet{Davies2024} reported 30 Na D absorption detections, while we only detected 21 (19 robust and 2 tentative). We find three of them does meet our requirement of Na D-surrounding continuum detection $\mathrm{SNR_{cont}>5}$, and another six have low Na D absorption significance. Some of this discrepancy might also be due to differences in data reduction between \citet{Davies2024} and DJA.

Among the 63 galaxies with robust Na D ISM absorption detection, we find that four are broad-line (BL) AGN reported by previous works \citep{Juodzbalis2026,Bugiani2025}. Given that, for bright BL AGN, the spectroscopic stellar population synthesis and Na D profile modeling can become unreliable when the spectrum is dominated by strong AGN power-law emission and broad-line features, we exclude them from the following analysis but only show their locations in Figure~\ref{fig:MS}. 

Depending on the geometry, it is also possible to see Na D emission rather than absorption, when absorbed photons are re-emitted along the line of sight due to resonant scattering  (\citealt{Prochaska2011}). Motivated by this possibility, we additionally fit a P-Cygni model, consisting of a blueshifted absorption component plus a redshifted emission component, to all galaxies with Na D ISM absorption detections. Again, we compare this model with our fiducial absorption-only model using the BIC. None of the Na D absorption detections satisfies $BIC_{abs,only}-BIC_{P-C}>10$, indicating that none shows a significant preference for the P-Cygni model. However, We indeed detect a pure Na D emission in one galaxy (SMILES-196290), which is identified as a BL AGN \citep{Sun2025} with strong extended ionized emission \citep{Zhu2025}. Therefore, the Na D emission might be attributed to the Na D-traced ISM ionized by the  strong AGN radiation, which thus also indicates the presence of Na D ISM in SMILES-196290. However, discussion of the Na D emission is beyond the scope of this work; therefore, in the following analysis, we focus only on galaxies with Na D ISM absorption.

To better quantify the measurement uncertainty of the Na D ISM profile of the 59 remaining robust Na D-ISM detected galaxies (four BL AGN are removed), we perform a Monte Carlo evaluation, i.e., generating 100 mock spectra based on the observed flux errors and rerun the pPXF and Na D ISM profile fitting described in Sections~\ref{sec:ppxf} and \ref{sec:nad_profile}. The best-fit Na D profile parameters and their 1$\sigma$ upper and lower errors are derived from the combined 100 MCMC posterior distributions. From this process, we are able to classify the flow direction of Na D ISM in those galaxies as: outflow if the 68th confidence interval of $\Delta V$ is negative; inflow if the 68th confidence interval of $\Delta V$ is positive; systemic, for all other cases. By this definition, we identify 25 Na D outflows, 8 inflows, and 26 systemic cases. We notice that three outflows identified by \citet{Davies2024} (BJ-10565, 11494, and 10021) are classified as systemic ISM based on our analysis. Such a kinematic discrepancy could be due to moderate spectral resolution (R$\sim$1000) and slightly different data reduction procedures. Therefore, we do not include those three galaxies in our Na D outflow census in Section~\ref{sec:nad_outf}. We confirm that our scientific conclusion will not change even if we include them in the analysis.

Finally, the best-fit Na D ISM profile of all the 59 ``robust" and 10 ``tentative" Na D ISM absorption-detected galaxies is displayed in Appendix~\ref{ap:Na D gallery}. The derived Na D profile parameters, as well as host galaxy properties, for the Na D inflows, systemic ISM, and outflows are shown in Appendix Table~\ref{Tab:inf}, \ref{Tab:sys} and \ref{Tab:outf}, respectively. 

\subsection{SED modeling}
\label{sec:SED}
We fit the galaxy photometry with the Prospector code \citep{Johnson2021}.
Our assumptions for the SEDs, including parameter priors, are similar to those used in \citet{Ji2024}. In brief,
we fix the redshift to the spectroscopic redshift modeled by pPXF (Section~\ref{sec:ppxf}). We use the Kroupa stellar IMF \citep{Kroupa2001}, and adopt the FSPS stellar synthesis code \citep{Conroy2009,Conroy2010} with the stellar isochrone libraries MIST \citep{Choi2016}
and the MILES stellar spectral libraries \citep{Sanchez-Blazquez2006,Falcon-Barroso2011}. We use the Madau intergalactic material (IGM) transmission model \citep{Madau1995}. We include the nebular emission model from \citet{Byler2017}. We leave the stellar metallicity ($Z_*$), the gas metallicity($Z_{\text{gas}}$), and the ionization parameter (U) as free parameters. We use a two-component dust attenuation model, where the attenuations toward nebular emission and young ($<10$ Myr) stellar
populations are modeled by a power law, and those towards old ($>10$ Myr) stellar populations are treated following the
parameterization of \citet{Noll2009}, a modified \citet{Calzetti2000} law with the 2175\AA ~dust feature. We also include active AGN dust torus templates from \citet{Nenkova2008a} and \citet{Nenkova2008b} to account for AGN contributions to the  galaxy SED.

Our default model adopts a nonparametric star formation history (SFH) with nine lookback time bins and the bursty continuity prior \citep{Leja2017,Tacchella2022}. The first two bins are fixed to be 0-30 and 30-100 Myr  to capture recent star formation activity. The final bin spans 
$0.85t_{\rm H} - t_{\rm H}$ where $t_H$ is the Hubble time at the representative redshift. The remaining six bins are evenly spaced in logarithmic time between 100 Myr and $0.85 t_{\rm H}$.

As a consistency check on our SED modeling, we compare the SFRs averaged over the most recent 30 Myr inferred from photometry-based SED fitting with slit-loss and dust-corrected H$\alpha$-based SFR (see Appendix~\ref{ap:SFR_comp}) for the subsample of galaxies with reliable H$\alpha$ detections and no evidence of ongoing AGN activity (i.e., predominantly star-forming galaxies). The two SFR estimates are broadly consistent in the median, with a median offset of only $\sim0.1$ dex, although there is substantial object-to-object scatter of $\sim0.7$ dex.

Secondly, we check the SED-inferred SFR for the quiescent galaxies, which are identified as having a position 0.5 dex below the main sequence as inferred from their SED-based SFR  (see details in Section~\ref{sec:nad_ism_census}). We cross-match them with the sub-millimeter source catalogs, including ASPECS-LP \citep{Aravena2020} and A3GOODS \citep{Adscheid2024} in GOODS-S, Super GOODS \citep{Cowie2017} in GOODS-N, and A3COSMOS \citep{Liu2019} in COSMOS, to see if any of them may have significant dust-obscured star formation. While $\sim10$ quiescent galaxies have sub-millimeter detections, we find, in most of the cases, those galaxies have AGNs; for the rest without AGNs, the SFRs derived from SED modeling including sub-millimeter data are not significantly larger than our estimate with only UV-to-NIR photometry and in no case do they shift these galaxies back onto the main sequence. Only one galaxy, SMILES-207739 at z$=$1.04, shows a notable difference: our SED fitting based on UV-to-NIR data alone returns an almost zero SFR, whereas including the sub-millimeter data increases the inferred SFR to $18~M_\odot\,\mathrm{yr}^{-1}$ \citep{Aravena2020}. However, even with this higher SFR, the galaxy still lies below the main sequence by about 0.5 dex. 

Our tests therefore indicate that our UV-to-NIR SED modeling yields robust SFR estimates for the population analysis.
In the following analysis, we adopt the SED-inferred SFR over the past 30 Myr as the fiducial SFR.

\section{Results} \label{sec:result}

\subsection{The incidence of Na D ISM at $0.6<z<4$}
\label{sec:nad_ism_census}

In this section, we investigate the incidence of Na D ISM as functions of star-formation properties (Section~\ref{sec:nad_rate}), dust attenuation (Section~\ref{sec:nad_dust}), star formation history (Section~\ref{sec:nad_sfh}) and the presence of AGN (Section~\ref{sec:BPT}).

\subsubsection{Dependence on star-formation properties}
\label{sec:nad_rate}

\begin{figure*}
\centering
\includegraphics[width=0.7\textwidth]{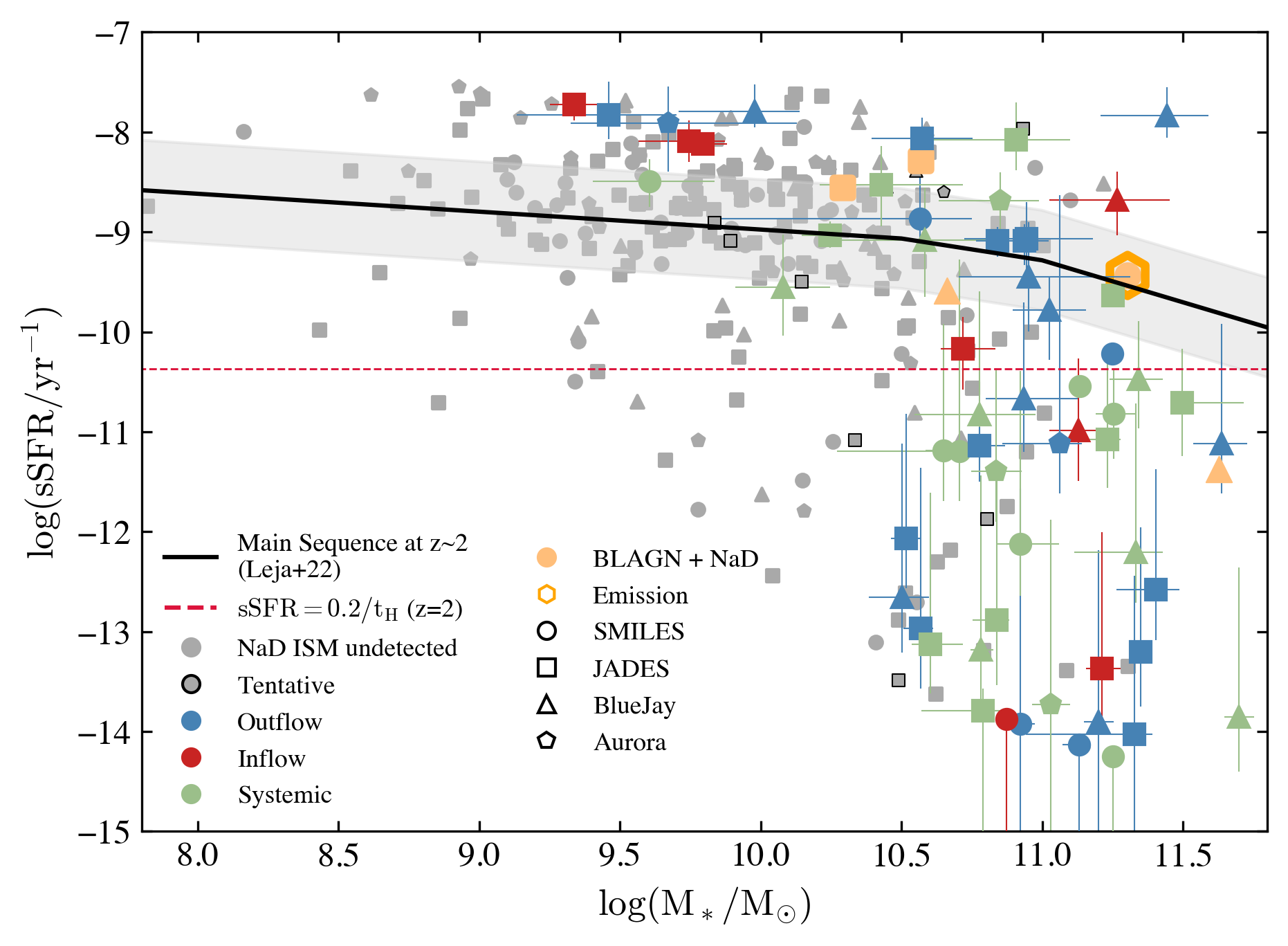}
\caption{SED-derived SFR versus galaxy stellar mass diagram. Galaxy sample from SMILES, JADES, Blue Jay, and Aurora are marked by circles, squares, triangles, and pentagons, respectively. Naked gray points are galaxies without Na D ISM detection, and those with black outline are galaxies with ``tentative" detection. Na D outflow, inflow, and systemic ISM are represented by blue, red, and green colors, respectively. Five broad-line AGN with Na D ISM detections are shown in yellow; one of them -- SMILES-196290, which exhibits Na D emission -- is highlighted with an orange hexagon outline. The black line with gray-shaded regions shows the star-forming main sequence (SFMS) at z$\sim$2 from \citet{Leja2022}, with a scatter of 0.5 dex. The massive quiescent galaxies selected by $\Delta \text{MS}<-0.5$ dex are also generally located below $\mathrm{sSFR} < 0.2/t_{H}$ at z$\sim$2 (red dashed line), another commonly used quiescent selection criterion \citep{Carnall2018,Baker2025b}.}
\label{fig:MS}
\end{figure*}

\begin{figure*}
\centering
\includegraphics[width=1\textwidth]{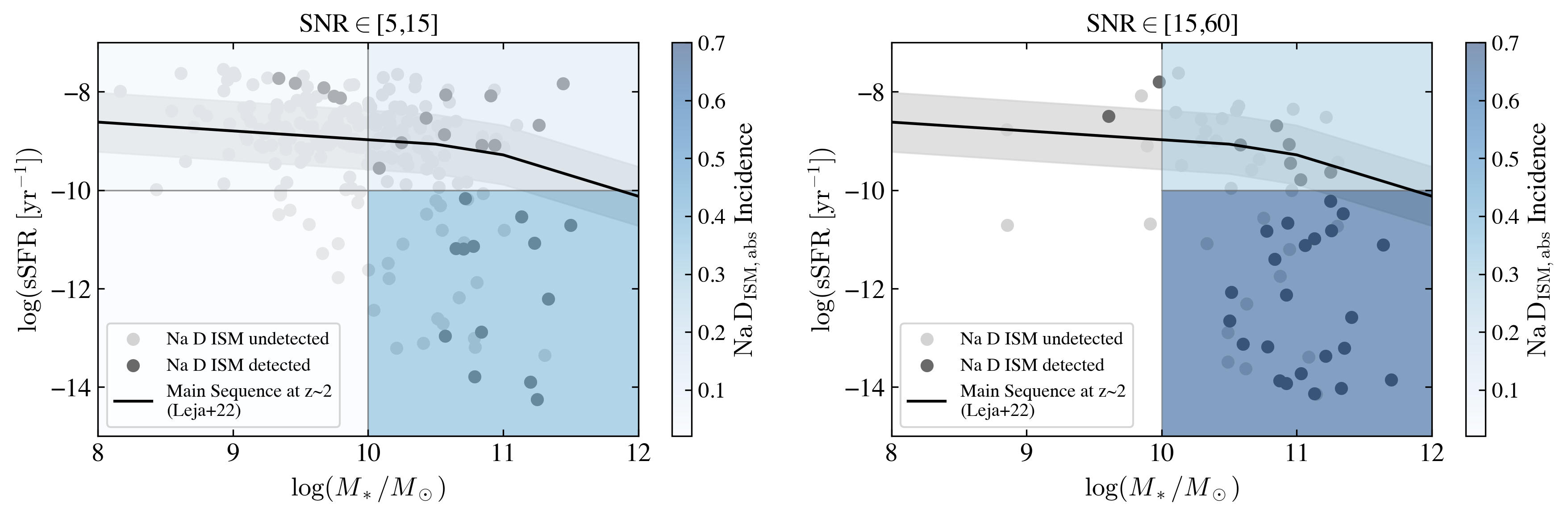}
\caption{The Na D ISM incidence across the $M_*$-sSFR plane. Galaxies with low-SNR and high-SNR continuum detections around Na D are shown in the left and right panels, respectively. The light-gray and dark-gray points represent the galaxies with and without Na D ISM detection, respectively. The black line with gray shaded regions shows the star-forming main sequence (SFMS) at z$\sim$2, same as the line shown in Figure~\ref{fig:MS}. The whole $M_*$-sSFR plane is split into four regions by $\mathrm{\log(M_*/M_\odot)=10}$ and $\mathrm{\log(sSFR[yr^{-1}])=-10}$, and then color-coded by the Na D ISM incidence within the region.}
\label{fig:NaD_incidence}
\end{figure*}

Figure~\ref{fig:MS} shows the distribution of the sample galaxies on the $M_*$--sSFR diagram. We find that most Na D ISM (52/59, 88\%) are detected in galaxies with $\log(M_*/M_\odot) > 10$, spanning the full range from above to below the star-forming main sequence. In contrast, at the lower-mass end ($\log(M_*/M_\odot) \leq 10$), Na D ISM detections are found only in galaxies on or above the main sequence. We divide our sample galaxies into four subgroups based on their locations on the $M_*$--sSFR plane, separated by $\mathrm{\log(M_*/M_\odot)=10}$ and $\mathrm{\log(sSFR/yr^{-1})=-10}$. To control for observational bias in the detectability of Na D absorption, which depends strongly on the continuum SNR around Na D, ${\rm SNR}_{\rm Na\,D,cont}$, we further divide the sample into two subsamples at ${\rm SNR}_{\rm Na\,D,cont}=15$, above which we find the detection rate approximately plateaus.

As Figure~\ref{fig:NaD_incidence} shows, in the low-SNR subsample, the Na D detection rate is 37$\pm$8\% in the high-$M_*$ \& low-sSFR region and 12$\pm$4\% in the high-$M_*$ \& high-sSFR region, while much lower in the ``low-$M_*$ \& high sSFR" region (5$\pm$2\%). A qualitatively similar pattern is seen in the high-SNR subsample, where the Na D detection rate is highest in the high-$M_*$ \& low-sSFR region (66$\pm$8\%) and decreases to 26$\pm$9\% in the high-$M_*$ \& high-sSFR region. In the low-$M_*$ subgroups, the sample sizes are too small to constrain the incidence. The similar incidence pattern in the high-$M_*$ regions across the high- and low-SNR subsamples suggests that the dependence of Na D detection on $M_*$ and sSFR is not driven by continuum-SNR-related observational bias. Instead, these results indicate that Na D-traced neutral ISM is much more prevalent in galaxies with high $M_*$, and that -- among massive galaxies -- the Na D detection rate is higher below the main sequence than along it. 

\begin{figure*}[t]
\centering
\includegraphics[width=0.9\textwidth]{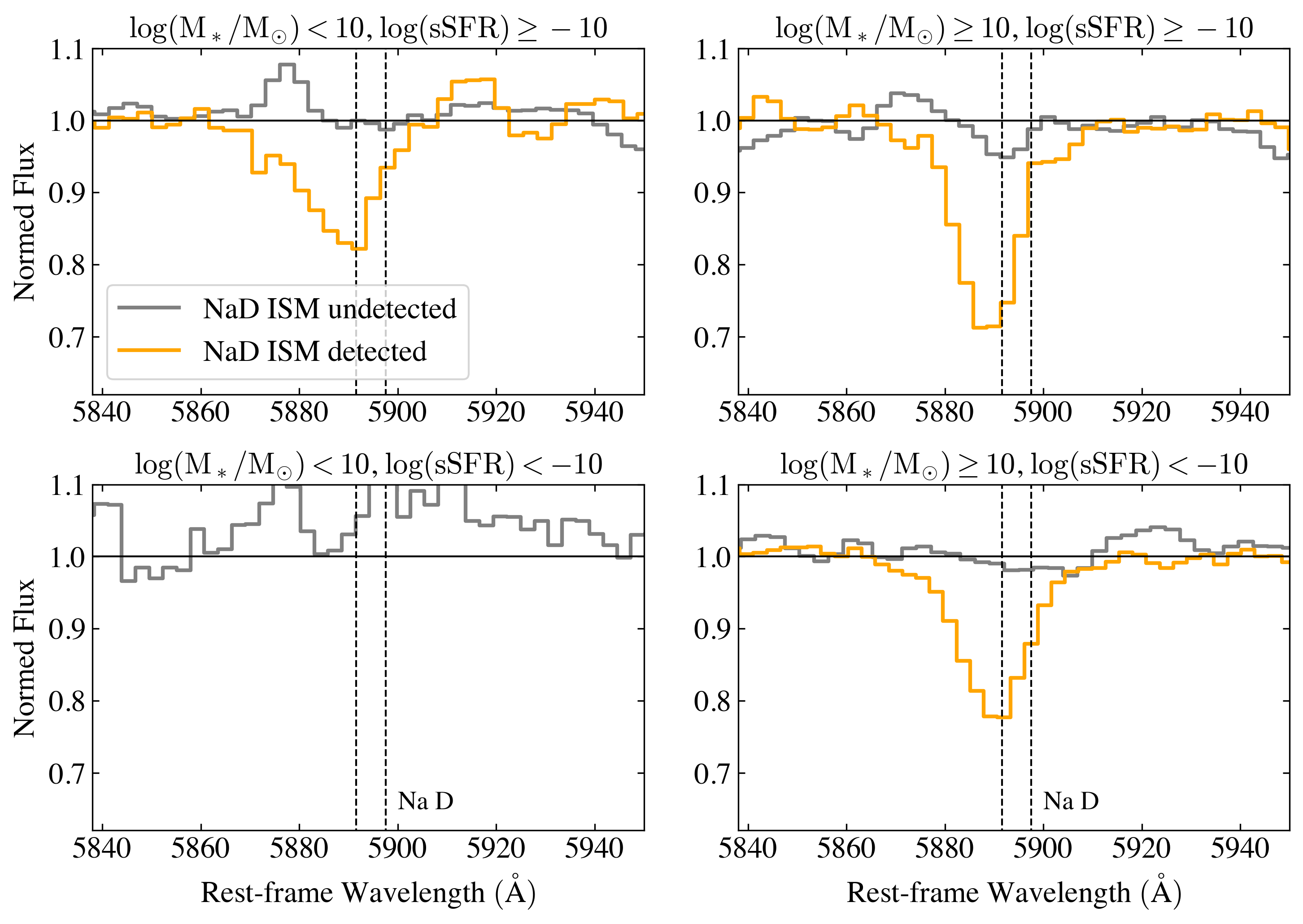}
\caption{The stacked Na D residual spectrum (stellar component removed) of galaxies at four different $M_*$-sSFR regions. The spectra of Na D ISM-detected and undetected galaxies are shown as orange and gray lines, respectively.}
\label{fig:NaD_stacked}
\end{figure*}

We also perform a spectral stacking analysis for galaxies with and without Na D detections. Each galaxy spectrum is first normalized by its pPXF-derived stellar continuum (Section~\ref{sec:ppxf}). When present, the nearby \hei\ $\lambda\lambda5877$ emission is subtracted before stacking using the best-fit model from Section~\ref{sec:nad_profile}. We then construct the stacked Na D spectra using inverse-variance weighting. We have verified that median stacking yields the same qualitative conclusions.

Figure~\ref{fig:NaD_stacked} shows the stacked spectra of Na D-detected and undetected galaxies in the four subgroups. Overall, the stacked spectra of Na D-undetected galaxies do not show clear Na D ISM absorption, except in the high-$M_*$ and high-sSFR subgroup, where the stacked spectrum of 77 galaxies shows tentative Na D ISM absorption at $\sim 2\sigma$, above the threshold we adopted for defining ``tentative" detection ($1.5\sigma$) while not across the ``robust" threshold ($2\sigma$) significantly. These results indicate that our measured Na D ISM incidence rates are broadly robust, with no clear evidence that galaxies with strong Na D ISM absorption are being missed by our analysis.

Taking a closer look at the stacked line profiles of the Na~D ISM detected galaxies, we find that the two high-$M_*$ subgroups show deeper absorption than the ``low-$M_*$ \& high-sSFR'' subgroup, suggesting that the neutral ISM is more abundant in massive galaxies than in low-mass galaxies. Moreover, between the two high-$M_*$ subgroups, both stacked profiles are blueshifted, but the profile of the high-sSFR subgroup (FWHM$\mathrm{=490\pm33\,km\,s^{-1}}$) is narrower than that of the low-sSFR subgroup (FWHM$\mathrm{=641\pm14\,km\,s^{-1}}$). 

The narrower Na D absorption profile seen in the ``high-$M_*$ \& high-sSFR'' subgroup can be explained in two ways. First, the Na D ISM absorptions in this subgroup may be more predominantly systemic or blueshifted than those in the ``high-$M_*$ \& low-sSFR'' subgroup. This interpretation is supported by our observations: we find only one inflow in the ``high-$M_*$ \& high-sSFR'' subgroup, whereas 4 out of 36 ``robust" Na D ISM detections in the ``high-$M_*$ \& low-sSFR'' subgroup are redshifted. These redshifted absorbers therefore contribute substantially to the red absorption wing of the stacked profile in the ``high-$M_*$ \& low-sSFR'' subgroup, making the absorption appear broader. Alternatively, the spectra considered here have only moderate resolution. If redshifted Na D emission is present at a non-negligible level, it may partially fill in an intrinsically broad absorption profile, causing the observed Na D profile to appear narrower. This second scenario is also plausible, as \citet{Concas2019} showed that the stacked Na D profile of massive star-forming galaxies in the local Universe can be reproduced only by including a redshifted emission component in addition to blueshifted Na D absorption.

\subsubsection{Dependence on dust}
\label{sec:nad_dust}

\begin{figure}
\centering
\includegraphics[width=\columnwidth]{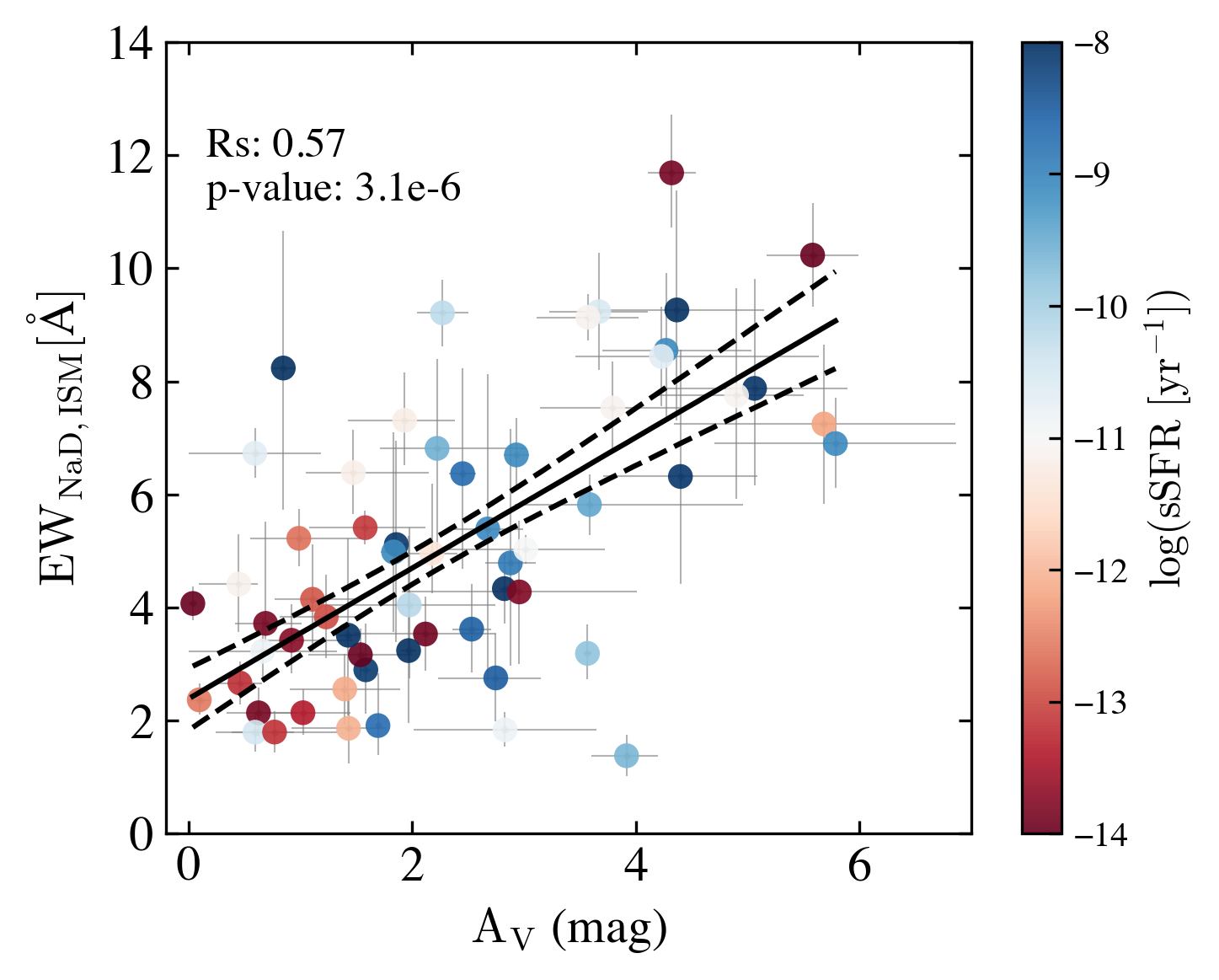}
\caption{Equivalent width of the Na D ISM absorption component, $\mathrm{EW}_{\mathrm{NaD,ISM}}$, as a function of $A_V$. The points are color-coded by galaxies' sSFR. The solid line shows the best-fit log-linear relation and the dashed lines show the 90\% confidence intervals. The strong correlation between $\mathrm{EW}_{\mathrm{NaD,ISM}}$ and $A_V$ is confirmed by a Spearman rank test with $R_s=0.57$ and $p\ll0.05$.}
\label{fig:ISM_AV_trend}
\end{figure}

We now examine how the incidence of Na D ISM absorption depends on dust attenuation, a physical property that has been found to govern the amount of Na D-traced neutral gas in the local Universe \citep[e.g.,][]{Chen2010, Poznanski2012}.

Considering the full sample, the median $A_V$ of the Na D ISM-detected galaxies is higher than that of the undetected ones, with median values of 2 and 1 mag, respectively. For the Na D ISM-detected subsample, we test the correlation between $A_V$ and the equivalent width of the Na D ISM absorption, $\mathrm{EW}_\mathrm{NaD, ISM}$. As shown in Figure~\ref{fig:ISM_AV_trend}, we find a strong positive correlation between these two quantities. A Spearman rank test confirms this correlation to be statistically significant with $p \ll 0.05$. These results indicate that the presence of Na D-traced ISM strongly depends on dust attenuation.

\begin{figure*}
\centering
\includegraphics[width=1\textwidth]{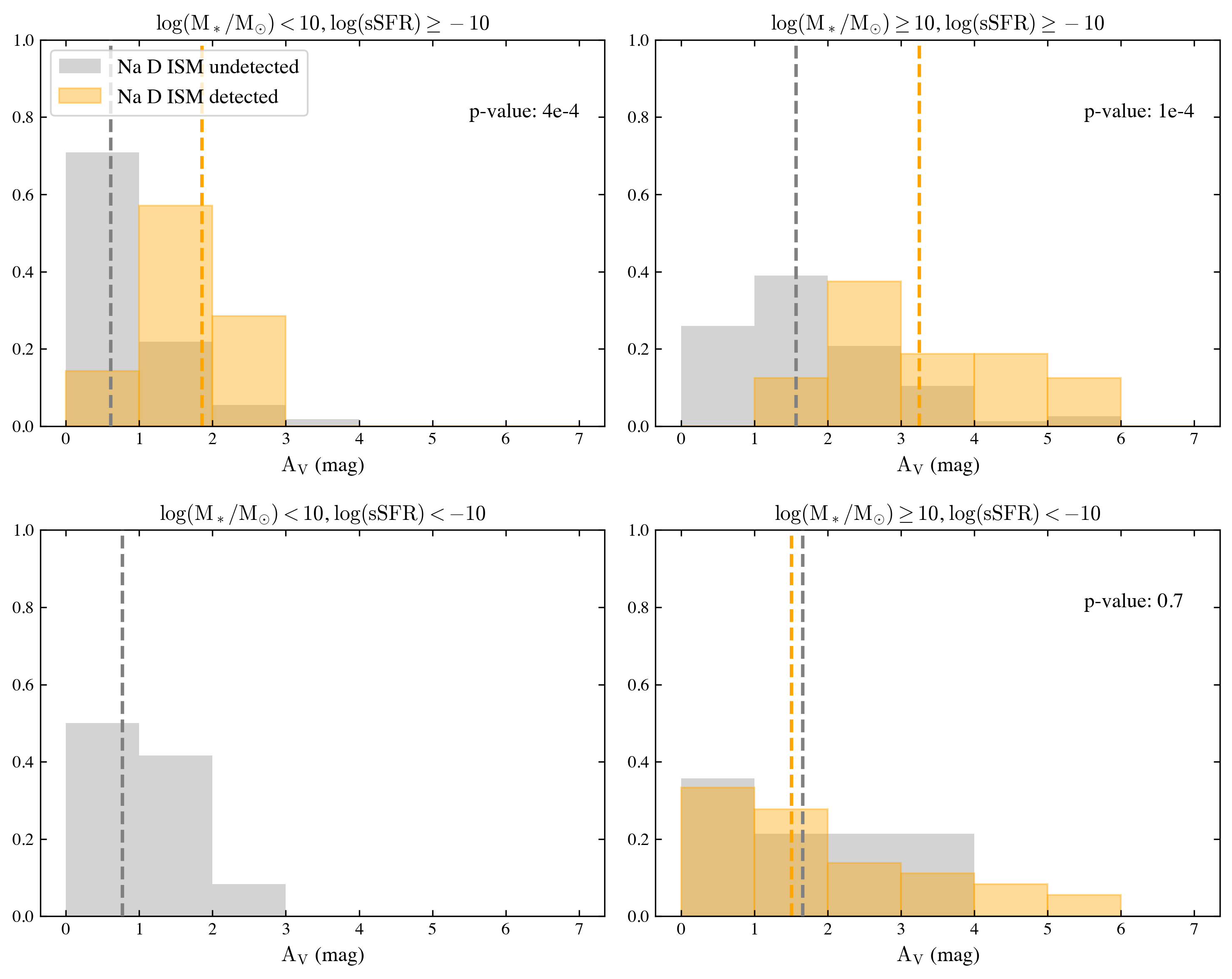}
\caption{The $A_V$ distribution of Na D ISM detected (orange) and undetected (gray) galaxies at four different $M_*$-sSFR regions. The median of each distribution is marked by a dashed line. The p-value shown in each panel represents the significance of the K-S test for the difference between the two distributions.}
\label{fig:NaD_AV_comp}
\end{figure*}

In Figure~\ref{fig:NaD_AV_comp}, we further compare the $A_V$ distributions of the four subgroups across different regions of the $M_*$--sSFR plane, where we also perform K-S tests on the $A_V$ distributions of the detected and undetected subsamples. For the two high-sSFR subgroups, regardless of stellar mass, the median $A_V$ of the Na D ISM-detected galaxies is higher than that of the undetected galaxies by about 1 mag. In both cases, the K-S test returns $p<0.05$, indicating that the two $A_V$ distributions are different. These results suggest that dust attenuation is a key factor governing the detectability of Na D ISM in star-forming galaxies. This can be understood because Na D has a low ionization potential of 5.14 eV and requires substantial dust shielding to survive radiation from intense star-formation or AGN activity.

In contrast to the high-sSFR subgroups, the $A_V$ distributions of the Na D ISM-detected and undetected galaxies in the ``high-$M_*$ \& low-sSFR subgroup" are similar. The median $A_V$ values are 1.7 and 1.5 mag for the detected and undetected subsamples, respectively, and the K-S test indicates that the two distributions are statistically indistinguishable. In addition, the $A_V$ of Na D-detected massive quiescent galaxies is also comparable to that of Na D ISM-undetected massive star-forming galaxies (see the right column of Figure~\ref{fig:NaD_AV_comp}). Taken together, these results suggest that, unlike in star-forming galaxies, the detectability of Na D ISM absorption in massive quiescent systems is not primarily controlled by dust attenuation, but instead by other physical factors that we examine in the next section (Section~\ref{sec:nad_sfh}). It is also worth noting that, although the ``high-$M_*$ \& low-sSFR" subgroup generally has lower $A_V$ than the ``high-$M_*$ \& high-sSFR" subgroup, its $A_V$ distribution exhibits a long tail toward high values. This suggests that a non-negligible fraction of cosmic-noon quiescent galaxies still retain substantial amounts of dust, consistent with recent studies reporting dusty quiescent systems in the early Universe \citep{Siegel2025,Ji2024b}.

\subsubsection{Dependence on star formation history}
\label{sec:nad_sfh}

\begin{figure*}
\centering
\includegraphics[width=0.9\textwidth]{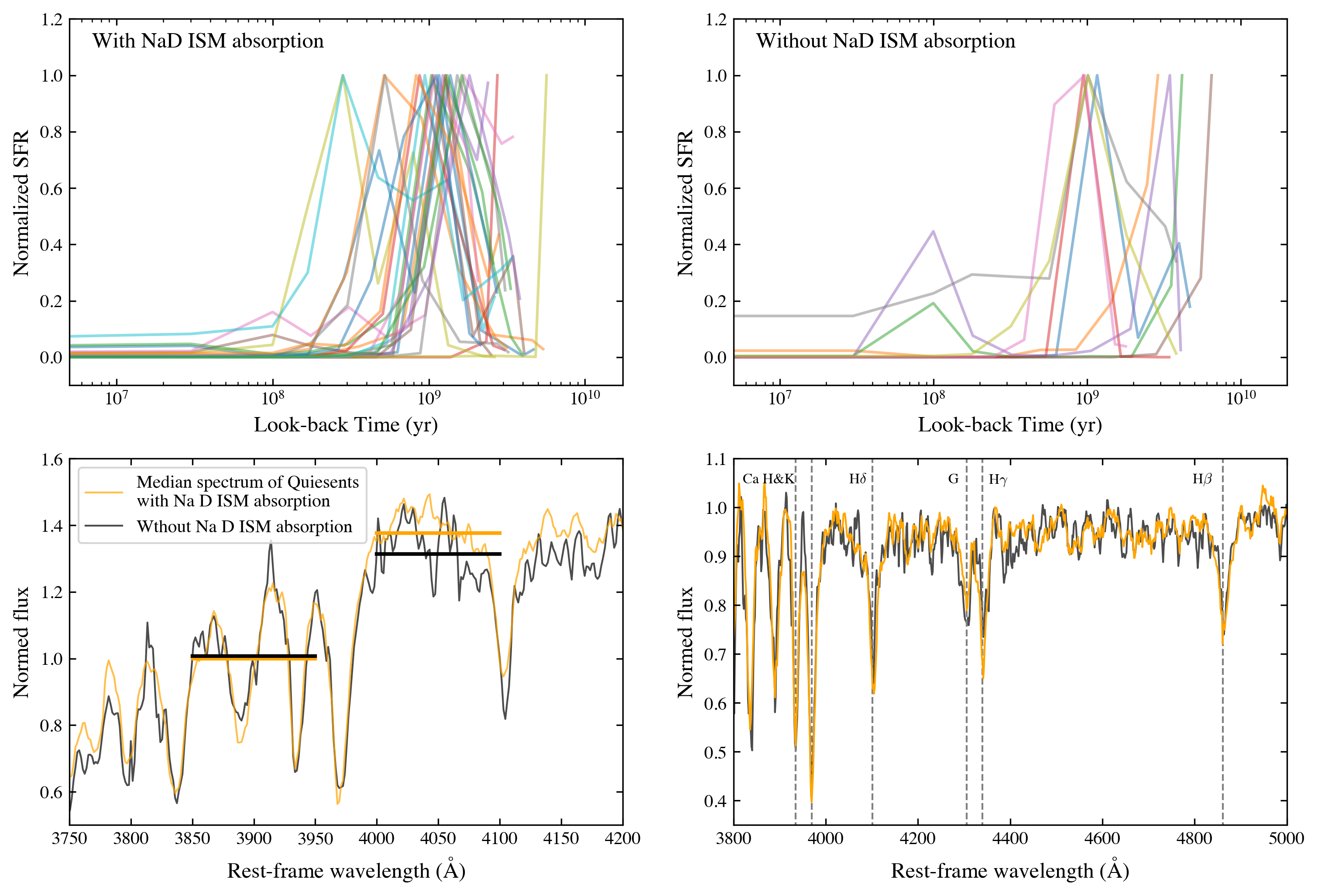}
\caption{Comparison of stellar population properties between Na D ISM absorption detected (left) and undetected (right) massive quiescent galaxies. {\bf Top:} Reconstructed non-parametric SFH of each individual galaxy. Each SFH is normalized to its own maximum SFR across the nine time bins. {\bf Bottom Left:} Medium observed spectra around 4000\AA~break of Na D ISM absorption detected (orange) and undetected (black) massive quiescent galaxies. The horizontal lines represent the average fluxes within the two continuum windows (3850--3950\AA~and 4000--4100\AA) for $D_n4000$ calculation. Since each spectrum is normalized to its average continuum flux on rest-frame 3850--3950\AA before median-average, a larger median flux difference between the two continuum windows means a stronger 4000\AA~break. {\bf Bottom Right:} Median spectra of H$\delta$-to-H$\beta$ absorptions of Na D ISM absorption detected (orange) and undetected (black) massive quiescent galaxies. Each individual spectrum is normalized by a smoothing-spline fit before calculating median average. The dashed grey lines mark the locations of Ca H\&K, H$\delta$, G-band, H$\gamma$, and H$\beta$. Overall, Na D–detected quiescents span both older systems with stronger 4000\AA~breaks, Ca H\&K, and G-band features, and younger, rapidly quenched (PSB-like) systems with stronger Balmer absorption.}
\label{fig:SFH_NaD_detect_QG}
\end{figure*}

\begin{figure}
\includegraphics[width=1\columnwidth]{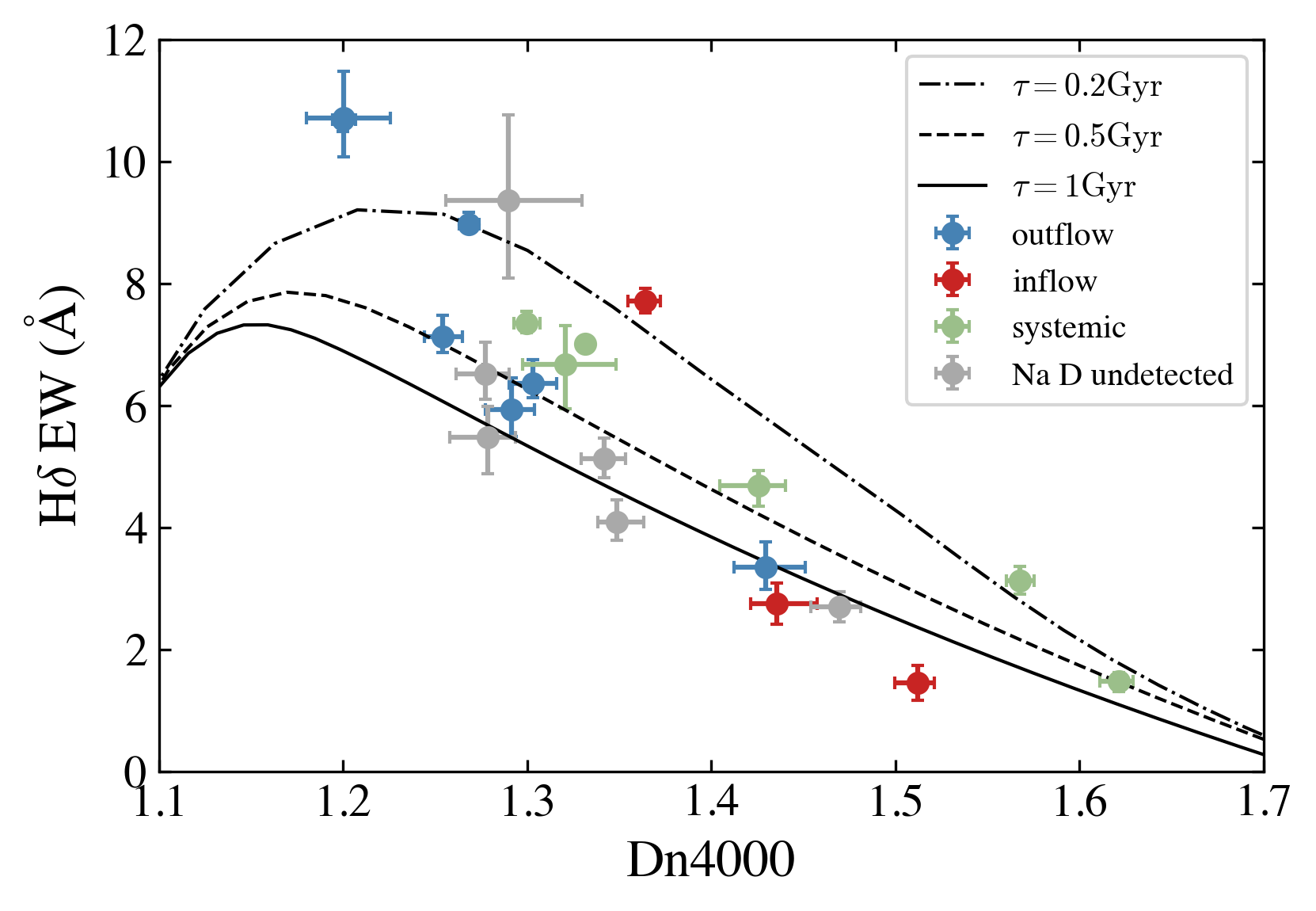}
\caption{Massive quiescent galaxies in the $H\delta_A$ vs. $D_n4000$ diagram. The symbols are the same as Figure~\ref{fig:MS}. The black reference lines from top to bottom are tau-model SFH with $\tau=$0.2, 0.5, and 1 Gyr, respectively. The Na D ISM is preferentially detected in either old quiescent galaxies with $D_n4000>1.4$, or in young, rapid-quenching ($\tau=$0.2 and 0.5 Gyr) galaxies with larger $H\delta_A$ and behaves as an outflow.}
\label{fig:QG_burst_NaD}
\end{figure}

Since dust attenuation alone does not explain the presence of Na D ISM absorption in the ``high-$M_*$ \& low-sSFR" subgroup, we next examine whether quiescent galaxies with and without Na D detections differ in their stellar population properties. We begin with the spectral differences, which are more directly observed and therefore less model dependent, and then interpret them in the context of  star-formation history reconstructed from SED fitting (Section \ref{sec:SED}).

To enable a reliable comparison of age-sensitive features, we restrict the sample to galaxies in the ``high-$M_*$ \& low-sSFR'' group with $\mathrm{SNR}>15$. 
The middle and bottom rows of Figure~\ref{fig:SFH_NaD_detect_QG} show that, on average, the Na D-detected galaxies have a stronger 4000~\AA\ break, as well as stronger Ca H\&K and G-band absorption, than the undetected subsample. These features indicate a larger contribution from evolved stellar populations. At the same time, the Na D-detected galaxies also tend to exhibit stronger Balmer absorption. This combination suggests that the Na D-detected population is not uniform: it includes both older systems with prominent metal-line absorption and systems with signatures of relatively recent, rapid quenching.

We quantify these trends using the $D_n4000$ and $H\delta_A$ indices, following \citet{Balogh1999} and \citet{Worthey1997}. Both indices are measured from the best-fit stellar component derived with pPXF (Section~\ref{sec:ppxf}), and their uncertainties are propagated using MCMC. Owing to wavelength-coverage and detector-gap limitations, only 22 of the 32 galaxies in this subgroup have usable measurements of both indices. Figure~\ref{fig:QG_burst_NaD} shows the distribution of these 22 galaxies on the $D_n4000$--$H\delta_A$ plane, together with FSPS evolutionary tracks for exponentially declining SFHs with $\tau = 0.2$, 0.5, and 1 Gyr.

Two trends emerge. First, Na D ISM seems to be more common among the quiescent galaxies with larger $D_n4000$: 6 of 7 (86\%) systems with $D_n4000>1.4$ show Na D absorption, compared with 10 of 15 (67\%) with $D_n4000\leq1.4$. Second, within the quiescent population with smllar $D_n4000$ ($\leq1.4$), the Na D-detected galaxies tend to have larger $H\delta_A$ than the undetected galaxies, placing them closer to tracks with more rapidly declining SFHs. By contrast, the undetected systems generally occupy regions consistent with more gradual decline.

These spectral diagnostics suggest that Na D ISM is preferentially detected in massive quiescent galaxies with two types of star-formation histories: older systems with larger $D_n4000$, and younger systems with smaller $D_n4000$ but stronger $H\delta_A$, indicative of a more rapid shutdown of star formation. This implies that Na D detectability depends not only on the time since quenching, but also on the manner in which quenching occurred.

This interpretation is broadly consistent with the reconstructed SFHs shown in Figure~\ref{fig:SFH_NaD_detect_QG}. The Na D-detected galaxies more often show an earlier decline in star formation, typically beginning $\sim100$--1000 Myr ago, whereas the undetected systems tend to show a more recent drop, usually within $\lesssim100$ Myr. The SFHs also reinforce the diversity implied by the spectral diagnostics, particularly within the detected population.

Finally, the Na D kinematics show a similar dependence on stellar population properties. Most Na D outflows (6/7) are found in the younger massive quiescent galaxies with high $H\delta_A$; and among young quiescent galaxies with Na D detections, more than half (6/10) of the absorption is tracing outflowing gas. This suggests that Na D kinematics, especially outflows, are also linked to recent quenching history. We discuss the physical origin of the dependence of Na D detectability and kinematics on dust and quenching history in Section~\ref{sec:nad_quenching_path}.

\subsubsection{Dependence on AGN presence}
\label{sec:BPT}

\begin{figure*}
\centering
\includegraphics[width=0.8\textwidth]{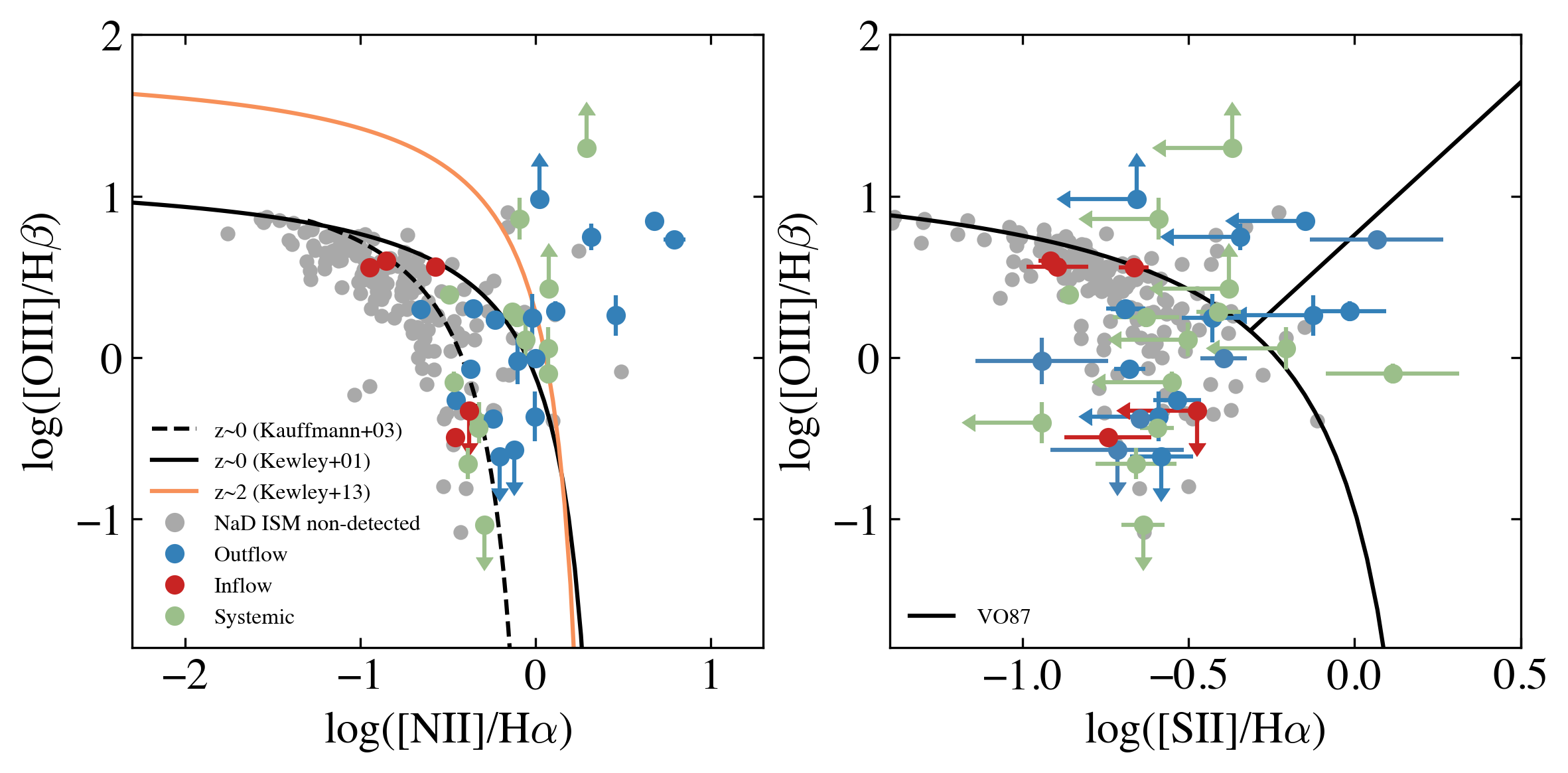}
\caption{Distribution of parent galaxies in the $[\NII]$-BPT diagram (left) and $[\SII]$-VO87 diagram (right). As in  Figure~\ref{fig:MS}, gray points are galaxies without Na D ISM detection, and Na D outflow, inflow, and systemic ISM are represented by blue, red, and green colors, respectively. The black solid, black dashed, and orange line in the left panel represents the SF-AGN boundary from \citet{Kauffmann2003}, \citet{Kewley2001}, and \citet{Kewley2013}, respectively; The black lines in the right panel are from \citet{Veilleux1987}.}
\label{fig:BPT}
\end{figure*}

We next examine whether the detectability and kinematics of Na D-traced ISM depend on AGN activity. We identify AGN primarily using optical emission-line diagnostics based on the [\NII]-BPT \citep{Baldwin1981} and [\SII]-VO87 \citep{Veilleux1987} diagrams, using the line ratios [\OIII]/H$\beta$, [\NII]/H$\alpha$, and [\SII]/H$\alpha$. Line ratios are treated as upper or lower limits when the numerator or denominator line has SNR$<3$, respectively, and galaxies with only tentative detections in both lines are excluded from the diagnostic diagrams. 

We classify a galaxy as AGN-hosting if its line ratios lie in the AGN region of either diagram. We do not classify galaxies in the ``composite'' region of the [\NII]-BPT diagram or the ``LINER'' region of the [\SII]-VO87 diagram as AGN, although these systems may still contain some AGN contribution. In addition, since AGN can also be identified through multiwavelength signatures beyond optical line-ratio diagnostics, we further supplement the optical classification with broad-line identifications \citep{Davies2024,Sun2025,Juodzbalis2026}, X-ray detections \citep{Xue2016,Luo2017}, and mid-IR AGN selections \citep{Lyu2022,Lyu2024} from the literature, thereby constructing a more complete AGN census.

Under this selection, the AGN incidence among Na D ISM-detected galaxies is 41\% (24/59), higher than the 23\% (53/227) in the undetected sample. A stronger trend emerges with Na D kinematics: the AGN incidence rises from 13\% (1/8) in the inflow subsample to 38\% (10/26) in the systemic subsample and 52\% (13/25) in the outflow subsample. This suggests that AGN activity is more strongly associated with the kinematic state of the Na D-traced ISM than with its mere detectability, particularly with the presence of outflows.  We will further examine the role of AGN in driving outflows across different galaxy populations in Section~\ref{sec:nad_outf}.

\subsection{Neutral outflow in cosmic-noon galaxies}
\label{sec:nad_outf}

In this section, we investigate the Na D outflow-detected galaxies in our sample, focusing on their properties, the mechanisms that may drive them, and their dependence on host-galaxy and AGN characteristics.

\subsubsection{Na D outflow properties}
\label{sec:nad_outf_prop}

Following \citet{Davies2024} and \citet{Sun2026}, we derive the physical properties of the Na D outflows based on the Na D profile modeling described in Section~\ref{sec:nad_profile}. These include the outflow velocity ($V_{\rm out}$), mass and mass outflow rate ($M_{\rm out}$ and $\dot{M}_{\rm out}$), momentum and momentum outflow rate ($p_{\rm out}$ and $\dot{p}_{\rm out}$), and energy and energy outflow rate ($E_{\rm out}$ and $\dot{E}_{\rm out}$). We define the outflow velocity as $V_{\rm out}=|\Delta v|+2\sigma$, where $\sigma=b/\sqrt{2}$ is the velocity dispersion of the Na D ISM line.
We caution that, however, we do not apply an inclination correction to the measured velocities because (1) the outflow geometry and inclination cannot be robustly constrained without spatially resolved IFU observations, and (2) we aim to remain consistent with other JWST-based Na D outflow studies and enable a fair comparison with their reported velocities.

We compute the mass outflow rate as
\begin{align*}
\dot{M}_{\rm out} &= \frac{M_{\text{\rm out}}v_{\text{\rm out}}}{r_{\text{\rm out}}}\\
&= 11.45\left(C_{\Omega}\frac{C_f}{0.4}\right)\left(\frac{\text{N(H I)}}{10^{21}\,\mathrm{cm^{-2}}}\right)\\
&\times\left(\frac{r_{\text{\rm out}}}{1\,\mathrm{kpc}}\right)\left(\frac{v_{\text{\rm out}}}{200\,\mathrm{km\,s^{-1}}}\right)\,\mathrm{M_{\odot}\,yr^{-1}},
\end{align*}
assuming outflow extent $r_{\rm out}=1\,\mathrm{kpc}$ and covering factor related to the wind opening angle $C_{\Omega}=0.5$. The hydrogen column density $\mathrm{N(H\,I)}$ is inferred from $\mathrm{N(Na\,I)}$ assuming Milky-Way-like Na abundance and dust depletion, together with a neutral fraction of 10\% \citep{Rupke2005a}. We derive $\mathrm{N(Na\,I)}$ from the best-fit central optical depth ($\tau_{r,0}$) of the red Na D line:
\begin{align*}
\text{N(Na I)} &= 10^{13}\left(\frac{\tau_{r, 0}}{0.758}\right)\left(\frac{0.4164}{f_{lu}}\right)\\
&\times\left(\frac{1215\mathrm{\AA}}{\lambda_{lu}}\right)\left(\frac{b}{10\,\mathrm{km\,s^{-1}}}\right)\,\mathrm{cm^{-2}},
\end{align*}
where $f_{lu}=0.32$ and $\lambda_{lu}=5897.55\,\mathrm{\AA}$.

From $M_{\rm out}$ and $\dot{M}_{\rm out}$, we further derive the momentum and energy outflow rates. The measured properties of the 25 Na D outflows are listed in Table~\ref{Tab:outf_detail}.

Uncertainties in the outflow properties are propagated from the MCMC fitting of the Na~D profile parameters. We caution that, in reality, the systematic uncertainties in the outflow mass, momentum, and energy, as well as their corresponding rates, can be substantially larger than those estimated from the MCMC analysis. These additional uncertainties arise in part from assumptions about the neutral fraction and $C_{\Omega}$ and can be as large as $\sim0.6$ dex \citep{Davies2024,Sun2026}. In addition, because the NIRSpec/MSA data probe only the central regions of the galaxies, the outflow radius, $r_{\rm out}$, cannot be directly constrained. We therefore conservatively assume an outflow extent of $r_{\rm out}=1,\mathrm{kpc}$ when calculating the outflow rates, which introduces substantial additional uncertainty into the derived values. For example, IFU data for JADES-197911 from the GA-NIFS survey\footnote{\url{https://ga-nifs.github.io/}} reveal a larger Na D outflow extent than we assumed, with $r_{\rm out}\sim2.7 \,\mathrm{kpc}$. Consequently, the inferred mass-outflow rate, $\dot{M}_{\rm out}\sim60\,\mathrm{M_\odot\,yr^{-1}}$ \citep{DEugenio2024,Scholtz2026}, is also higher compared with our estimate of $\sim16\,\mathrm{M_\odot\,yr^{-1}}$. Therefore, we emphasize that the reported mass, momentum, and energy outflow rates in this work should be regarded only as lower limits, and that more accurate measurements will require future IFU observations.

With those aforementioned systematic uncertainties in mind, we examine how $V_{\rm out}$, $\dot{M}_{\rm out}$, $\dot{E}_{\rm out}$, and mass-loading factor ($\eta=\dot{M}_{\rm out}/\mathrm{SFR}$) correlate with host-galaxy properties. For this purpose, we divide the outflow-detected sample into star-forming and quiescent subsamples based on their relative distance from the star-forming sequence, $\Delta\mathrm{MS}$: galaxies with $\Delta\mathrm{MS}>-0.5$ dex are classified as star-forming, while those with $\Delta\mathrm{MS}\leq-0.5$ dex are classified as quiescent.\footnote{We verify that this quiescent subsample is also consistent with the commonly adopted criterion $\mathrm{sSFR}<0.2/t_{\rm H}$ \citep{Franx2008,Carnall2018,Baker2025b,Ji2026}, and recover the spectroscopically confirmed quiescent galaxies in JADES \citep{Baker2025} and Blue Jay \citep{Park2024}.} The host classification is listed in Table~\ref{Tab:outf_detail}. In addition to $M_*$, SFR, and sSFR, we also consider the SFR surface density, $\Sigma_{\rm SFR}=\mathrm{SFR}/(2\pi R_{\rm eff}^2)$, and the circular velocity, $V_{\rm cir}=\sqrt{(GM_*)/(2R_{\rm eff})}$, which trace the intensity of star formation and the depth of the gravitational potential, respectively. Here, we use the $R_{\rm eff}$ measurements reported by \citet{Genin2025}, based on JWST/NIRCam imaging.

\subsubsection{Outflow velocity and mass outflow rate}
\label{sec:nad_outf_V}

\begin{figure*}
\centering
\includegraphics[width=0.7\textwidth]{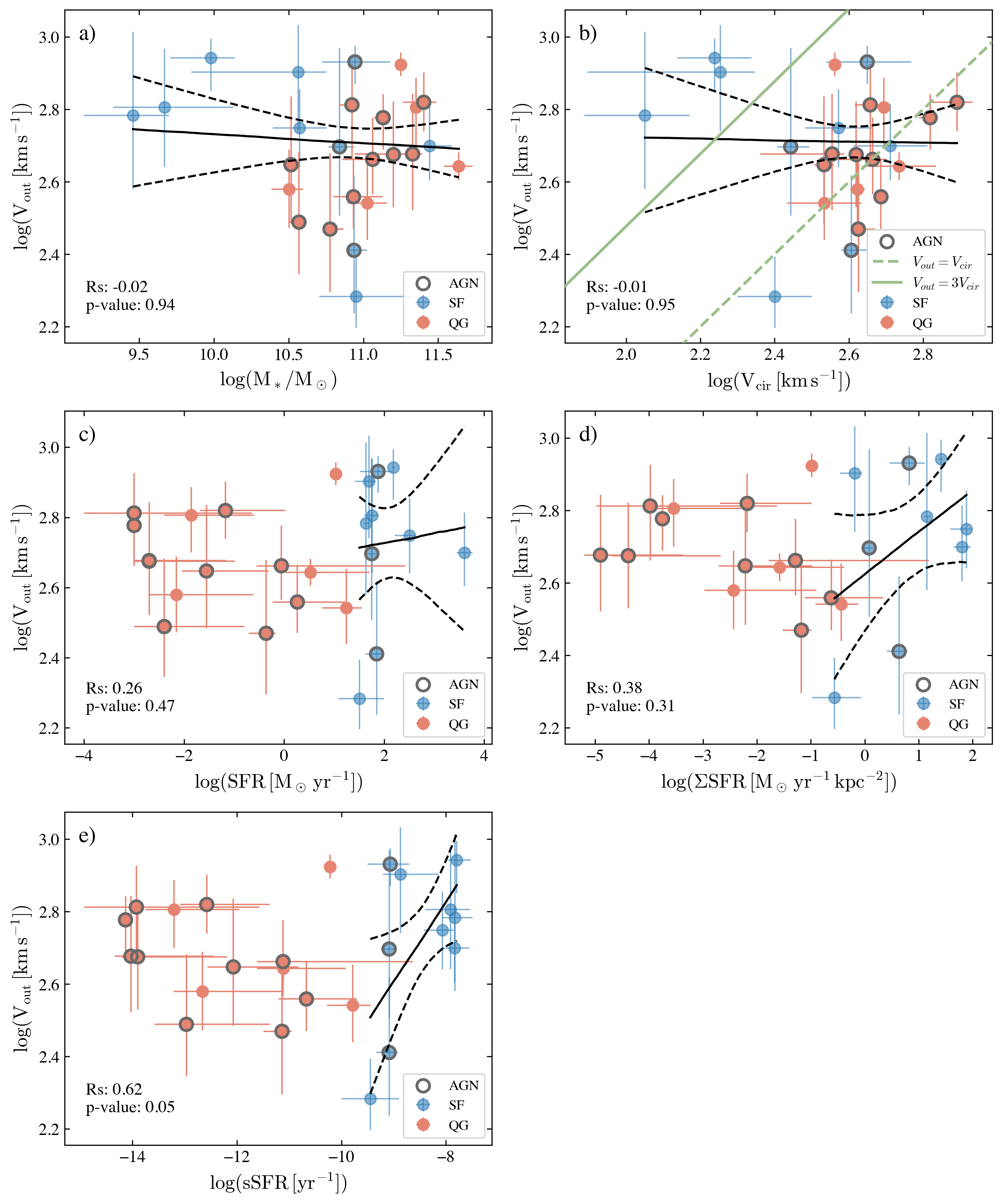}
\caption{Outflow velocity $V_{\rm out}$ as a function of $M_*$, $V_{\rm cir}$, SFR, $\Sigma$SFR, and sSFR. The Na D outflow sample is divided into the quiescent ($\Delta {\rm MS}\leq$ -0.5 dex, red) and star-forming galaxies ($\Delta {\rm MS}>$ -0.5 dex, blue). AGN hosts are outlined in grey. In each panel, the solid black line shows the best-fit log-linear relation and the dashed lines indicate the 90\% confidence interval. The regression is performed using the full outflow sample for the $M_*$, $V_{\rm cir}$ trends, while using only the star-forming subsample for the star-formation indicators (SFR, $\Sigma$SFR, and sSFR). The Spearman coefficient and the corresponding p-value are shown in each panel.}
\label{fig:Vout_trend}
\end{figure*}

\begin{figure*}[t!]
\centering
\includegraphics[width=0.7\textwidth]{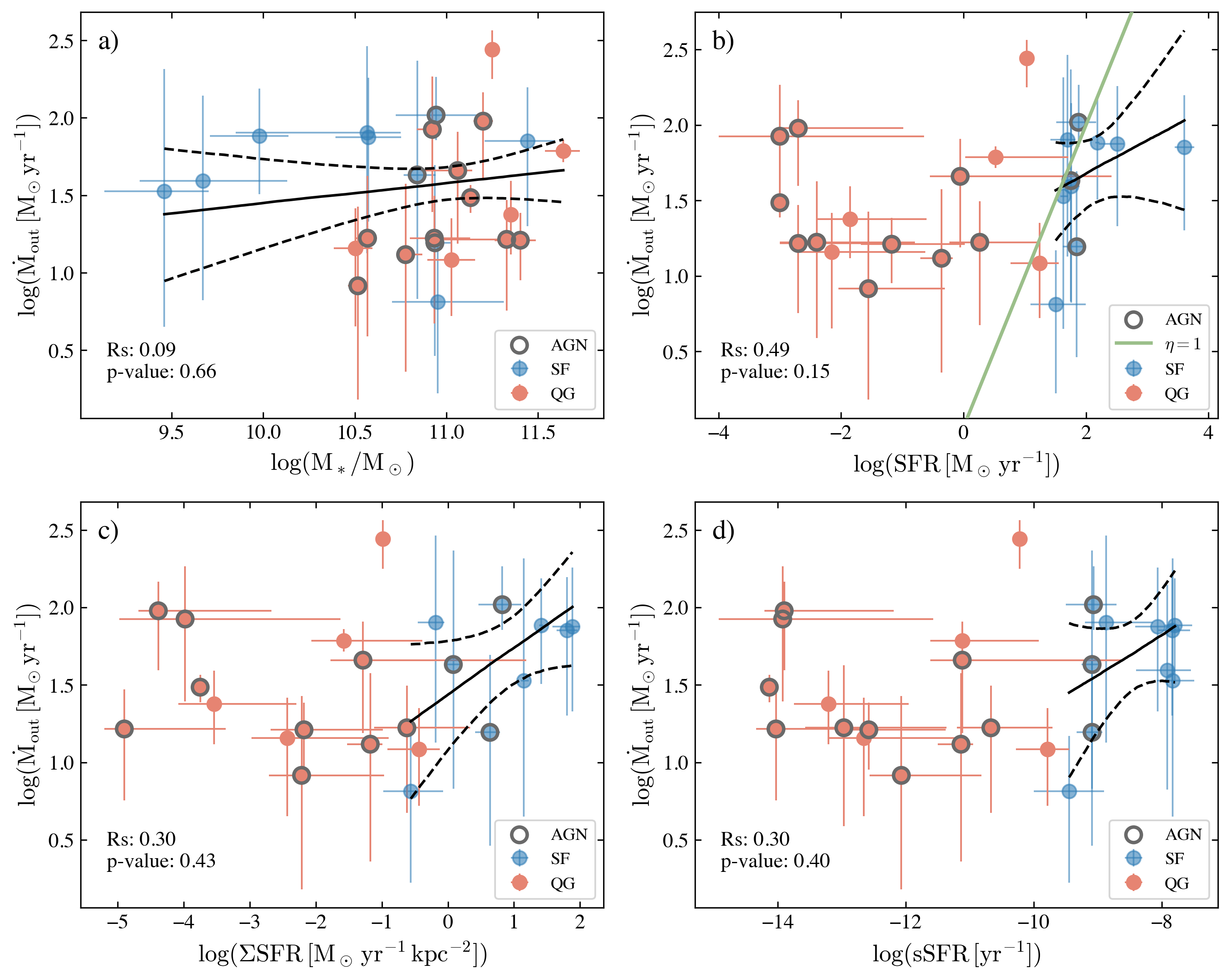}
\caption{Mass outflow rate $\dot{M}_{\rm out}$ as a function of $M_*$, SFR, $\Sigma$SFR, and sSFR. The symbols are the same as Figure~\ref{fig:Vout_trend}. Again, the regression is performed using the full outflow sample for the $M_*$ trend, while using only the star-forming subsample for the star formation indicators (SFR, $\Sigma$SFR, and sSFR). The green line in panel b represents the case when mass loading factor $\eta=1$ (i.e., $\dot{M}_{\rm out}$), which is the maximum value predicted for ``SF-driven" outflows.}
\label{fig:Mdot_out_trend}
\end{figure*}

Figure~\ref{fig:Vout_trend} shows the relation between $V_{\rm out}$ and the aforementioned host-galaxy properties, together with the Spearman correlation coefficients and $p$-values. 

To begin, we find no clear correlation between $V_{\rm out}$ and either $M_*$ (panel a) or $V_{\rm cir}$ (panel b). Nevertheless, nearly all Na D outflows satisfy $V_{\rm out}<3V_{\rm cir}$, where $3V_{\rm cir}$ is commonly adopted as an approximation to the escape velocity \citep{Veilleux2005}. This suggests that most Na D outflows, particularly those in galaxies with deep potential wells, are unlikely to escape and instead remain gravitationally confined, whereas the outflows in the three systems with relatively shallow potential wells in our sample ($\log(V_{\rm cir}/{\rm km\,s^{-1}})<2.4$) may reach the CGM/IGM.

Panels c--e of Figure~\ref{fig:Vout_trend} show how $V_{\rm out}$ varies with star-formation activity. In the quiescent subsample, $V_{\rm out}$ shows no clear relation with SFR, sSFR, or $\Sigma_{\rm SFR}$. By contrast, the star-forming subsample shows weak positive trends between $V_{\rm out}$ and all three quantities, with Spearman coefficients of $R_S\sim0.3$--0.6. Among these, only the  $V_{\rm out}$--sSFR  relation reaches the nominal ($p\sim0.05$) level, while the trends with SFR and $\Sigma_{\rm SFR}$ are not statistically significant. Given the number of correlations explored in this section, we regard this result as tentative, and larger samples will be needed to assess these trends more robustly.

Similarly, Figure~\ref{fig:Mdot_out_trend} shows no clear dependence of $\dot{M}_{\rm out}$ on $M_*$, nor any obvious correlation with star-formation properties in the quiescent subsample. In contrast, the star-forming subsample exhibits moderately positive trends of $\dot{M}_{\rm out}$ with SFR, $\Sigma_{\rm SFR}$, and sSFR ($R_S\sim0.3-0.5$), although none reaches the nominal $p\sim0.05$ significance level. Again, larger samples will therefore be needed to determine whether these trends are robust. A similar distinction is seen in the mass-loading factor $\eta$, shown in panel b of Figure~\ref{fig:Mdot_out_trend}. In star-forming galaxies, $\eta$ is generally around or below unity, consistent with expectations for star formation-driven outflows \citep[e.g.,][]{Dave2011,Somerville2015}. By contrast, the quiescent subsample mostly lies above $\eta=1$, implying that these outflows can remove cool gas faster than it is consumed by star formation. Such elevated mass-loading factors are more commonly associated with AGN-driven outflows, for which $\eta$ can be enhanced by one or even two orders of magnitude \citep[e.g.,][]{Gonzalez-Alfonso2017}.

Taken together, the trends in $V_{\rm out}$ and $\dot{M}_{\rm out}$ are consistent with Na D outflows in star-forming galaxies being primarily driven by ongoing star formation, whereas those in quiescent galaxies likely require a different driving mechanism (e.g. AGN).

\subsubsection{Energy and momentum outflow rates}
\label{sec:Eout_pout}
\begin{figure*}
\centering
\includegraphics[width=0.8\textwidth]{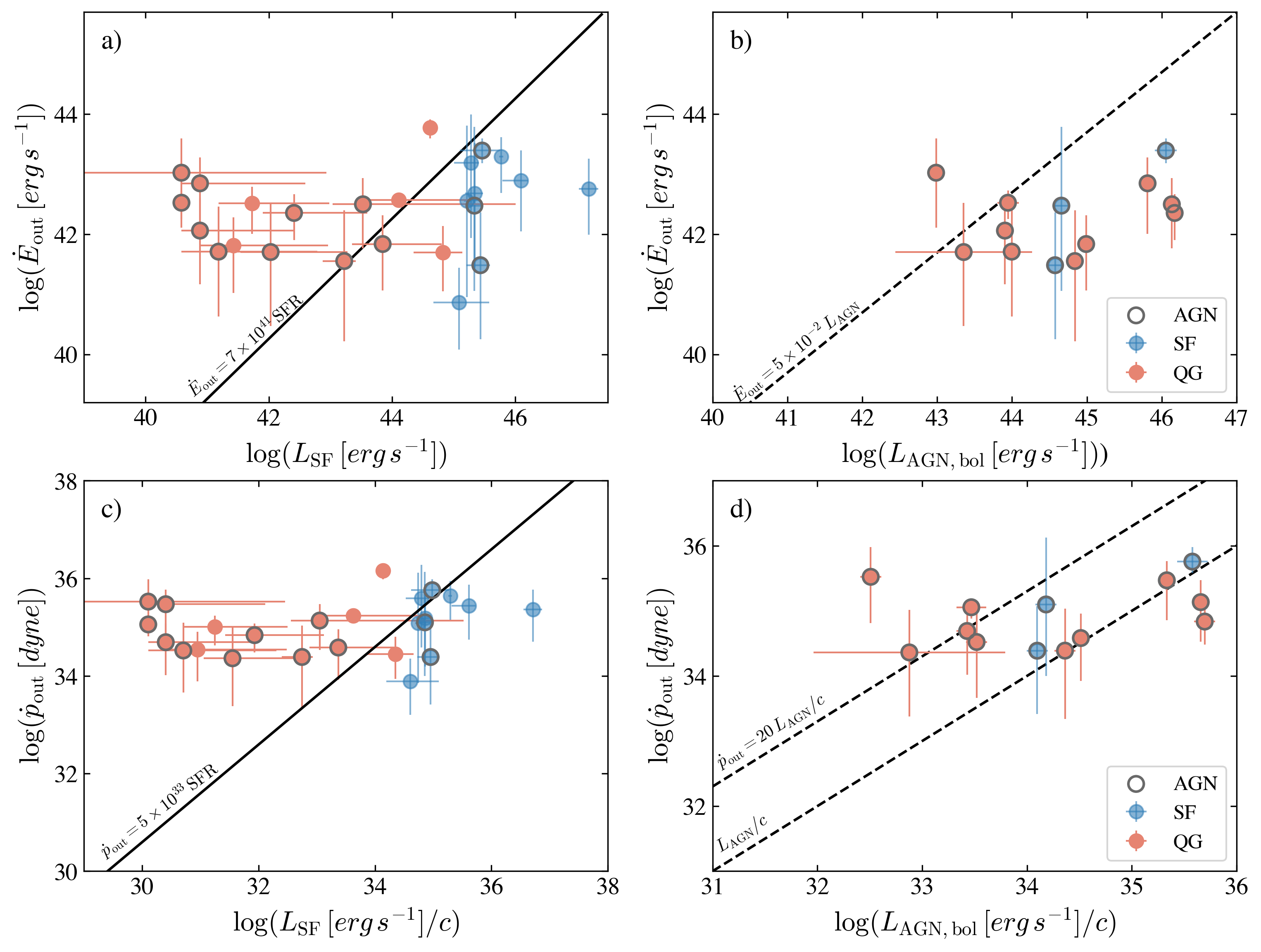}
\caption{Energy outflow rate $\dot{E}_{\rm out}$ and momentum outflow rate $\dot{p}_{\rm out}$ of $z\sim$2 Na D outflows compared to the SF (left) and AGN luminosity (right). The symbols are the same as Figure~\ref{fig:Vout_trend} and \ref{fig:Mdot_out_trend}. The black solid lines on the left column represent the expected outflow kinetic energy driven by star formation, and the black dashed lines on the right column represent the typical $\dot{E}_{\rm out}$ values for AGN-driven winds, and $\dot{p}_{\rm out}$ for AGN energy-driven (20$L_{\rm AGN}/c$) and momentum driven winds ($L_{\rm AGN}/c$).}
\label{fig:Edot_pdot}
\end{figure*}

The different trends seen for star-forming and quiescent galaxies in the previous section suggest that their Na D outflows may arise from different launching mechanisms. We therefore examine this issue in terms of energy and momentum outflow rates, by comparing the observed Na D outflows with the energy and momentum that can be supplied by ongoing star formation and AGN activity. Since these are the two main processes capable of driving strong galactic outflows \citep{King2015,Rupke2018}, we estimate their expected output rates in this section.

We estimate the energy available from current star formation from the star-formation luminosity, $L_{\rm SF}\approx {\rm SFR}_{\rm SED}\times10^{10}L_\odot$, and the corresponding momentum input as $L_{\rm SF}/c$. For galaxies classified as AGN, we estimate the AGN luminosity, $L_{\rm AGN}$, from X-ray fluxes when available, following \citet{Duras2020}. For X-ray-undetected AGN, we instead adopt the [\OIII]-based bolometric correction of \citet{Netzer2009}, $L_{\rm bol}=600L_{[\OIII]}$, while noting that such estimates are uncertain because $L_{[\OIII]}$ depends on obscuration and the physical conditions of the narrow-line region \citep{Diamond-Stanic2009,Pennell2017,Netzer2019}. The corresponding AGN momentum input is taken to be $L_{\rm AGN}/c$.

Figure~\ref{fig:Edot_pdot} compares the measured energy and momentum rates of the Na D outflows with the expected contributions from star formation and AGN. For star formation, we adopt the reference relations for supernova-driven winds from \citet{Veilleux2005}, based on solar-metallicity Starburst99 models \citep{Leitherer1999}: $\dot{E}_{\rm SN}=7\times10^{41}\,{\rm SFR}$ erg s$^{-1}$ and $\dot{p}_{\rm SN}=5\times10^{33}\,{\rm SFR}$ dyne. For AGN-driven outflows, theoretical expectations suggest that, in the energy-conserving case, the kinetic power can reach $\sim5\%$ of $L_{\rm AGN}$ and the momentum rate up to $\sim20L_{\rm AGN}/c$, whereas in the momentum-conserving case the momentum rate is expected to be comparable to $L_{\rm AGN}/c$ \citep{King2015}.

The left column of Figure~\ref{fig:Edot_pdot} shows that the Na D outflows in star-forming galaxies all have $\dot{E}_{\rm out}$ below the energy injection rate expected from their current star formation, indicating that ongoing star formation can plausibly power these neutral outflows. The same conclusion is supported by the momentum-rate comparison. In contrast, most quiescent galaxies lie above the maximum energy and momentum that their current star formation could provide, implying that their weak ongoing star formation is insufficient to drive the observed Na D outflows.

AGN are a plausible power source for many of the Na D outflows in quiescent galaxies. AGN are present in 10/15 (67\%) of the quiescent outflow hosts, much higher than 3/10 (30\%) of the outflow-detected star-forming subsample. The outflows in those AGN hosts typically have $\dot{E}_{\rm out}<0.05L_{\rm AGN}$, and their momentum rates lie between the expectations for momentum-conserving and energy-conserving AGN-driven winds, indicating that ongoing AGN activity can plausibly power the outflows in these cases. However, five quiescent galaxies with Na D outflows are not identified as AGN, including SMILES-206183, which hosts the strongest Na D outflow currently known at $z>1$ \citep{Sun2026}. Among them, only BJ-10314 could plausibly be powered by its present-day star formation, perhaps because it lies just below our quiescent threshold ($\Delta {\rm MS}=-0.5$ dex). For the remaining four systems, neither instantaneous star formation nor ongoing AGN activity appear sufficient to explain the observed neutral outflows, implying that they likely reflect energy or momentum injected by past activity. We discuss possible driving mechanisms in these systems in Section~\ref{sec: outf_mech}.

\section{Discussion} \label{sec:discuss}

With the largest compilation to date of JWST/NIRSpec observations covering Na D at $0.6<z<4$, we have carried out a census of Na D-traced neutral ISM and outflows across a broad range of galaxy populations. They are detected primarily in massive galaxies; lower-mass detections are confined to star-forming systems. Behind the mass dependence of Na D ISM incidence, we find that the Na D detectability shows a clear dependence on dust attenuation. In massive quiescent galaxies, we find that it also depends on stellar population properties and quenching history, being preferentially associated with both older systems and younger, rapidly quenched galaxies. We further identify 25 Na D outflows, and the sources of these outflows appear to differ between star-forming and quiescent hosts.

Motivated by these findings, we further discuss outflow driving mechanisms in detail in Section~\ref{sec: outf_mech}. In Section~\ref{sec:multiphase} we compare neutral and ionized outflows; in Section~\ref{sec:lowz_comp} we place the cosmic-noon Na D outflows in the context of local observations; and in Section~\ref{sec:nad_quenching_path} we focus on the implications of Na D-traced neutral ISM for quiescent galaxies.

\subsection{Na D Outflow Driving Mechanisms}
\label{sec: outf_mech}

The physics of Na D-traced neutral outflows at cosmic noon remains poorly constrained. In the local universe, Na D outflows are commonly observed in star-forming galaxies \citep[e.g.,][]{Rupke2005b,Chen2010,Concas2019}, while their connection to AGN activity remains less clear \citep[e.g.,][]{Sarzi2016,Concas2019,Baron2022}. At $z\gtrsim1$, Na D outflows have been identified in quiescent galaxies both with and without clear AGN signatures \citep{Davies2024,D'Eugenio2024,Perez-Gonzalez2025,Valentino2025,Sun2026,Taylor2026,ZhuP2026}, raising the question of whether they are driven by ongoing star formation, current AGN activity, or past feedback episodes. Our census provides important constraints on these possibilities by distinguishing the behavior of Na D outflows in star-forming and quiescent galaxies at $0.6<z<4$.

The contrast between the two populations presented in Section \ref{sec:nad_outf} suggests that Na D outflows at cosmic noon are \textit{not} powered by a single mechanism. 
In star-forming galaxies, the observed outflow energetics are broadly consistent with the energy and momentum available from ongoing star formation. We also find weak positive trends of $V_{\rm out}$ and $\dot{M}_{\rm out}$ with star-formation activity, although most of these trends are not statistically significant and only the $V_{\rm out}$--sSFR relation reaches the nominal $p\sim0.05$ level.
This implies that the neutral phase traced by Na D is coupled to the same feedback processes that regulate actively star-forming galaxies, and that ongoing star formation can plausibly launch the observed cool outflows.

The situation is very likely different in quiescent galaxies. There, the Na D outflows generally require more energy and momentum than  can be supplied by the low level of ongoing star formation, implying that residual star formation is unlikely to be their dominant driver (Figure \ref{fig:Edot_pdot}). Many of these quiescent outflows are in galaxies hosting AGN, and in those cases, the observed outflow energetics are broadly consistent with AGN power. This suggests that Na D outflows in quiescent galaxies trace a mode of feedback that is physically distinct from the star formation-driven winds seen in star-forming systems, and are more directly connected to black hole activity. In this sense, the Na D-traced neutral phase appears able to preserve signatures of AGN feedback even after star formation has already largely ceased. This finding of AGN-driven feedback on cosmic noon quiescent galaxies are in line with the local finding of Na D outflows in AGN-like post-starburst galaxies \citep{Baron2022,Sun2024}, as well as local ``Red Geysers" galaxies -- a population of quiescent galaxies hosting low-level radio AGN and ionized outflows \citep{Cheung2016,Roy2018,Moghni2026}.

A particularly interesting result is the presence of five quiescent galaxies without AGN but still with strong Na D outflows (Figure~\ref{fig:Edot_pdot}), including SMILES-206183, the strongest Na D outflow at $z>1$ as reported already in \citet{Sun2026}. Except for BJ-10314, these outflows are difficult to explain with present-day activity inside the galaxies. \citet{Sun2026} propose the ``AGN fossil outflow'' scenario to explain the extremely strong Na D outflow in SMILES-206183, in which the observed neutral outflow was launched during an earlier episode of intense AGN activity that has since faded. The same interpretation is plausible for the three newly identified non-AGN quiescent outflows in our sample. A ``star-formation fossil outflow'' scenario is disfavored because the typical Na D outflow timescale is $\sim10$ Myr, assuming an outflow extent of a few kpc and a velocity of $\sim500\,\mathrm{km\,s^{-1}}$. This timescale lies well within the youngest 30 Myr bin of our non-parametric SED fitting, so the SED-inferred SFR should already largely reflect the star-formation level at the time when the outflow was launched. By contrast, AGN duty cycles can be significantly shorter than 10 Myr \citep[e.g.,][]{Schawinski2015,Zubovas2022}, leaving sufficient time for the accretion to shut down while the outflow remains observable. These non-AGN quiescent systems may therefore represent fossil outflows powered by past AGN activity.

This interpretation is in line with other recent studies. \citet{Taylor2026} identified three Na D outflows in quiescent galaxies at $z>2$, and all three have outflow energy rates that exceed the maximum energy available from current star formation or AGN activity. The possibility of fossil outflows is also supported by theoretical modeling, which shows that a substantial fraction of AGN-driven outflows is detectable after the AGN has already shut down \citep{Zubovas2022}. 

We note that, despite compiling the largest Na D sample at cosmic noon to date, constraining outflow timescales remains difficult, primarily because slit-integrated spectroscopy provides only weak constraints on outflow extent. As a result, the inferred outflow mass, energetics, and timescale remain uncertain. Future spatially resolved IFU observations of Na D outflows in cosmic-noon quiescent galaxies will therefore be essential for clarifying their physical properties, duty cycles, and connection to galaxy quenching.

In summary, our census supports a picture in which Na D-traced neutral outflows at cosmic noon are state dependent: in star-forming galaxies, their kinematics and energetics are broadly consistent with being powered by ongoing star formation, whereas in quiescent galaxies they are more often linked to AGN feedback, either synchronous in AGN hosts or fossil in systems lacking current AGN signatures. AGN fossil outflows may be common among cosmic-noon quiescent galaxies rather than rare phenomena.

\subsection{Comparison with ionized-phase outflows at cosmic noon}\label{sec:multiphase}

This highlights the importance of comparing different gas phases to fully understand how feedback redistributes mass, momentum, and energy through galaxies. At cosmic noon, however, population-level multiphase comparisons remain limited, largely because neutral outflow phases have been much harder to observe. 

Ionized outflows at cosmic noon have been studied by  \citet{Forster2019}, who  presented a census of [\NII]+H$\alpha$-traced outflows at $0.6<z<2.7$ from the ${\rm KMOS^{3D}}$ survey, distinguishing between ``SF-driven'' and ``AGN-driven'' systems. For the SF-driven outflows, they found that the outflow incidence correlates most strongly with  $\Sigma_{\rm SFR}$. A similar trend is seen in our Na D outflows in star-forming galaxies, where for our star-forming galaxies, the outflow velocity depends on star-formation activity (Figure \ref{fig:Vout_trend}). For the AGN-driven outflows, \citet{Forster2019} found that the incidence depends more strongly on stellar mass, largely because AGN incidence itself increases with stellar mass. Consistent with this, Na D outflows accompanied by AGN signatures in our sample are predominantly hosted by massive galaxies ($\log(M_*/M_\odot)>10.5$, see Figure~\ref{fig:Vout_trend}).

Despite these broad similarities among star-forming galaxies, the neutral and ionized phases differ markedly in the massive quiescent regime. Adopting a consistent quiescent selection of $\log(M_*/M_\odot)>10$ and $\Delta{\rm MS}\leq-0.5$ dex, we find 15 neutral outflows among 38 massive quiescent galaxies (39\%), whereas \citet{Forster2019} reported 11 ionized outflows among 57 galaxies (19\%). This suggests that neutral outflows are more readily detected, and possibly more prevalent, than ionized outflows in massive quiescent galaxies at $z\sim2$. A likely explanation is tracer-dependent detectability: ionized-outflow identification relies on nebular emission lines and therefore requires sufficiently bright ionized gas, which is often weak in quiescent systems unless they host an AGN or residual star formation. By contrast, Na D absorption uses the stellar continuum as the background source and can therefore trace cooler outflows even in line-weak galaxies. In this sense, Na D is sensitive to a broader range of feedback states, including systems in which the original driving source may already have faded. Indeed, we identify five Na D outflows in quiescent galaxies without AGN signatures, suggesting that neutral outflows can remain observable even when standard ionized-gas diagnostics become ineffective, as discussed in Section~\ref{sec: outf_mech}.

We stress, however, that a fair comparison between neutral and ionized outflows ultimately requires a unified galaxy sample with measurements of both phases. Such a census is still missing at $z\sim2$. One early example comes from the JWST Blue Jay survey, where seven of 16 massive quiescent galaxies show Na D outflows while only three show [\OIII]-traced ionized outflows \citep{Bugiani2025}. With the improved statistics of our Na D census, a future joint analysis of neutral and ionized outflows in a common sample will allow us to quantify their relative incidence, overlap, and correlations across different galaxy populations.

\subsection{Comparison with lower-redshift Na D outflows}\label{sec:lowz_comp}

\begin{figure*}
\centering
\includegraphics[width=1\textwidth]{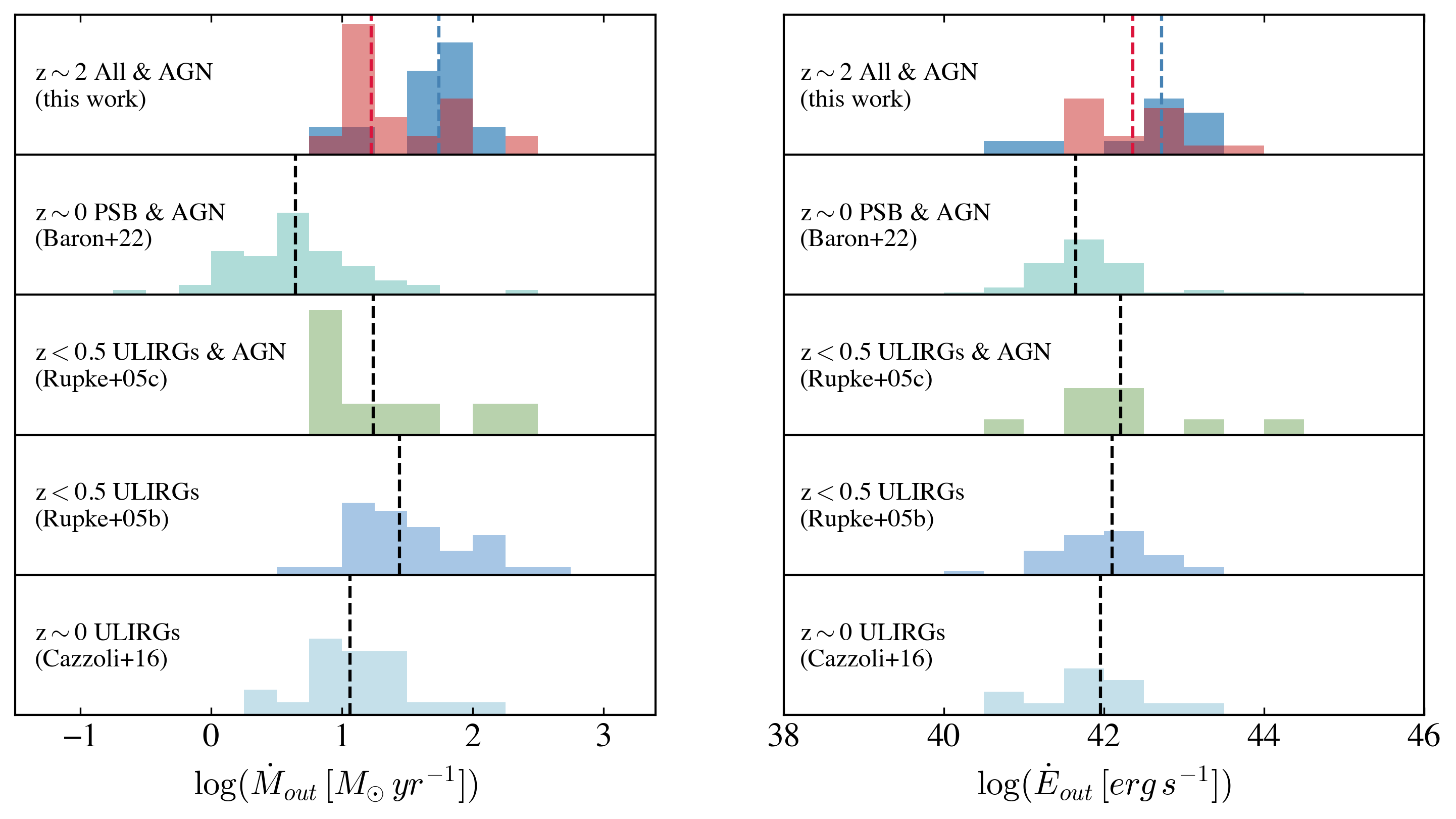}
\caption{Comparison of neutral Na D outflow properties ($\log(\dot{M}_{\rm out})$ on the left and $\log(\dot{E}_{\rm out})$ on the right) in different samples, including 1) the outflows detected in the star-forming (blue) and quiescent (red) at $0.6<z<4$ galaxies; 2) the local PSB AGN hosts from \citet{Baron2022}; 3) the $z<0.5$ Seyfert-2 ULIRGs from \citet{Rupke2005c}; 4) the $z<0.5$ SF/LINER ULIRGs from \citet{Rupke2005b}; 5) the local pure-star forming ULIRGs from \citet{Cazzoli2016}. The median values of each sample are shown by a dashed line. The $\log(\dot{M}_{\rm out})$ and $\log(\dot{E}_{\rm out})$ of the z$\sim$2 Na D outflows appear to be higher than those of the local Na D outflows, but such differences can be mostly attributed to the Na D profile modeling method difference.}
\label{fig:local_comparison}
\end{figure*}

To explore possible cosmic evolution in Na D outflows, we compare $\dot{M}_{\rm out}$ and $\dot{E}_{\rm out}$ of our $0.6<z<4$ sample with those reported in the local Universe. The local comparison samples include Na D outflows in ULIRGs with AGN \citep{Rupke2005c}, ULIRGs without AGN \citep{Rupke2005b,Cazzoli2016}, and post-starburst galaxies with AGN \citep{Baron2022}. Except for \citet{Cazzoli2016}, who used spatially resolved spectroscopy, these local measurements were also derived from limited-aperture or long-slit data probing only the central galaxy regions, making them broadly comparable to our JWST/NIRSpec MSA-based measurements.

Figure~\ref{fig:local_comparison} compares $\dot{M}_{\rm out}$ and $\dot{E}_{\rm out}$ for the Na D outflows at cosmic noon and in the local Universe. The cosmic-noon sample spans $\log(\dot{M}_{\rm out}[M_{\odot}\,{\rm yr^{-1}}])\sim0.5$--2.5, with a lower median in quiescent galaxies ($\sim1.2$) than in star-forming galaxies ($\sim1.7$). The corresponding $\log(\dot{E}_{\rm out}[{\rm erg\,s^{-1}}])$ ranges from $\sim40$ to 44, with a smaller difference between the two galaxy populations.

Overall, the cosmic-noon Na D outflows appear offset to higher $\dot{M}_{\rm out}$ and $\dot{E}_{\rm out}$ by roughly 0.3--0.5 dex relative to local ULIRGs \citep{Rupke2005b,Rupke2005c,Cazzoli2016} and local PSBs \citep{Baron2022}. However, we do not interpret this as strong evidence for redshift evolution, because the offset may largely arise from differences in Na D profile modeling. The local studies generally used two ISM kinematic components when required, separating a narrow systemic component from a broader blueshifted outflow component, whereas our analysis adopts a single ISM component because of the limited resolution of the NIRSpec medium gratings. If a substantial fraction of the absorption in our spectra arises from gas near the systemic velocity, this simplified modeling would overestimate the outflowing mass and energetics. In the limiting case where the systemic and outflowing Na D components contribute comparably, the inferred $\dot{M}_{\rm out}$ would decrease by a factor of two, comparable to the observed offset. We therefore do not regard the current comparison as strong evidence for redshift evolution in Na D outflow properties from $z\sim0$ to 4. Our conclusion is consistent with the findings of \citet{Lyu2026}, who used Mg II $\lambda\lambda2796,2803$ as an alternative tracer of cool gas and found no strong evolution in the velocity of Mg II-traced outflows out to $z\sim6$. However, a more robust test will require future observations with higher spectral and spatial resolution to better separate the systemic ISM and outflowing components, characterize the outflow geometry, and thereby improve constraints on the redshift evolution of outflow energetics.

\subsection{Na~D detectability and outflows in cosmic noon massive quiescent galaxies}
\label{sec:nad_quenching_path}

In Section~\ref{sec:nad_rate}, we showed that Na D-traced neutral ISM is widespread in massive quiescent galaxies at cosmic noon. Although larger samples will be needed to confirm the trend, Section~\ref{sec:nad_dust} suggests that its detectability also depends on stellar-population properties, with Na D absorption preferentially detected both in older systems with larger $D_n4000$ and in younger, rapidly quenched galaxies with stronger Balmer absorption (Figures~\ref{fig:SFH_NaD_detect_QG} and \ref{fig:QG_burst_NaD}).

A possible interpretation is that the detectability of Na D ISM absorption is governed by a balance between ionization and shielding. In massive quiescent galaxies, after quenching, the cool gas and dust content are often reduced, but the ionizing field also evolves as they age. In recently quenched systems, residual young stars may still produce  sufficiently hard UV photons to suppress neutral sodium unless enough shielding remains. As the stellar population ages and the radiation field softens, Na D absorption can emerge even at relatively modest $A_V$. This picture naturally explains why Na D is more readily detected in older quiescent galaxies.

However, the presence of Na D ISM in younger quiescent galaxies with strong H$\delta_A$ (Figure~\ref{fig:QG_burst_NaD}) suggests that age is not the only factor determining the Na D ISM absorption detectability. In these younger systems, while Na D is generally harder to detect because of their residual UV fields and typically reduced shielding, galaxies that experienced more rapid quenching may still retain or redistribute enough cool neutral gas for Na D absorption to remain visible. In particular, strong feedback associated with rapid quenching may increase the covering fraction of absorbing material, thereby enhancing Na D detectability and producing the outflow signatures we observe.  This behavior is broadly consistent with previous studies of post-starburst galaxies at both cosmic noon and in the local Universe: \citet{Taylor2024,Taylor2026} found that Na D outflows are preferentially detected in cosmic-noon quiescent galaxies quenched within the past $\sim1$ Gyr, while \citet{Sun2024} showed that both the Na D outflow fraction and velocity decline from young to old local post-starburst populations.

Taken together, these results suggest that Na D absorption in massive quiescent galaxies is favored under two conditions: either the stellar population is old enough that the ionizing field has weakened, or quenching was sufficiently recent and rapid that substantial cool gas remains, often in an outflowing configuration. The high incidence of Na D outflows in young, post-starburst-like systems, together with the evidence that AGN may be the primary driver of outflows in quiescent galaxies, points to a close connection between AGN-driven gas removal or redistribution and the quenching process. However, because AGN activity and galaxy quenching operate on very different timescales, our observation cannot establish a direct causal link between them. Still, our identification of fossil AGN outflows suggests that AGN feedback may play an important role in maintaining quiescence after the AGN itself has faded. We detect four potential fossil AGN outflows among 15 quiescent galaxies with Na D outflows, indicating that the impact of AGN feedback can persist well beyond the active phase and may be common among quiescent galaxies at cosmic noon. The continued detectability of these fossil outflows after AGN turn-off in quiescent galaxies suggests that they may keep the gas in a disturbed or heated state, preventing new star formation. 

Finally, we caution that the current constraints remain limited by the small number of quiescent galaxies with robust $D_n4000$ and $H\delta_A$ measurements. Larger spectroscopic samples at and beyond cosmic noon are needed to better constrain the stellar ages and quenching histories of $z>1$ quiescent galaxies. In addition, direct measurements of cold gas in quiescent systems beyond the local Universe remain scarce (see Figure~3 of \citealt{Scholtz2026}). Establishing the connection among feedback, cool-gas content, and quenching will therefore require a larger $z\sim2$ quiescent sample with joint constraints on outflows, molecular gas, and well-measured star-formation histories.

\section{Conclusion} \label{sec:concl}

In this work, we present the largest statistical census to date of Na D-traced neutral interstellar gas and outflows at cosmic noon, based on JWST/NIRSpec medium-resolution spectroscopy of 309 galaxies at $0.6 < z < 4$ drawn from the SMILES, JADES, Blue Jay, and Aurora surveys. Our sample spans a wide range of SFR and $M_*$, with robust continuum detections in the vicinity of the Na~D doublet. By subtracting the stellar photospheric contribution to the Na~D doublet, we isolate the neutral ISM component and use it to investigate both the incidence of cool gas absorption and the prevalence and kinematics of neutral outflows across a broad range of galaxy evolutionary stages.

We have ``robust" Na D-traced ISM absorption detection in 59 galaxies and additional 10 ``tentative" detections. Among the robust Na D ISM absorption detection, the vast majority (88\%) of them occur in massive systems with $\log(M_*/M_\odot) > 10$, whereas lower-mass detections are confined to star-forming galaxies. Our analysis shows that the detectability of Na D absorption depends on different physical factors in different galaxy populations. In star-forming galaxies, Na D strength correlates strongly with dust attenuation, indicating that dust shielding plays a central role in allowing neutral sodium-bearing gas to survive and remain observable. In massive quiescent galaxies, however, dust alone does not explain the presence of Na D. Instead, Na D absorption is preferentially found in both older quiescent systems with larger 4000~\AA\ breaks and younger systems with stronger Balmer absorption H$\delta_A$, consistent with rapid recent quenching. This suggests that, in quiescent galaxies, Na D detectability depends not only on the amount of remaining cool gas, but also on the recent star-formation history.

We further classify the Na D ISM detections according to their kinematics. We identify 25 outflows, 8 inflows, and 26 systems whose Na D absorption is consistent with the systemic velocity. The AGN incidence is higher among galaxies with Na D outflows (52\%) than among those with systemic Na D absorption (38\%) or inflows (13\%), suggesting that AGN plays an important, though not exclusive, role in driving Na D outflows.

More broadly, our analysis suggests that Na D outflows are driven by different mechanisms in star-forming and quiescent galaxies. In star-forming galaxies, both the incidence and kinematics of neutral outflows correlate with star-formation properties, supporting a predominantly star-formation-driven origin. In quiescent galaxies, by contrast, the outflows show no clear connection to weak residual star formation, and in many cases the energy available from such star formation is insufficient to explain the inferred outflow energetics. Together with the relatively high AGN incidence (67\%) among quiescent galaxies hosting Na D outflows, these results suggest that AGN are the primary driver of neutral outflows in quiescent systems at cosmic noon.

However, five quiescent galaxies in our sample with Na D outflows do not show clear evidence for ongoing AGN activity, and for four of them, the energy output from their residual star formation is too low to account for the observed outflow strengths. We therefore identify these systems as candidates for fossil AGN outflows, in which a past AGN episode may have launched the outflow and the disturbed neutral gas remains observable after the active accretion phase has faded. This scenario remains tentative and will require larger samples and deeper multiwavelength constraints to be tested more directly.

Overall, our results highlight the importance of the cool neutral phase for understanding galaxy quenching at cosmic noon. Substantial neutral gas can persist not only in massive star-forming galaxies but also in quiescent galaxies with distinct quenching histories, while Na D outflows suggest that the processes regulating this gas differ between these populations. Future progress will require larger spectroscopic samples and, ideally, spatially resolved multiphase observations that connect neutral outflows to other gas phases (e.g., ionized and molecular gas), thereby providing a more comprehensive view of the role of feedback and its duty cycle in regulating star formation at cosmic noon.

\begin{acknowledgments}

The authors thank Dr. Sirio Belli and Dr. Weizhe Liu for useful discussions.

This work is based on
observations made with the NASA/ESA/CSA James Webb
Space Telescope. The JWST data were obtained from the Mikulski
Archive for Space Telescopes at the Space Telescope Science
Institute, which is operated by the Association of Universities for Research in Astronomy, Inc., under NASA contract
NAS 5-03127 for JWST. These observations are associated with programs 1180, 1181, 1207, 1210, 1286, 1287, 1810, 1837, 1914, and 3215.
The authors acknowledge the Blue Jay and Aurora teams for developing their observing programs. A fraction of spectroscopic and photometric data used in this work were retrieved
from the Dawn JWST Archive (DJA)\footnote{\url{https://dawn-cph.github.io/dja/}}. DJA is an initiative of the Cosmic Dawn Center (DAWN).

YS, ZJ, YZ, and CNAW acknowledge support from the NIRCam Science Team contract to the University of Arizona, NAS5-02015. 
FDE acknowledges support by the Science and Technology Facilities Council (STFC), by the ERC through Advanced Grant 695671 ``QUENCH'', and by the UKRI Frontier Research grant RISEandFALL.
GHR acknowledges support from the JWST Mid-Infrared Instrument (MIRI) Science Team Lead, grant 80NSSC18K0555, from NASA Goddard Space Flight Center to the University of Arizona.
WMB gratefully acknowledges support from DARK via the DARK fellowship and from a research grant (VIL54489) from VILLUM FONDEN.
AJB acknowledges funding from the "FirstGalaxies" Advanced Grant from the European Research Council (ERC) under the European Union's Horizon 2020 research and innovation programme (Grant agreement No. 789056).
SC acknowledges support by European Union’s HE ERC Starting Grant No. 101040227 - WINGS.
JMH is supported by JWST Program 8544.
MP acknowledges support through the grants PID2021-130918NB-I00, PID2024-159902NA-I00, and RYC2023-044853-I, funded by the Spain Ministry of Science and Innovation/State Agency of Research MCIN/AEI/10.13039/501100011033 and El Fondo Social Europeo Plus FSE+. 
PGP-G acknowledges support from grant PID2022-139567NB-I00 funded by Spanish Ministerio de Ciencia e Innovaci\'on MCIN/AEI/10.13039/501100011033, FEDER, UE.
H\"U thanks the Max Planck Society for support through the Lise Meitner Excellence Program. H\"U acknowledges funding by the European Union (ERC APEX, 101164796). Views and opinions expressed are however those of the authors only and do not necessarily reflect those of the European Union or the European Research Council Executive Agency. Neither the European Union nor the granting authority can be held responsible for them.

We respectfully acknowledge the University of Arizona is on the land and territories of Indigenous peoples. Today, Arizona is home to 22 federally recognized tribes, with Tucson being home to the O'odham and the Yaqui. The University strives to build sustainable relationships with sovereign Native Nations and Indigenous communities through education offerings, partnerships, and community service.

\end{acknowledgments}

\facilities{JWST}

\software{\texttt{AstroPy}\citep{astropy2013,astropy2018,astropy2022},  
\texttt{emcee}\citep{Foreman-Mackey2013},
\texttt{Prospector}\citep{Johnson2021}, \texttt{SciPy}\citep{Virtanen2020},
\texttt{pPXF}\citep{cappellari2017,Cappellari2023}
}

\appendix

\section{Na D ISM Absorption detected sample}
\label{ap:Na D gallery}

\renewcommand{\thetable}{A\arabic{table}}
\setcounter{table}{0}

\begin{deluxetable*}{cccccccccccc}
\tablecaption{Properties of galaxies with Na D outflows}
\label{Tab:outf}
\tablewidth{0pt}
\tabletypesize{\scriptsize}
\tablehead{\colhead{1) Survey} &
  \colhead{2) ID} &
  \colhead{3) R.A.} &
  \colhead{4) Decl.} &
  \colhead{5) z} &
  \colhead{6) $\log(M_*)$} &
  \colhead{7) $\log({\rm SFR})$} &
  \colhead{8) $A_V$} &
  \colhead{9) ${\rm EW_{Na D, ISM}}$} &
  \colhead{10) ${
\rm SNR_{Na D, ISM}}$} &
  \colhead{11) $\Delta v$} &
  \colhead{12) $b$}
  \\
  \colhead{} &
  \colhead{} &
  \colhead{(deg)} &
  \colhead{(deg)} &
  \colhead{} &
  \colhead{$(M_\odot)$} &
  \colhead{$(M_\odot\,yr^{-1})$} &
  \colhead{$(mag)$} &
  \colhead{$(\AA)$} &
  \colhead{} &
  \colhead{$km\,s^{-1}$} &
  \colhead{$km\,s^{-1}$}
}
\startdata
SMILES &
  205818 &
  53.197600 &
  -27.786474 &
  1.098 &
  $10.57^{+0.19}_{-0.72}$ &
  $1.70^{+0.13}_{-0.27}$ &
  $2.88^{+0.23}_{-0.23}$ &
  $3.39^{+1.16}_{-0.96}$ &
  2.0 &
  $-473^{+206}_{-164}$ &
  $241^{+112}_{-158}$ \\
SMILES &
  206183 &
  53.176590 &
  -27.785521 &
  1.317 &
  $11.25^{+0.02}_{-0.02}$ &
  $1.03^{+0.02}_{-0.02}$ &
  $2.27^{+0.23}_{-0.23}$ &
  $9.16^{+0.45}_{-0.44}$ &
  7.2 &
  $-440^{+55}_{-57}$ &
  $273^{+66}_{-40}$ \\
SMILES &
  194473 &
  53.151470 &
  -27.825931 &
  1.612 &
  $10.92^{+0.05}_{-0.08}$ &
  $-3.00^{+2.35}_{-1.00}$ &
  $2.12^{+0.63}_{-0.63}$ &
  $3.58^{+0.44}_{-0.45}$ &
  2.9 &
  $-168^{+150}_{-161}$ &
  $346^{+39}_{-111}$ \\
SMILES &
  201306 &
  53.123138 &
  -27.803460 &
  2.344 &
  $11.13^{+0.04}_{-0.06}$ &
  $-3.00^{+0.00}_{-0.00}$ &
  $0.04^{+0.04}_{-0.04}$ &
  $4.08^{+0.21}_{-0.21}$ &
  7.3 &
  $-218^{+25}_{-25}$ &
  $275^{+62}_{-103}$ \\
Blue Jay &
  18668 &
  150.129272 &
  2.369564 &
  2.087 &
  $11.20^{+0.05}_{-0.05}$ &
  $-2.70^{+1.71}_{-0.30}$ &
  $4.32^{+0.21}_{-0.21}$ &
  $11.59^{+0.73}_{-0.72}$ &
  7.5 &
  $-196^{+42}_{-46}$ &
  $195^{+121}_{-124}$ \\
Blue Jay &
  18252 &
  150.098507 &
  2.365359 &
  2.552 &
  $11.44^{+0.15}_{-0.23}$ &
  $3.61^{+0.16}_{-0.16}$ &
  $4.37^{+0.78}_{-0.78}$ &
  $9.13^{+1.59}_{-1.46}$ &
  3.0 &
  $-326^{+75}_{-90}$ &
  $118^{+100}_{-80}$ \\
Blue Jay &
  11142 &
  150.073292 &
  2.293290 &
  2.445 &
  $10.93^{+0.20}_{-0.13}$ &
  $0.26^{+0.96}_{-0.49}$ &
  $0.59^{+0.59}_{-0.59}$ &
  $6.73^{+0.33}_{-0.32}$ &
  7.2 &
  $-233^{+22}_{-23}$ &
  $90^{+76}_{-62}$ \\
Blue Jay &
  11136 &
  150.109048 &
  2.294328 &
  1.863 &
  $10.95^{+0.36}_{-0.25}$ &
  $1.50^{+0.49}_{-0.41}$ &
  $3.58^{+1.37}_{-0.67}$ &
  $5.91^{+0.41}_{-0.40}$ &
  6.3 &
  $-139^{+27}_{-28}$ &
  $34^{+40}_{-24}$ \\
Blue Jay &
  10339 &
  150.093876 &
  2.284722 &
  2.364 &
  $10.50^{+0.09}_{-0.12}$ &
  $-2.15^{+1.53}_{-0.54}$ &
  $0.99^{+0.57}_{-0.44}$ &
  $5.27^{+0.37}_{-0.36}$ &
  5.7 &
  $-201^{+34}_{-32}$ &
  $126^{+78}_{-77}$ \\
Blue Jay &
  10314 &
  150.085725 &
  2.285124 &
  2.099 &
  $11.03^{+0.13}_{-0.13}$ &
  $1.24^{+0.31}_{-0.48}$ &
  $3.57^{+0.03}_{-0.03}$ &
  $2.98^{+0.35}_{-0.33}$ &
  4.4 &
  $-154^{+69}_{-56}$ &
  $144^{+92}_{-89}$ \\
Blue Jay &
  9871 &
  150.077798 &
  2.281137 &
  2.046 &
  $11.64^{+0.09}_{-0.10}$ &
  $0.52^{+1.19}_{-0.49}$ &
  $3.57^{+0.46}_{-0.46}$ &
  $9.12^{+0.33}_{-0.34}$ &
  10.5 &
  $-209^{+27}_{-26}$ &
  $163^{+28}_{-29}$ \\
Blue Jay &
  8002 &
  150.119934 &
  2.261021 &
  2.687 &
  $9.98^{+0.16}_{-0.27}$ &
  $2.19^{+0.03}_{-0.04}$ &
  $2.82^{+0.05}_{-0.05}$ &
  $4.09^{+0.43}_{-0.41}$ &
  4.0 &
  $-557^{+83}_{-75}$ &
  $209^{+96}_{-100}$ \\
JADES &
  42573 &
  53.178703 &
  -27.802702 &
  2.694 &
  $10.94^{+0.09}_{-0.05}$ &
  $1.85^{+0.15}_{-0.22}$ &
  $4.27^{+0.77}_{-0.57}$ &
  $8.29^{+0.87}_{-0.87}$ &
  5.0 &
  $-138^{+133}_{-111}$ &
  $73^{+110}_{-52}$ \\
JADES &
  44571 &
  53.140316 &
  -27.797596 &
  1.768 &
  $10.57^{+0.04}_{-0.06}$ &
  $-2.40^{+1.60}_{-0.60}$ &
  $1.23^{+0.28}_{-0.42}$ &
  $3.80^{+0.60}_{-0.56}$ &
  3.7 &
  $-160^{+65}_{-70}$ &
  $104^{+101}_{-71}$ \\
JADES &
  51236 &
  53.146172 &
  -27.779959 &
  2.582 &
  $10.94^{+0.24}_{-0.22}$ &
  $1.88^{+0.29}_{-0.36}$ &
  $5.78^{+1.08}_{-1.08}$ &
  $6.79^{+0.56}_{-0.57}$ &
  6.1 &
  $-389^{+90}_{-61}$ &
  $352^{+37}_{-89}$ \\
JADES &
  171147 &
  53.055626 &
  -27.874064 &
  2.563 &
  $11.35^{+0.02}_{-0.02}$ &
  $-1.85^{+1.25}_{-0.54}$ &
  $0.46^{+0.20}_{-0.21}$ &
  $2.62^{+0.24}_{-0.24}$ &
  6.1 &
  $-257^{+68}_{-70}$ &
  $281^{+89}_{-143}$ \\
JADES &
  197911 &
  53.165314 &
  -27.814140 &
  3.064 &
  $11.40^{+0.08}_{-0.14}$ &
  $-1.17^{+1.19}_{-0.50}$ &
  $0.10^{+0.10}_{-0.10}$ &
  $2.38^{+0.20}_{-0.20}$ &
  7.0 &
  $-389^{+40}_{-46}$ &
  $192^{+89}_{-92}$ \\
JADES &
  202925 &
  53.163022 &
  -27.797727 &
  1.999 &
  $10.52^{+0.05}_{-0.05}$ &
  $-1.55^{+1.25}_{-0.49}$ &
  $1.43^{+0.33}_{-0.51}$ &
  $1.81^{+0.44}_{-0.43}$ &
  2.2 &
  $-223^{+149}_{-156}$ &
  $156^{+145}_{-109}$ \\
JADES &
  215262 &
  53.141019 &
  -27.755119 &
  2.131 &
  $10.84^{+0.10}_{-0.07}$ &
  $1.75^{+0.12}_{-0.13}$ &
  $1.83^{+0.12}_{-0.12}$ &
  $4.63^{+1.88}_{-1.06}$ &
  2.7 &
  $-212^{+187}_{-312}$ &
  $226^{+121}_{-151}$ \\
JADES &
  199773 &
  53.163241 &
  -27.809057 &
  2.821 &
  $10.78^{+0.09}_{-0.03}$ &
  $-0.36^{+0.19}_{-0.35}$ &
  $0.45^{+0.17}_{-0.36}$ &
  $4.26^{+0.63}_{-0.62}$ &
  3.9 &
  $-128^{+82}_{-90}$ &
  $112^{+117}_{-77}$ \\
JADES &
  25055 &
  189.160382 &
  62.227502 &
  1.018 &
  $11.33^{+0.06}_{-0.39}$ &
  $-2.70^{+1.54}_{-0.30}$ &
  $1.54^{+0.47}_{-0.47}$ &
  $3.22^{+0.31}_{-0.31}$ &
  6.3 &
  $-182^{+141}_{-239}$ &
  $181^{+112}_{-112}$ \\
JADES &
  29711 &
  189.196910 &
  62.284104 &
  2.320 &
  $10.57^{+0.18}_{-0.18}$ &
  $2.51^{+0.11}_{-0.30}$ &
  $5.06^{+0.83}_{-0.83}$ &
  $7.60^{+1.34}_{-1.26}$ &
  3.1 &
  $-300^{+138}_{-117}$ &
  $187^{+131}_{-127}$ \\
JADES &
  64818 &
  189.026770 &
  62.289835 &
  1.360 &
  $9.46^{+0.24}_{-0.33}$ &
  $1.63^{+0.07}_{-0.05}$ &
  $1.97^{+0.07}_{-0.07}$ &
  $2.98^{+1.72}_{-0.95}$ &
  2.3 &
  $-299^{+229}_{-285}$ &
  $234^{+123}_{-158}$ \\
Aurora &
  50 &
  150.179276 &
  2.233684 &
  2.294 &
  $11.06^{+0.08}_{-0.20}$ &
  $-0.06^{+2.47}_{-0.49}$ &
  $3.79^{+0.65}_{-0.65}$ &
  $7.34^{+0.59}_{-0.54}$ &
  9.6 &
  $-290^{+48}_{-51}$ &
  $117^{+97}_{-78}$ \\
Aurora &
  5901 &
  150.189388 &
  2.238137 &
  2.396 &
  $9.67^{+0.46}_{-0.35}$ &
  $1.76^{+0.13}_{-0.16}$ &
  $1.43^{+0.10}_{-0.10}$ &
  $3.26^{+1.13}_{-0.90}$ &
  2.7 &
  $-345^{+202}_{-213}$ &
  $212^{+125}_{-140}$
\enddata
\tablecomments{1) Survey Name; 2) Source NIRSpec ID; 3) and 4) Sky coordinate; 5) pPXF-derived spectroscopic redshift; 6) SED-derived stellar mass; 7) SED-derived SFR within the recent 30 Myr; 8) Dust attenuation; 9) Equivalent width of ISM component of Na D; 10) peak SNR of ISM component of Na D; 11) Velocity offset of Na D ISM component relative to the systemic redshift; 12) Doppler width of the Na D ISM component.}
\end{deluxetable*}


\begin{deluxetable*}{cccccccccccc}
\tablecaption{Properties of galaxies with Na D inflows}
\label{Tab:inf}
\tablewidth{0pt}
\tabletypesize{\scriptsize}
\tablehead{\colhead{1) Survey} &
  \colhead{2) ID} &
  \colhead{3) R.A.} &
  \colhead{4) Decl.} &
  \colhead{5) z} &
  \colhead{6) $\log(M_*)$} &
  \colhead{7) $\log({\rm SFR})$} &
  \colhead{8) $A_V$} &
  \colhead{9) ${\rm EW_{Na D, ISM}}$} &
  \colhead{10) ${
\rm SNR_{Na D, ISM}}$} &
  \colhead{11) $\Delta v$} &
  \colhead{12) $b$}
  \\
  \colhead{} &
  \colhead{} &
  \colhead{(deg)} &
  \colhead{(deg)} &
  \colhead{} &
  \colhead{$(M_\odot)$} &
  \colhead{$(M_\odot\,yr^{-1})$} &
  \colhead{$(mag)$} &
  \colhead{$(\AA)$} &
  \colhead{} &
  \colhead{$km\,s^{-1}$} &
  \colhead{$km\,s^{-1}$}
}
\startdata
SMILES &
  191748 &
  53.101990 &
  -27.834628 &
  2.675 &
  $10.87^{+0.05}_{-0.04}$ &
  $-3.00^{+0.00}_{-0.00}$ &
  $0.63^{+0.34}_{-0.28}$ &
  $2.14^{+0.45}_{-0.42}$ &
  4.4 &
  $161^{+73}_{-107}$ &
  $117^{+166}_{-83}$ \\
Blue Jay &
  19572 &
  150.133196 &
  2.378721 &
  1.866 &
  $11.13^{+0.11}_{-0.10}$ &
  $0.14^{+0.71}_{-0.49}$ &
  $3.02^{+0.70}_{-0.75}$ &
  $5.03^{+0.26}_{-0.25}$ &
  10.1 &
  $150^{+18}_{-19}$ &
  $27^{+33}_{-19}$ \\
Blue Jay &
  13174 &
  150.112224 &
  2.313952 &
  2.099 &
  $11.27^{+0.19}_{-0.24}$ &
  $2.59^{+0.14}_{-0.30}$ &
  $2.45^{+0.12}_{-0.12}$ &
  $6.37^{+1.86}_{-1.69}$ &
  2.3 &
  $170^{+137}_{-120}$ &
  $179^{+119}_{-119}$ \\
JADES &
  215644 &
  53.214184 &
  -27.753964 &
  1.665 &
  $11.21^{+0.07}_{-0.06}$ &
  $-2.15^{+1.36}_{-0.54}$ &
  $1.02^{+0.42}_{-0.33}$ &
  $2.14^{+0.42}_{-0.40}$ &
  3.3 &
  $140^{+94}_{-90}$ &
  $269^{+75}_{-77}$ \\
JADES &
  10008071 &
  53.154460 &
  -27.771433 &
  2.227 &
  $9.34^{+0.15}_{-0.09}$ &
  $1.61^{+0.02}_{-0.05}$ &
  $0.85^{+0.10}_{-0.10}$ &
  $8.24^{+2.42}_{-2.50}$ &
  2.5 &
  $352^{+116}_{-114}$ &
  $138^{+139}_{-95}$ \\
JADES &
  25351 &
  189.173371 &
  62.230410 &
  3.128 &
  $9.75^{+0.13}_{-0.18}$ &
  $1.66^{+0.11}_{-0.17}$ &
  $1.86^{+0.09}_{-0.09}$ &
  $5.11^{+1.84}_{-1.73}$ &
  2.1 &
  $260^{+149}_{-170}$ &
  $162^{+140}_{-111}$ \\
JADES &
  27760 &
  189.174713 &
  62.249076 &
  0.850 &
  $10.72^{+0.12}_{-0.08}$ &
  $0.55^{+0.31}_{-0.39}$ &
  $1.97^{+0.77}_{-0.77}$ &
  $4.05^{+1.37}_{-1.30}$ &
  2.7 &
  $248^{+336}_{-150}$ &
  $175^{+181}_{-126}$ \\
JADES &
  30817 &
  189.197539 &
  62.291329 &
  2.244 &
  $9.79^{+0.09}_{-0.08}$ &
  $1.67^{+0.06}_{-0.06}$ &
  $1.59^{+0.05}_{-0.05}$ &
  $2.90^{+0.82}_{-0.78}$ &
  3.0 &
  $183^{+106}_{-89}$ &
  $96^{+116}_{-67}$
\enddata
\tablecomments{1)-12) Same as Table~\ref{Tab:outf}.}
\end{deluxetable*}

\begin{deluxetable*}{cccccccccccc}
\tablecaption{Properties of galaxies with Na D systemic ISM}
\label{Tab:sys}
\tablewidth{0pt}
\tabletypesize{\scriptsize}
\tablehead{\colhead{1) Survey} &
  \colhead{2) ID} &
  \colhead{3) R.A.} &
  \colhead{4) Decl.} &
  \colhead{5) z} &
  \colhead{6) $\log(M_*)$} &
  \colhead{7) $\log({\rm SFR})$} &
  \colhead{8) $A_V$} &
  \colhead{9) ${\rm EW_{Na D, ISM}}$} &
  \colhead{10) ${
\rm SNR_{Na D, ISM}}$} &
  \colhead{11) $\Delta v$} &
  \colhead{12) $b$}
  \\
  \colhead{} &
  \colhead{} &
  \colhead{(deg)} &
  \colhead{(deg)} &
  \colhead{} &
  \colhead{$(M_\odot)$} &
  \colhead{$(M_\odot\,yr^{-1})$} &
  \colhead{$(mag)$} &
  \colhead{$(\AA)$} &
  \colhead{} &
  \colhead{$km\,s^{-1}$} &
  \colhead{$km\,s^{-1}$}
}
\startdata
SMILES &
  209480 &
  53.149098 &
  -27.774387 &
  1.093 &
  $10.92^{+0.14}_{-0.08}$ &
  $-1.20^{+1.73}_{-0.50}$ &
  $1.40^{+0.49}_{-0.49}$ &
  $2.56^{+0.62}_{-0.64}$ &
  4.0 &
  $-29^{+93}_{-87}$ &
  $106^{+101}_{-73}$ \\
SMILES &
  212434 &
  53.151409 &
  -27.766742 &
  0.895 &
  $9.61^{+0.23}_{-0.20}$ &
  $1.11^{+0.10}_{-0.12}$ &
  $2.75^{+0.40}_{-0.51}$ &
  $2.75^{+0.79}_{-0.77}$ &
  2.1 &
  $-10^{+80}_{-87}$ &
  $90^{+112}_{-63}$ \\
SMILES &
  214075 &
  53.196909 &
  -27.760532 &
  3.603 &
  $10.65^{+0.12}_{-0.06}$ &
  $-0.53^{+1.27}_{-0.50}$ &
  $1.47^{+0.67}_{-0.42}$ &
  $6.38^{+0.77}_{-0.74}$ &
  5.5 &
  $-4^{+51}_{-49}$ &
  $123^{+95}_{-78}$ \\
SMILES &
  207739 &
  53.179110 &
  -27.780619 &
  1.038 &
  $11.25^{+0.03}_{-0.03}$ &
  $-3.00^{+0.00}_{-0.00}$ &
  $5.58^{+0.41}_{-0.41}$ &
  $10.23^{+0.92}_{-0.91}$ &
  6.7 &
  $-56^{+29}_{-29}$ &
  $53^{+53}_{-37}$ \\
SMILES &
  198545 &
  53.161414 &
  -27.811144 &
  2.827 &
  $11.13^{+0.03}_{-0.03}$ &
  $0.59^{+0.09}_{-0.09}$ &
  $3.67^{+0.44}_{-0.44}$ &
  $9.23^{+1.04}_{-1.04}$ &
  4.2 &
  $-4^{+28}_{-28}$ &
  $37^{+41}_{-26}$ \\
SMILES &
  201027 &
  53.126930 &
  -27.804682 &
  2.870 &
  $10.71^{+0.05}_{-0.44}$ &
  $-0.49^{+1.86}_{-0.50}$ &
  $1.93^{+0.45}_{-0.50}$ &
  $7.31^{+0.84}_{-0.80}$ &
  5.5 &
  $-6^{+49}_{-47}$ &
  $69^{+76}_{-48}$ \\
SMILES &
  204520 &
  53.125113 &
  -27.790849 &
  1.551 &
  $11.26^{+0.08}_{-0.06}$ &
  $0.44^{+0.44}_{-0.44}$ &
  $2.83^{+0.82}_{-0.82}$ &
  $1.84^{+0.31}_{-0.30}$ &
  3.0 &
  $-11^{+58}_{-60}$ &
  $146^{+111}_{-89}$ \\
Blue Jay &
  11494 &
  150.073883 &
  2.297977 &
  2.093 &
  $11.70^{+0.06}_{-0.05}$ &
  $-2.15^{+1.49}_{-0.54}$ &
  $0.69^{+0.32}_{-0.27}$ &
  $3.72^{+1.79}_{-0.34}$ &
  3.1 &
  $-269^{+408}_{-22}$ &
  $30^{+366}_{-22}$ \\
Blue Jay &
  10565 &
  150.094200 &
  2.287267 &
  2.441 &
  $10.78^{+0.20}_{-0.23}$ &
  $-0.05^{+1.21}_{-0.50}$ &
  $0.66^{+0.66}_{-0.66}$ &
  $3.23^{+0.59}_{-0.54}$ &
  4.4 &
  $84^{+68}_{-103}$ &
  $121^{+123}_{-86}$ \\
Blue Jay &
  10021 &
  150.089377 &
  2.282296 &
  1.811 &
  $10.08^{+0.17}_{-0.17}$ &
  $0.53^{+0.49}_{-0.46}$ &
  $2.22^{+0.79}_{-0.51}$ &
  $6.81^{+1.58}_{-1.91}$ &
  3.5 &
  $-414^{+521}_{-164}$ &
  $383^{+14}_{-90}$ \\
Blue Jay &
  19705 &
  150.107752 &
  2.379484 &
  2.467 &
  $11.33^{+0.10}_{-0.22}$ &
  $-0.87^{+1.47}_{-0.49}$ &
  $5.68^{+1.17}_{-1.34}$ &
  $7.25^{+1.40}_{-1.41}$ &
  4.8 &
  $-79^{+86}_{-100}$ &
  $153^{+141}_{-108}$ \\
Blue Jay &
  16419 &
  150.095610 &
  2.350069 &
  1.925 &
  $11.34^{+0.09}_{-0.10}$ &
  $0.87^{+0.57}_{-0.48}$ &
  $0.59^{+0.35}_{-0.35}$ &
  $1.79^{+0.33}_{-0.34}$ &
  2.6 &
  $-243^{+255}_{-143}$ &
  $366^{+27}_{-101}$ \\
Blue Jay &
  9395 &
  150.125656 &
  2.275250 &
  2.127 &
  $10.78^{+0.04}_{-0.04}$ &
  $-2.40^{+1.74}_{-0.60}$ &
  $0.77^{+0.55}_{-0.39}$ &
  $1.80^{+0.37}_{-0.36}$ &
  3.9 &
  $-60^{+181}_{-115}$ &
  $218^{+132}_{-146}$ \\
Blue Jay &
  9180 &
  150.135692 &
  2.272714 &
  1.603 &
  $10.58^{+0.29}_{-0.38}$ &
  $1.50^{+0.37}_{-0.49}$ &
  $2.93^{+0.11}_{-0.11}$ &
  $6.70^{+0.66}_{-0.59}$ &
  7.1 &
  $-54^{+60}_{-72}$ &
  $118^{+69}_{-75}$ \\
JADES &
  8375 &
  53.144518 &
  -27.779103 &
  1.551 &
  $10.25^{+0.10}_{-0.10}$ &
  $1.21^{+0.09}_{-0.07}$ &
  $2.67^{+0.32}_{-0.32}$ &
  $5.39^{+2.73}_{-2.68}$ &
  2.2 &
  $-256^{+267}_{-234}$ &
  $199^{+140}_{-139}$ \\
JADES &
  44832 &
  53.132648 &
  -27.732383 &
  1.548 &
  $10.43^{+0.29}_{-0.22}$ &
  $1.90^{+0.32}_{-0.29}$ &
  $2.53^{+0.18}_{-0.18}$ &
  $3.62^{+0.80}_{-0.77}$ &
  3.3 &
  $-103^{+120}_{-92}$ &
  $90^{+113}_{-63}$ \\
JADES &
  138717 &
  53.119187 &
  -27.765842 &
  1.884 &
  $10.60^{+0.12}_{-0.07}$ &
  $-2.52^{+1.51}_{-0.48}$ &
  $1.58^{+0.54}_{-0.50}$ &
  $5.41^{+0.30}_{-0.29}$ &
  16.1 &
  $31^{+17}_{-17}$ &
  $86^{+42}_{-50}$ \\
JADES &
  163028 &
  53.099629 &
  -27.890669 &
  1.588 &
  $11.25^{+0.01}_{-0.01}$ &
  $1.62^{+0.04}_{-0.04}$ &
  $3.92^{+0.28}_{-0.31}$ &
  $1.38^{+0.37}_{-0.36}$ &
  2.4 &
  $-48^{+167}_{-289}$ &
  $332^{+53}_{-115}$ \\
JADES &
  189524 &
  53.131446 &
  -27.841383 &
  1.614 &
  $11.50^{+0.22}_{-0.04}$ &
  $0.79^{+0.54}_{-0.48}$ &
  $4.23^{+1.41}_{-0.77}$ &
  $8.44^{+0.88}_{-0.88}$ &
  6.8 &
  $37^{+53}_{-56}$ &
  $292^{+72}_{-102}$ \\
JADES &
  22456 &
  189.035722 &
  62.243154 &
  3.129 &
  $10.91^{+0.19}_{-0.34}$ &
  $2.83^{+0.16}_{-0.23}$ &
  $4.40^{+0.69}_{-0.69}$ &
  $6.32^{+2.25}_{-1.90}$ &
  2.1 &
  $14^{+226}_{-147}$ &
  $188^{+140}_{-140}$ \\
JADES &
  28428 &
  189.235679 &
  62.253620 &
  2.953 &
  $11.23^{+0.05}_{-0.09}$ &
  $0.15^{+0.76}_{-0.48}$ &
  $4.90^{+0.60}_{-0.71}$ &
  $7.75^{+1.90}_{-1.82}$ &
  2.5 &
  $47^{+135}_{-152}$ &
  $289^{+81}_{-137}$ \\
JADES &
  72396 &
  189.263048 &
  62.170385 &
  2.761 &
  $10.84^{+0.04}_{-0.08}$ &
  $-2.05^{+1.45}_{-0.65}$ &
  $1.11^{+0.42}_{-0.34}$ &
  $4.15^{+0.97}_{-0.93}$ &
  3.4 &
  $50^{+87}_{-91}$ &
  $99^{+123}_{-69}$ \\
JADES &
  72719 &
  189.199697 &
  62.172290 &
  1.736 &
  $10.79^{+0.09}_{-0.22}$ &
  $-3.00^{+0.00}_{-0.00}$ &
  $2.96^{+1.05}_{-1.05}$ &
  $4.28^{+1.25}_{-1.28}$ &
  3.0 &
  $31^{+84}_{-91}$ &
  $71^{+95}_{-50}$ \\
Aurora &
  100231 &
  189.224830 &
  62.251141 &
  2.957 &
  $10.84^{+0.09}_{-0.09}$ &
  $-0.56^{+1.03}_{-0.49}$ &
  $2.18^{+0.23}_{-0.23}$ &
  $4.96^{+1.24}_{-0.71}$ &
  9.7 &
  $-70^{+64}_{-135}$ &
  $210^{+132}_{-154}$ \\
Aurora &
  30564 &
  189.204075 &
  62.294341 &
  2.482 &
  $10.85^{+0.14}_{-0.22}$ &
  $2.17^{+0.18}_{-0.36}$ &
  $1.69^{+0.02}_{-0.02}$ &
  $1.91^{+0.92}_{-0.52}$ &
  2.9 &
  $-194^{+222}_{-374}$ &
  $240^{+111}_{-170}$ \\
Aurora &
  100238 &
  189.178228 &
  62.285188 &
  2.487 &
  $11.03^{+0.07}_{-0.07}$ &
  $-2.70^{+1.85}_{-1.30}$ &
  $0.92^{+0.55}_{-0.29}$ &
  $3.42^{+0.64}_{-0.59}$ &
  3.2 &
  $42^{+84}_{-87}$ &
  $213^{+111}_{-117}$
\enddata
\tablecomments{1)-12) Same as Table~\ref{Tab:outf}.}
\end{deluxetable*}

\begin{deluxetable*}{cccccccccccc}
\tablecaption{Properties of galaxies with tentative Na D ISM absorption detection}
\label{Tab:tentative}
\tablewidth{0pt}
\tabletypesize{\scriptsize}
\tablehead{\colhead{1) Survey} &
  \colhead{2) ID} &
  \colhead{3) R.A.} &
  \colhead{4) Decl.} &
  \colhead{5) z} &
  \colhead{6) $\log(M_*)$} &
  \colhead{7) $\log({\rm SFR})$} &
  \colhead{8) $A_V$} &
  \colhead{9) ${\rm EW_{Na D, ISM}}$} &
  \colhead{10) ${
\rm SNR_{Na D, ISM}}$} &
  \colhead{11) $\Delta v$} &
  \colhead{12) $b$}
  \\
  \colhead{} &
  \colhead{} &
  \colhead{(deg)} &
  \colhead{(deg)} &
  \colhead{} &
  \colhead{$(M_\odot)$} &
  \colhead{$(M_\odot\,yr^{-1})$} &
  \colhead{$(mag)$} &
  \colhead{$(\AA)$} &
  \colhead{} &
  \colhead{$km\,s^{-1}$} &
  \colhead{$km\,s^{-1}$}
}
\startdata
Blue Jay &
  18071 &
  150.135268 &
  2.363612 &
  2.822 &
  $10.45^{+0.45}_{-0.30}$ &
  $1.91^{+0.25}_{-0.30}$ &
  $2.53^{+0.03}_{-0.03}$ &
  $2.20^{+0.77}_{-0.79}$ &
  1.6 &
  $-115^{+128}_{-144}$ &
  $125^{+164}_{-90}$ \\
Blue Jay &
  10245 &
  150.089814 &
  2.284797 &
  1.808 &
  $10.55^{+0.19}_{-0.27}$ &
  $2.17^{+0.11}_{-0.35}$ &
  $2.49^{+0.73}_{-0.68}$ &
  $2.76^{+0.37}_{-0.35}$ &
  1.6 &
  $14^{+34}_{-33}$ &
  $46^{+49}_{-32}$ \\
JADES &
  192316 &
  53.168046 &
  -27.832569 &
  0.671 &
  $10.34^{+0.11}_{-0.18}$ &
  $-0.75^{+0.27}_{-0.48}$ &
  $0.83^{+0.57}_{-0.36}$ &
  $0.73^{+0.39}_{-0.39}$ &
  1.6 &
  $371^{+218}_{-283}$ &
  $177^{+145}_{-122}$ \\
JADES &
  33159 &
  189.194706 &
  62.307222 &
  2.050 &
  $10.49^{+0.15}_{-0.08}$ &
  $-3.00^{+0.00}_{-0.00}$ &
  $0.72^{+0.59}_{-0.43}$ &
  $2.22^{+0.88}_{-0.82}$ &
  1.6 &
  $-261^{+250}_{-273}$ &
  $255^{+108}_{-159}$ \\
JADES &
  34145 &
  189.121246 &
  62.314985 &
  1.778 &
  $10.80^{+0.13}_{-0.11}$ &
  $-1.07^{+1.60}_{-0.50}$ &
  $2.44^{+0.80}_{-0.86}$ &
  $2.86^{+1.06}_{-1.14}$ &
  1.9 &
  $-103^{+138}_{-132}$ &
  $78^{+100}_{-55}$ \\
JADES &
  173387 &
  53.109875 &
  -27.870195 &
  2.697 &
  $10.14^{+0.09}_{-0.09}$ &
  $0.65^{+0.46}_{-0.48}$ &
  $2.05^{+0.06}_{-0.06}$ &
  $0.80^{+0.64}_{-0.56}$ &
  1.5 &
  $120^{+168}_{-258}$ &
  $121^{+151}_{-86}$ \\
JADES &
  180575 &
  53.085472 &
  -27.858170 &
  3.473 &
  $10.93^{+0.10}_{-0.19}$ &
  $2.97^{+0.11}_{-0.11}$ &
  $2.35^{+0.10}_{-0.10}$ &
  $4.16^{+2.27}_{-2.21}$ &
  1.5 &
  $-106^{+188}_{-282}$ &
  $164^{+142}_{-112}$ \\
JADES &
  64734 &
  189.021773 &
  62.296589 &
  1.382 &
  $9.89^{+0.15}_{-0.07}$ &
  $0.80^{+0.08}_{-0.13}$ &
  $1.05^{+0.92}_{-0.92}$ &
  $0.67^{+0.63}_{-0.47}$ &
  1.8 &
  $-2^{+330}_{-231}$ &
  $149^{+146}_{-105}$ \\
JADES &
  71914 &
  189.253780 &
  62.167011 &
  2.459 &
  $9.84^{+0.46}_{-0.26}$ &
  $0.93^{+0.58}_{-0.38}$ &
  $2.42^{+0.91}_{-1.00}$ &
  $6.60^{+2.55}_{-2.50}$ &
  1.9 &
  $-15^{+208}_{-158}$ &
  $134^{+162}_{-96}$ \\
Aurora &
  37 &
  150.181709 &
  2.239991 &
  3.459 &
  $10.65^{+0.06}_{-0.09}$ &
  $2.05^{+0.20}_{-0.08}$ &
  $1.53^{+0.15}_{-0.15}$ &
  $2.69^{+1.64}_{-1.53}$ &
  1.7 &
  $139^{+212}_{-329}$ &
  $155^{+156}_{-110}$
\enddata
\tablecomments{1)-12) Same as Table~\ref{Tab:outf}.}
\end{deluxetable*}

\begin{deluxetable*}{ccccccccc}
\tablecaption{Na D outflow properties}
\label{Tab:outf_detail}
\tablewidth{0pt}
\tabletypesize{\scriptsize}
\tablehead{\colhead{1) Survey} &
  \colhead{2) ID} &
  \colhead{3) $\log(V_{\rm out})$} &
  \colhead{4) $\log(\dot{M}_{\rm out})$} &
  \colhead{5) $\log(\dot{E}_{\rm out})$} &
  \colhead{6) $\log(\dot{p}_{\rm out})$} &
  \colhead{7) $\eta$} &
  \colhead{8) Class} &
  \colhead{9) AGN}
  \\
  \colhead{} &
  \colhead{} &
  \colhead{$km\,s^{-1}$} &
  \colhead{$(M_\odot\,yr^{-1})$} &
  \colhead{$(erg\,s^{-1})$} &
  \colhead{$(dyne)$} &
  \colhead{} &
  \colhead{} &
  \colhead{}
}
\startdata
SMILES   & 205818 & $2.90^{+0.13}_{-0.16}$ & $1.90^{+0.56}_{-0.78}$ & $43.19^{+0.80}_{-1.26}$ & $35.59^{+0.68}_{-1.01}$ & $0.21^{+0.63}_{-0.79}$  & SF & 0 \\
SMILES   & 206183 & $2.92^{+0.03}_{-0.03}$ & $2.44^{+0.12}_{-0.19}$ & $43.77^{+0.14}_{-0.18}$ & $36.16^{+0.13}_{-0.18}$ & $1.41^{+0.12}_{-0.19}$  & QG & 0 \\
SMILES   & 194473 & $2.81^{+0.11}_{-0.15}$ & $1.93^{+0.34}_{-0.53}$ & $43.03^{+0.57}_{-0.92}$ & $35.52^{+0.46}_{-0.71}$ & $4.93^{+4.34}_{-2.41}$  & QG & 1 \\
SMILES   & 201306 & $2.78^{+0.07}_{-0.09}$ & $1.49^{+0.08}_{-0.10}$ & $42.53^{+0.20}_{-0.27}$ & $35.06^{+0.14}_{-0.17}$ & $4.49^{+0.08}_{-0.10}$  & QG & 1 \\
Blue Jay & 18668  & $2.68^{+0.15}_{-0.15}$ & $1.98^{+0.19}_{-0.38}$ & $42.85^{+0.43}_{-0.84}$ & $35.47^{+0.30}_{-0.62}$ & $4.68^{+0.28}_{-1.75}$  & QG & 1 \\
Blue Jay & 18252  & $2.70^{+0.11}_{-0.10}$ & $1.85^{+0.34}_{-0.55}$ & $42.76^{+0.51}_{-0.76}$ & $35.36^{+0.41}_{-0.66}$ & $-1.75^{+0.37}_{-0.57}$ & SF & 0 \\
Blue Jay & 11142  & $2.56^{+0.11}_{-0.09}$ & $1.22^{+0.27}_{-0.55}$ & $41.84^{+0.48}_{-0.77}$ & $34.58^{+0.37}_{-0.66}$ & $0.96^{+0.52}_{-1.10}$  & QG & 1 \\
Blue Jay & 11136  & $2.28^{+0.11}_{-0.09}$ & $0.81^{+0.36}_{-0.59}$ & $40.87^{+0.58}_{-0.79}$ & $33.89^{+0.47}_{-0.69}$ & $-0.69^{+0.53}_{-0.77}$ & SF & 0 \\
Blue Jay & 10339  & $2.58^{+0.11}_{-0.11}$ & $1.16^{+0.26}_{-0.50}$ & $41.82^{+0.47}_{-0.79}$ & $34.54^{+0.36}_{-0.65}$ & $3.31^{+0.55}_{-1.61}$  & QG & 0 \\
Blue Jay & 10314  & $2.54^{+0.11}_{-0.10}$ & $1.08^{+0.27}_{-0.36}$ & $41.70^{+0.44}_{-0.65}$ & $34.45^{+0.36}_{-0.51}$ & $-0.16^{+0.50}_{-0.48}$ & QG & 0 \\
Blue Jay & 9871   & $2.64^{+0.04}_{-0.04}$ & $1.79^{+0.07}_{-0.07}$ & $42.57^{+0.12}_{-0.13}$ & $35.23^{+0.09}_{-0.10}$ & $1.26^{+0.32}_{-1.19}$  & QG & 0 \\
Blue Jay & 8002   & $2.94^{+0.05}_{-0.09}$ & $1.88^{+0.30}_{-0.38}$ & $43.30^{+0.32}_{-0.61}$ & $35.64^{+0.31}_{-0.48}$ & $-0.30^{+0.31}_{-0.38}$ & SF & 0 \\
JADES    & 42573  & $2.41^{+0.21}_{-0.17}$ & $1.19^{+0.50}_{-0.73}$ & $41.49^{+0.88}_{-1.23}$ & $34.39^{+0.69}_{-0.98}$ & $-0.66^{+0.55}_{-0.75}$ & SF & 1 \\
JADES    & 44571  & $2.49^{+0.19}_{-0.14}$ & $1.22^{+0.40}_{-0.63}$ & $41.71^{+0.75}_{-1.08}$ & $34.52^{+0.57}_{-0.86}$ & $3.62^{+0.77}_{-1.72}$  & QG & 1 \\
JADES    & 51236  & $2.93^{+0.04}_{-0.06}$ & $2.02^{+0.25}_{-0.16}$ & $43.40^{+0.20}_{-0.21}$ & $35.76^{+0.22}_{-0.18}$ & $0.14^{+0.38}_{-0.33}$  & SF & 1 \\
JADES    & 171147 & $2.81^{+0.08}_{-0.11}$ & $1.38^{+0.22}_{-0.26}$ & $42.52^{+0.27}_{-0.51}$ & $35.00^{+0.23}_{-0.38}$ & $3.23^{+0.51}_{-1.28}$  & QG & 0 \\
JADES    & 197911 & $2.82^{+0.08}_{-0.08}$ & $1.21^{+0.18}_{-0.26}$ & $42.36^{+0.31}_{-0.45}$ & $34.84^{+0.24}_{-0.35}$ & $2.38^{+0.43}_{-1.22}$  & QG & 1 \\
JADES    & 202925 & $2.65^{+0.19}_{-0.16}$ & $0.92^{+0.51}_{-0.74}$ & $41.71^{+0.82}_{-1.23}$ & $34.36^{+0.65}_{-0.98}$ & $2.47^{+0.75}_{-1.45}$  & QG & 1 \\
JADES    & 215262 & $2.70^{+0.27}_{-0.19}$ & $1.63^{+0.74}_{-0.80}$ & $42.48^{+1.31}_{-1.42}$ & $35.10^{+1.03}_{-1.10}$ & $-0.12^{+0.75}_{-0.81}$ & SF & 1 \\
JADES    & 199773 & $2.47^{+0.21}_{-0.17}$ & $1.12^{+0.46}_{-0.76}$ & $41.56^{+0.84}_{-1.34}$ & $34.39^{+0.65}_{-1.05}$ & $1.48^{+0.58}_{-0.78}$  & QG & 1 \\
JADES    & 25055  & $2.68^{+0.17}_{-0.16}$ & $1.22^{+0.25}_{-0.46}$ & $42.06^{+0.54}_{-0.89}$ & $34.69^{+0.39}_{-0.67}$ & $3.91^{+0.35}_{-1.61}$  & QG & 1 \\
JADES    & 29711  & $2.75^{+0.11}_{-0.11}$ & $1.88^{+0.38}_{-0.55}$ & $42.90^{+0.50}_{-0.85}$ & $35.44^{+0.43}_{-0.70}$ & $-0.63^{+0.47}_{-0.56}$ & SF & 0 \\
JADES    & 64818  & $2.78^{+0.23}_{-0.20}$ & $1.53^{+0.79}_{-0.88}$ & $42.56^{+1.26}_{-1.60}$ & $35.09^{+1.03}_{-1.23}$ & $-0.11^{+0.79}_{-0.88}$ & SF & 0 \\
Aurora   & 50     & $2.66^{+0.11}_{-0.10}$ & $1.66^{+0.25}_{-0.47}$ & $42.50^{+0.44}_{-0.74}$ & $35.13^{+0.34}_{-0.60}$ & $1.71^{+0.49}_{-2.52}$  & QG & 1 \\
Aurora   & 5901   & $2.81^{+0.16}_{-0.16}$ & $1.59^{+0.55}_{-0.77}$ & $42.68^{+0.84}_{-1.24}$ & $35.19^{+0.69}_{-1.00}$ & $-0.16^{+0.57}_{-0.78}$ & SF & 0
\enddata
\tablecomments{1) Survey Name; 2) Source NIRSpec ID; 3) Outflow Velocity, defined as $\Delta v+2\sigma$; 4) Outflow Mass rate; 5) Outflow energy rate; 6) Outflow momentum rate; 7) mass loading factor, definded as $\dot{M}_{\rm out}/{\rm SFR}$; 8) Host galaxy type: ``SF" and ``QG" means star-forming and quiescent, respectively; 9) 1 (0) if host has (does not have) an AGN.}
\end{deluxetable*}

\begin{figure*}
\figurenum{A1}
\centering
\includegraphics[width=0.8\textwidth]{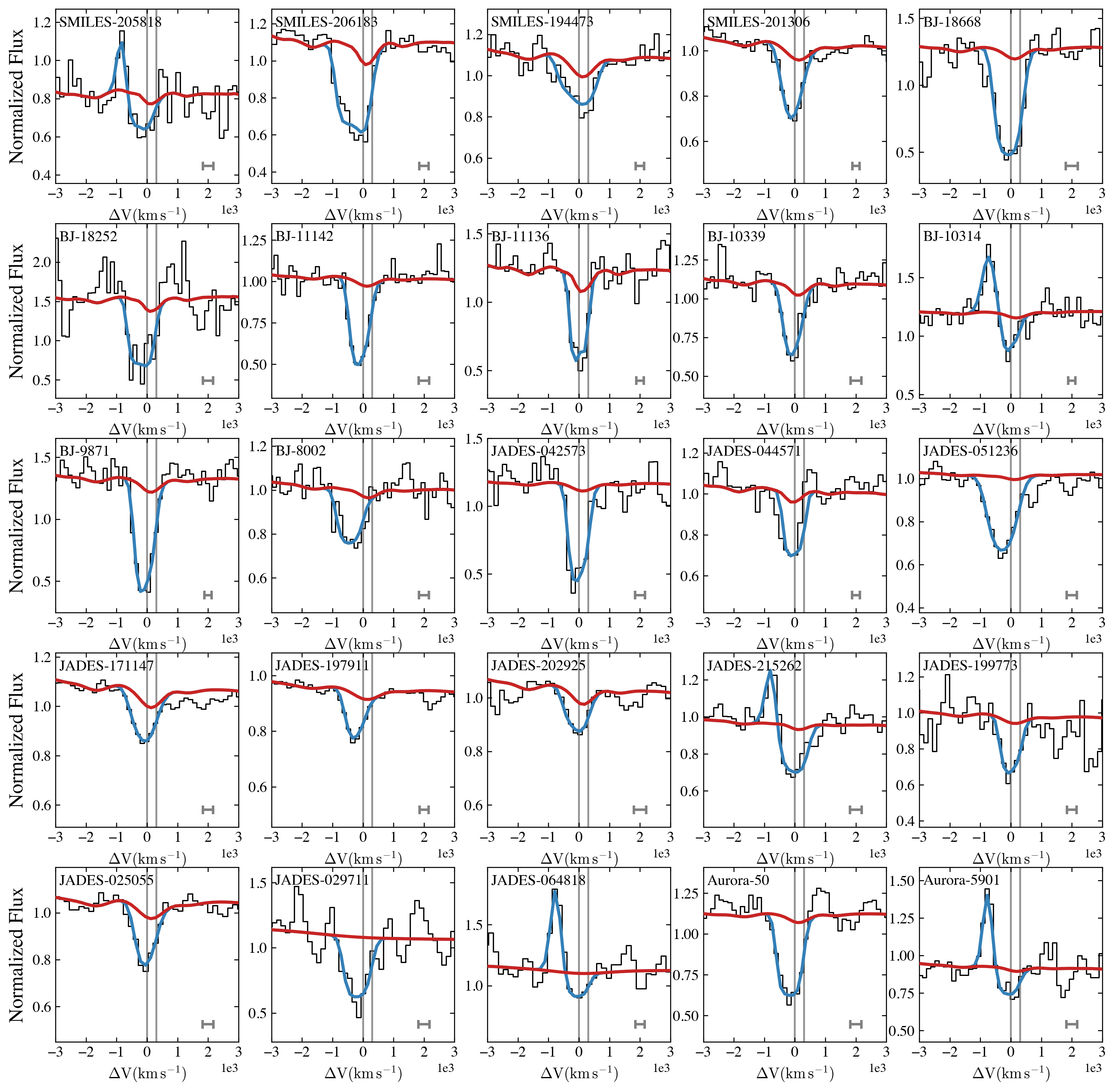}
\caption{Best-fit Na D profile of 25 outflows. The black, red, and blue lines show the 1D JWST/NIRSpec MSA spectrum, the best-fit stellar continuum by pPXF, and the best-fit ISM Na D absorption ($+$ He I 5877\AA~emission if present), respectively. The two gray vertical lines illustrate the systemic velocity of Na D$\lambda\lambda$ 5890, 5896. The gray bar represents the FWHM of the spectral resolution.}
\label{fig:NaD_outf_all}
\end{figure*}

\begin{figure*}
\figurenum{A2}
\centering
\includegraphics[width=0.8\textwidth]{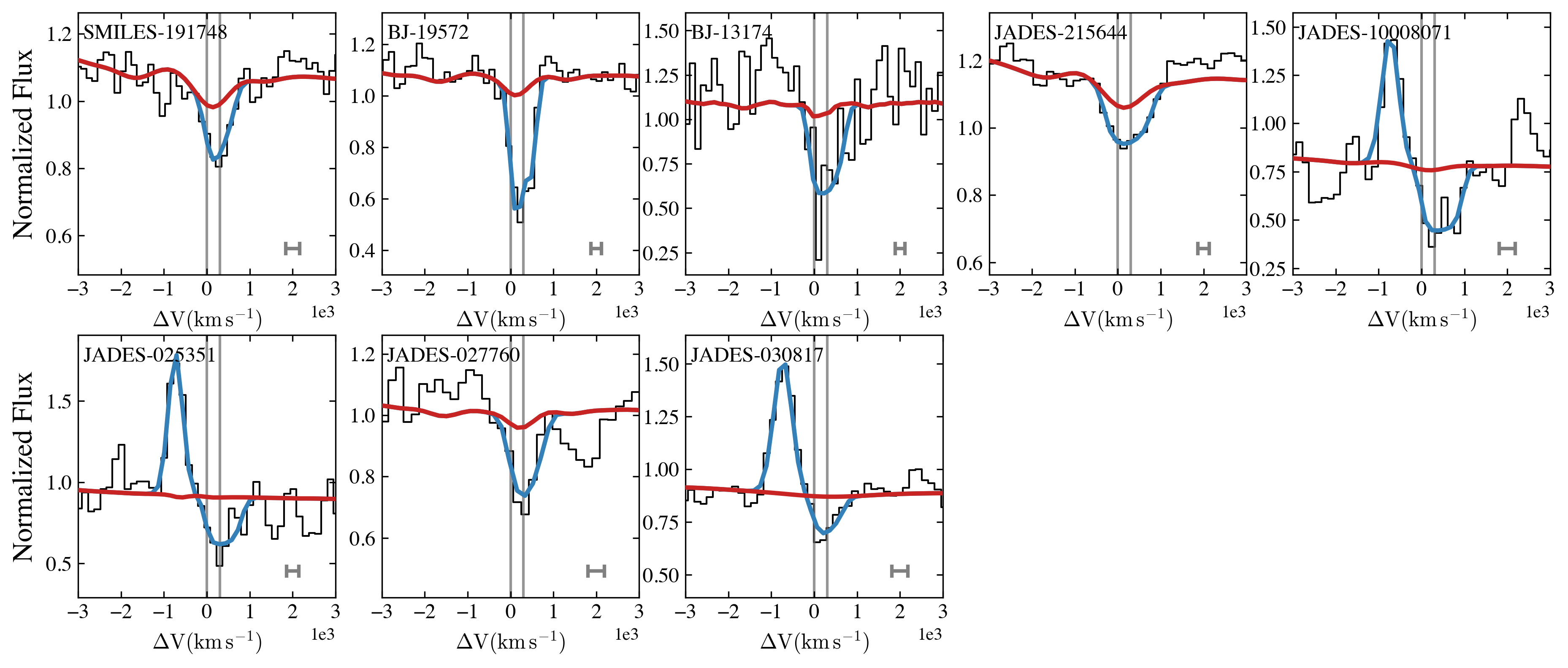}
\caption{Best-fit Na D profile of 8 inflows. The marks are the same as Figure~\ref{fig:NaD_outf_all}.}
\label{fig:NaD_inf_all}
\end{figure*}

\begin{figure*}
\figurenum{A3}
\centering
\includegraphics[width=0.8\textwidth]{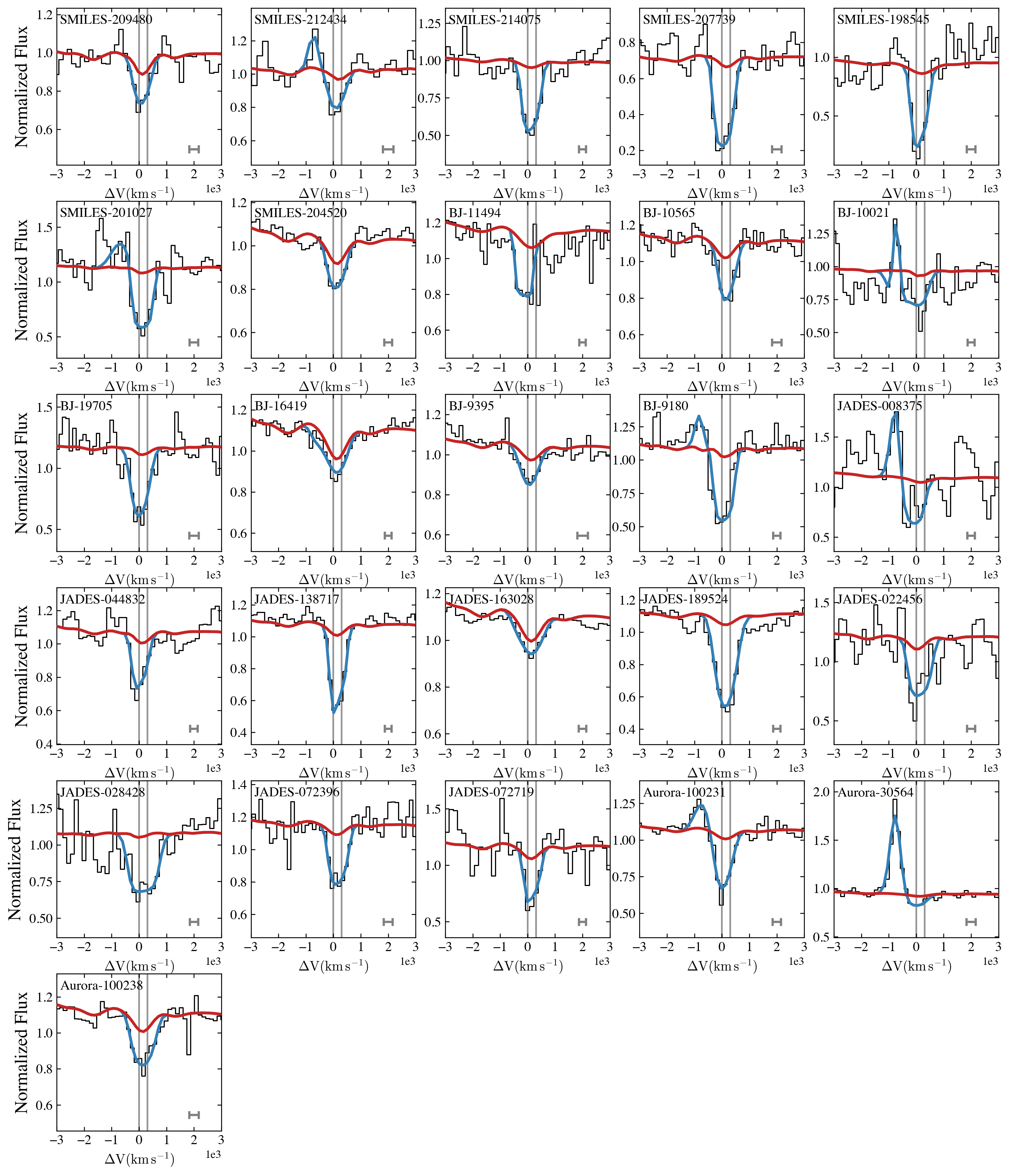}
\caption{Best-fit Na D profile of 26 galaxies with Na D-traced systemic ISMs. The marks are the same as Figure~\ref{fig:NaD_outf_all}.}
\label{fig:NaD_sys_all}
\end{figure*}

\begin{figure*}
\figurenum{A4}
\centering
\includegraphics[width=0.8\textwidth]{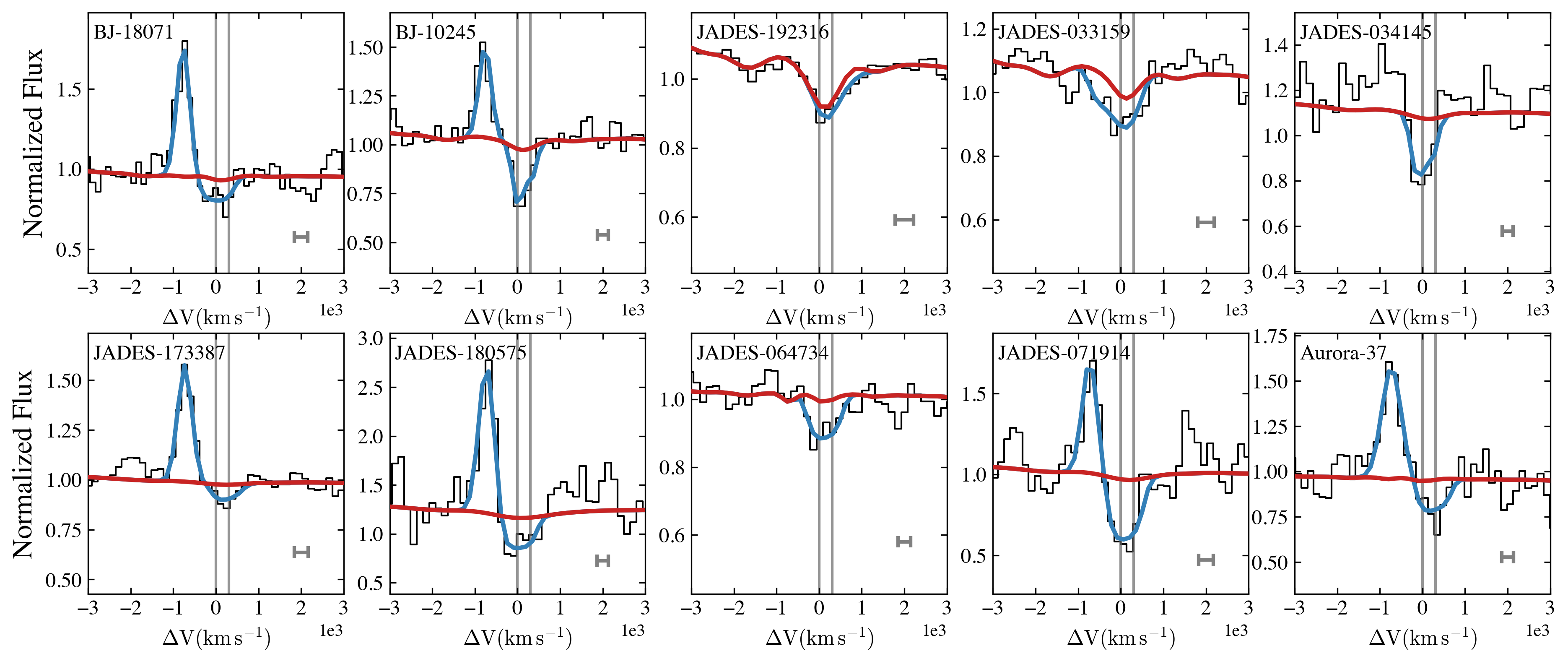}
\caption{Best-fit Na D profile of 10 galaxies with ``tentative" Na D ISM detection. The marks are the same as Figure~\ref{fig:NaD_outf_all}.}
\label{fig:NaD_tentative_all}
\end{figure*}

In this section, we first present the host galaxy properties and their best-fit Na D profiles of 25 Na D outflows, 8 inflows, 26 systemic ISMs, and 10 ``tentative" Na D ISMs. The fitting procedure is described in detail in Section~\ref{sec:nad_profile}. The profiles of detected outflows, inflows, systemic ISMs, and tentative detections are shown in Figure~\ref{fig:NaD_outf_all}, \ref{fig:NaD_inf_all}, \ref{fig:NaD_sys_all}, and \ref{fig:NaD_tentative_all} respectively, and their host galaxy properties and best-fit profile parameters are shown in Table~\ref{Tab:outf}, \ref{Tab:inf}, \ref{Tab:sys}, and \ref{Tab:tentative} respectively.

Finally, in Table~\ref{Tab:outf_detail}, we list the outflow properties ($V_{\rm{out}}$, $\dot{M}_{\rm{out}}$, $\dot{E}_{\rm{out}}$, $\dot{p}_{\rm{out}}$, and $\eta$), and their host galaxy type and AGN classifications. The definition for each outflow property is described in Section~\ref{sec:nad_outf_prop}.

\section{pPXF stellar continuum modeling examination}\label{ap: Mgbtest}

In this section, we present a robustness test of pPXF stellar fitting to assess the reliability of Na D ISM detection, especically on the quiescent galaxy sample. We examine the pPXF-subtracted residual spectrum of Mg b (5167, 5173, and 5184 \AA), a pure stellar feature, to test whether pPXF models it well. First, we measure the averaged significance of the flux difference between the original spectrum and the pPXF best-fit stellar continuum around Mg b (5150 -- 5200\AA), i.e., $(F_{\rm obs}-F_{\rm *, pPXF})/F_{\rm err}$. Meanwhile, we also perform median stacking on the Mg b residual for the entire Na D ISM-detected sample, as well as for the ISM-detected quiescent subsample only, to examine whether Mg b stellar features remain in the residuals of the Na D ISM-detected sample. 

\begin{figure*}
\figurenum{B1}
\centering
\includegraphics[width=0.8\textwidth]{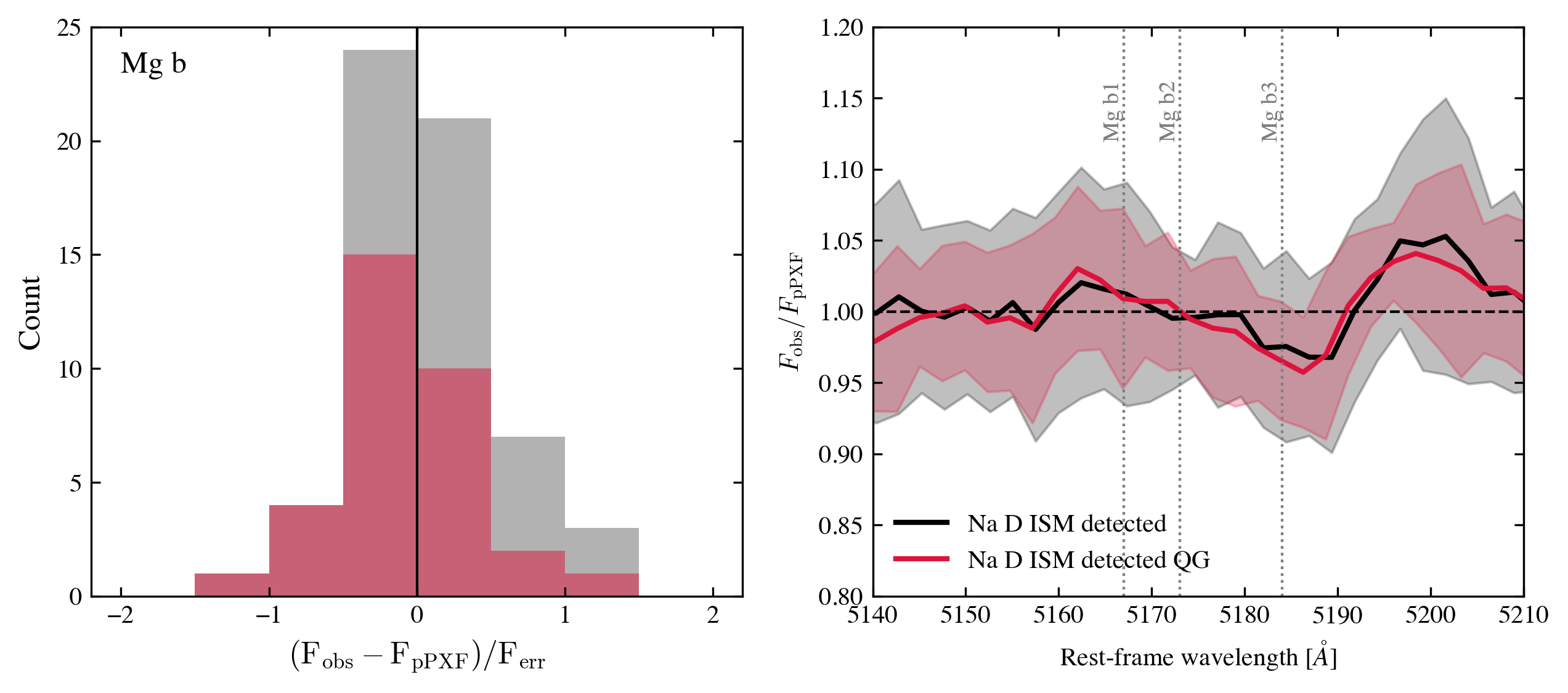}
\caption{Distribution of the flux residuals (left) and stacked residual spectra (right) for the purely stellar Mg b feature in the Na D ISM-detected sample. Results for the full Na D ISM-detected sample are shown in dark grey, while those for the Na D ISM-detected quiescent galaxies are shown in red. Both tests confirm that Mg b is generally well modeled by pPXF, suggesting that the stellar contribution around Na D has been properly removed and that the Na D ISM detections are robust.}
\label{fig:Mgb_check}
\end{figure*}

The left panel of Figure~\ref{fig:Mgb_check} shows that the Mg b residual distributions for both the full ISM-detected sample and the ISM-detected quiescent subsample are generally centered around zero. The right panel of Figure~\ref{fig:Mgb_check} further illustrates that the stacked Mg b residual spectra for these two samples no longer show any obvious Mg b absorption. These results indicate that pPXF generally models the stellar continuum well, and therefore support the robustness of the Na D ISM detections, especially for quiescent galaxies.

\section{SED-derived and H$\alpha$-derived SFR comparison}
\label{ap:SFR_comp}

\begin{figure*}
\figurenum{C1}
\centering
\includegraphics[width=0.7\textwidth]{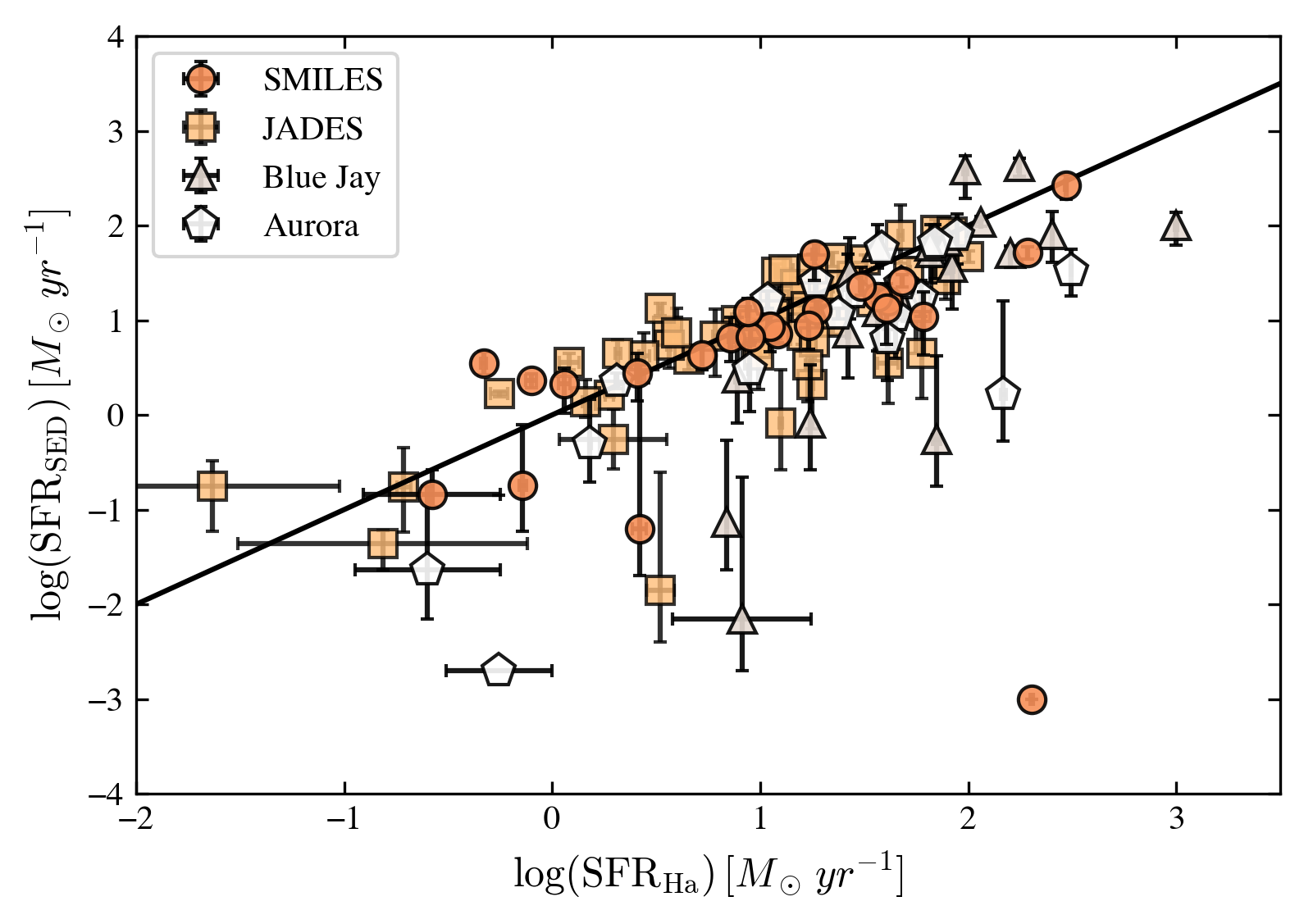}
\caption{Comparison between H$\alpha$-based SFR and SED-based SFR. Galaxies from the SMILES, JADES, Blue Jay, and Aurora surveys are marked by circles, squares, triangles, and pentagon, respectively. Overall the fiducial SFR from SED modeling agrees well with the H$\alpha$ measurements.}
\label{fig:SFR_comp}
\end{figure*}

In this section, we test the robustness of our SFR estimation through SED modeling (Section~\ref{sec:SED}) by comparing it with H$\alpha$-inferred SFR. As we mentioned in Section~\ref{sec:ppxf}
H$\alpha$ flux and uncertainty are estimated from pPXF stellar$+$nebular emission modeling using Monte Carlo, and then dust-corrected using $A_V$ derived from the Balmer Decrement, or SED if Balmer Decrement is unavailable. Furthermore, as we pointed out in Section~\ref{sec:data}, we correct the slit-loss of the MSA spectrum by matching with the broad-band NIRCam photometries for individual galaxies to derive the total H$\alpha$ luminosity of the entire galaxy. Finally, to ensure reliable H$\alpha$-based SFR, we exclude the galaxies with low-SNR H$\alpha$ (SNR$<$1) or classified as AGN or lie in the ``composite" region of the [\NII]-BPT diagram (see Section~\ref{sec:BPT}).
Finally, we convert $L_{H\alpha}$ to SFR using the relation from \citet{Kennicutt2012}.

Figure~\ref{fig:SFR_comp} illustrates the comparison between SED-based and H$\alpha$-based SFR for the parent galaxies with reliable H$\alpha$-based SFR measurements. Overall, the two estimates agree well, with increasing scatter and larger SED uncertainties toward low SFR $\log({\rm SFR_{SED}}[M_\odot\,yr^{-1}])<0$, supporting the robustness of our SED modeling.

One notable outlier lies in the lower right corner of Figure~\ref{fig:SFR_comp}, whose SED-based SFR is much smaller than the H$\alpha$-based one. This galaxy is SMILES-207739 ($z=1.04$), a faint submillimeter galaxy with ALMA detection by the ASPECS-LP program \citep{Aravena2020}. Using multiwavelength photometry from the optical to millimeter, \citet{Aravena2020} derived its SED-based SFR with ${\rm\log(SFR[M_\odot\,yr^{-1}]})=1.25$, 4 dex higher than our SED-derived SFR. Adopting their IR/mm-constrained SED-based SFR reduces the discrepancy, bringing the SED- and $H\alpha$-based SFRs into agreement within the overall scatter. This suggests that for SMILES-207739, despite its general low SFR (${\rm\log(sSFR[yr^{-1}]})\sim-10$), our UV–to-NIR SED fitting underestimates its true SFR because it lacks the mid/far-IR and mm constraints which can robustly capture heavily obscured star formation.

\section{Comparison between Three-Nod versus Two-Nod local background subtraction}
\label{ap:nod_comp}

\begin{figure*}
\figurenum{D1}
\centering
\includegraphics[width=1\textwidth]{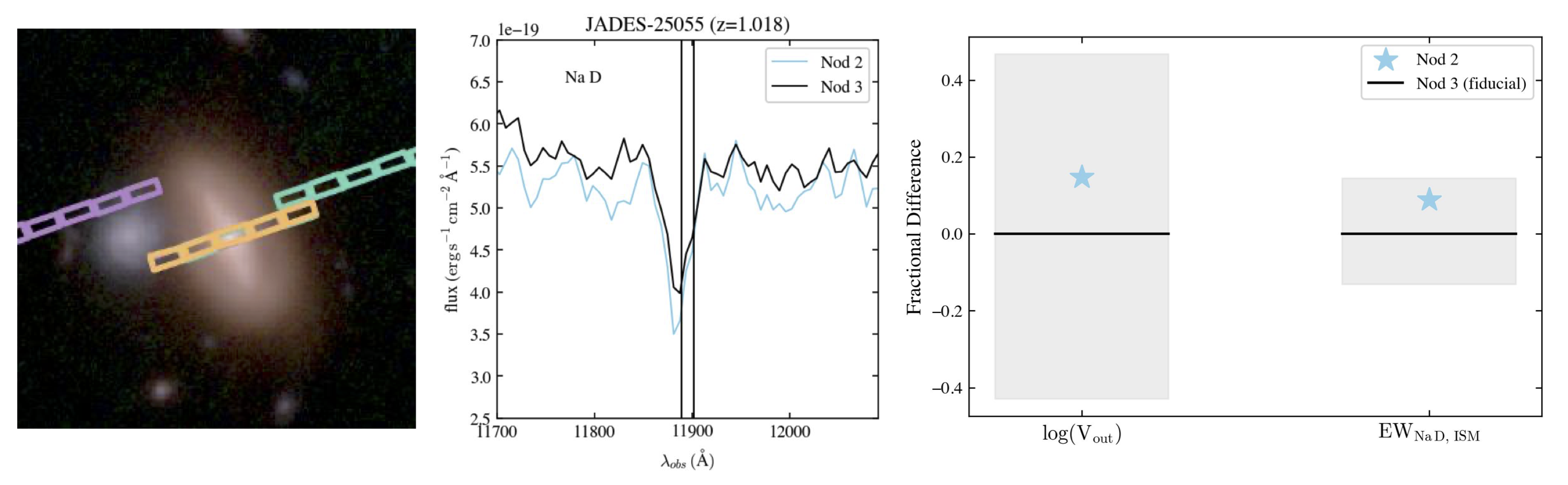}
\caption{Comparison of spectra and Na D ISM profile measurements between 3-nod and 2-nod data reduction versions for JADES-25055 at z$=$1.018.}
\label{fig:nod_comp}
\end{figure*}

We assess the potential self-subtraction problem when using the three-nod background subtraction for spatially extended galaxies. As a representative case, we select JADES-25055 at $z=1.018$, which has an effective radius of $r_{\rm e}=0.44''$ (Figure~\ref{fig:nod_comp}, left). The middle panel of Figure~\ref{fig:nod_comp} compares the spectrum adopted in this work, which is reduced with the three-nod local background subtraction, to the alternative two-nod reduction. We find that the spectra between the three-nod and two-nod versions are broadly consistent, although the three-nod spectrum shows a slightly shallower Na D absorption (most noticeably on the blue side), which may be due to modest self-subtraction in the local-background estimate for an extended source.

We further quantify the impact on the derived interstellar Na D equivalent width $\mathrm{EW}_{\mathrm{Na\,D,\,ISM}}$, and outflow velocity $V_{\rm out}$. As shown in the right panel of Figure~\ref{fig:nod_comp}, the values measured from the two-nod spectrum are slightly larger than those from the three-nod spectrum, consistent with the deeper and bluer Na D profile in the former. However, the differences are $<20\%$ and fully consistent within the $1\sigma$ uncertainties of our fiducial (three-nod) measurements. We therefore conclude that potential self-subtraction in the three-nod reduction does not significantly affect our Na D measurements and our scientific conclusions.

\bibliography{reference}{}

\begin{thebibliography}{}
\expandafter\ifx\csname natexlab\endcsname\relax\def\natexlab#1{#1}\fi
\providecommand{\url}[1]{\href{#1}{#1}}
\providecommand{\dodoi}[1]{doi:~\href{http://doi.org/#1}{\nolinkurl{#1}}}
\providecommand{\doeprint}[1]{\href{http://ascl.net/#1}{\nolinkurl{http://ascl.net/#1}}}
\providecommand{\doarXiv}[1]{\href{https://arxiv.org/abs/#1}{\nolinkurl{https://arxiv.org/abs/#1}}}

\bibitem[{S. {Adscheid} {et~al.}(2024){Adscheid}, {Magnelli}, {Liu}, {Bertoldi}, {Delvecchio}, {Gruppioni}, {Schinnerer}, {Traina}, {B{\'e}thermin}, \& {Gkogkou}}]{Adscheid2024}
{Adscheid}, S., {Magnelli}, B., {Liu}, D., {et~al.} 2024, \bibinfo{title}{{A$^{3}$COSMOS and A$^{3}$GOODSS: Continuum source catalogues and multi-band number counts},} \aap, 685, A1, \dodoi{10.1051/0004-6361/202348407}

\bibitem[{S. {Alberts} {et~al.}(2024){Alberts}, {Lyu}, {Shivaei}, {Rieke}, {P{\'e}rez-Gonz{\'a}lez}, {Bonaventura}, {Zhu}, {Helton}, {Ji}, {Morrison}, {Robertson}, {Stone}, {Sun}, {Williams}, \& {Willmer}}]{Alberts2024}
{Alberts}, S., {Lyu}, J., {Shivaei}, I., {et~al.} 2024, \bibinfo{title}{{SMILES Initial Data Release: Unveiling the Obscured Universe with MIRI Multiband Imaging},} \apj, 976, 224, \dodoi{10.3847/1538-4357/ad7396}

\bibitem[{M. {Aravena} {et~al.}(2020){Aravena}, {Boogaard}, {G{\'o}nzalez-L{\'o}pez}, {Decarli}, {Walter}, {Carilli}, {Smail}, {Weiss}, {Assef}, {Bauer}, {Bouwens}, {Cortes}, {Cox}, {da Cunha}, {Daddi}, {D{\'\i}az-Santos}, {Inami}, {Ivison}, {Novak}, {Popping}, {Riechers}, {van der Werf}, \& {Wagg}}]{Aravena2020}
{Aravena}, M., {Boogaard}, L., {G{\'o}nzalez-L{\'o}pez}, J., {et~al.} 2020, \bibinfo{title}{{The ALMA Spectroscopic Survey in the Hubble Ultra Deep Field: The Nature of the Faintest Dusty Star-forming Galaxies},} \apj, 901, 79, \dodoi{10.3847/1538-4357/ab99a2}

\bibitem[{ {Astropy Collaboration} {et~al.}(2013){Astropy Collaboration}, {Robitaille}, {Tollerud}, {Greenfield}, {Droettboom}, {Bray}, {Aldcroft}, {Davis}, {Ginsburg}, {Price-Whelan}, {Kerzendorf}, {Conley}, {Crighton}, {Barbary}, {Muna}, {Ferguson}, {Grollier}, {Parikh}, {Nair}, {Unther}, {Deil}, {Woillez}, {Conseil}, {Kramer}, {Turner}, {Singer}, {Fox}, {Weaver}, {Zabalza}, {Edwards}, {Azalee Bostroem}, {Burke}, {Casey}, {Crawford}, {Dencheva}, {Ely}, {Jenness}, {Labrie}, {Lim}, {Pierfederici}, {Pontzen}, {Ptak}, {Refsdal}, {Servillat}, \& {Streicher}}]{astropy2013}
{Astropy Collaboration}, {Robitaille}, T.~P., {Tollerud}, E.~J., {et~al.} 2013, \bibinfo{title}{{Astropy: A community Python package for astronomy},} \aap, 558, A33, \dodoi{10.1051/0004-6361/201322068}

\bibitem[{ {Astropy Collaboration} {et~al.}(2018){Astropy Collaboration}, {Price-Whelan}, {Sip{\H{o}}cz}, {G{\"u}nther}, {Lim}, {Crawford}, {Conseil}, {Shupe}, {Craig}, {Dencheva}, {Ginsburg}, {VanderPlas}, {Bradley}, {P{\'e}rez-Su{\'a}rez}, {de Val-Borro}, {Aldcroft}, {Cruz}, {Robitaille}, {Tollerud}, {Ardelean}, {Babej}, {Bach}, {Bachetti}, {Bakanov}, {Bamford}, {Barentsen}, {Barmby}, {Baumbach}, {Berry}, {Biscani}, {Boquien}, {Bostroem}, {Bouma}, {Brammer}, {Bray}, {Breytenbach}, {Buddelmeijer}, {Burke}, {Calderone}, {Cano Rodr{\'\i}guez}, {Cara}, {Cardoso}, {Cheedella}, {Copin}, {Corrales}, {Crichton}, {D'Avella}, {Deil}, {Depagne}, {Dietrich}, {Donath}, {Droettboom}, {Earl}, {Erben}, {Fabbro}, {Ferreira}, {Finethy}, {Fox}, {Garrison}, {Gibbons}, {Goldstein}, {Gommers}, {Greco}, {Greenfield}, {Groener}, {Grollier}, {Hagen}, {Hirst}, {Homeier}, {Horton}, {Hosseinzadeh}, {Hu}, {Hunkeler}, {Ivezi{\'c}}, {Jain}, {Jenness}, {Kanarek}, {Kendrew}, {Kern}, {Kerzendorf}, {Khvalko}, {King}, {Kirkby}, {Kulkarni},
  {Kumar}, {Lee}, {Lenz}, {Littlefair}, {Ma}, {Macleod}, {Mastropietro}, {McCully}, {Montagnac}, {Morris}, {Mueller}, {Mumford}, {Muna}, {Murphy}, {Nelson}, {Nguyen}, {Ninan}, {N{\"o}the}, {Ogaz}, {Oh}, {Parejko}, {Parley}, {Pascual}, {Patil}, {Patil}, {Plunkett}, {Prochaska}, {Rastogi}, {Reddy Janga}, {Sabater}, {Sakurikar}, {Seifert}, {Sherbert}, {Sherwood-Taylor}, {Shih}, {Sick}, {Silbiger}, {Singanamalla}, {Singer}, {Sladen}, {Sooley}, {Sornarajah}, {Streicher}, {Teuben}, {Thomas}, {Tremblay}, {Turner}, {Terr{\'o}n}, {van Kerkwijk}, {de la Vega}, {Watkins}, {Weaver}, {Whitmore}, {Woillez}, {Zabalza}, \& {Astropy Contributors}}]{astropy2018}
{Astropy Collaboration}, {Price-Whelan}, A.~M., {Sip{\H{o}}cz}, B.~M., {et~al.} 2018, \bibinfo{title}{{The Astropy Project: Building an Open-science Project and Status of the v2.0 Core Package},} \aj, 156, 123, \dodoi{10.3847/1538-3881/aabc4f}

\bibitem[{ {Astropy Collaboration} {et~al.}(2022){Astropy Collaboration}, {Price-Whelan}, {Lim}, {Earl}, {Starkman}, {Bradley}, {Shupe}, {Patil}, {Corrales}, {Brasseur}, {N{\"o}the}, {Donath}, {Tollerud}, {Morris}, {Ginsburg}, {Vaher}, {Weaver}, {Tocknell}, {Jamieson}, {van Kerkwijk}, {Robitaille}, {Merry}, {Bachetti}, {G{\"u}nther}, {Aldcroft}, {Alvarado-Montes}, {Archibald}, {B{\'o}di}, {Bapat}, {Barentsen}, {Baz{\'a}n}, {Biswas}, {Boquien}, {Burke}, {Cara}, {Cara}, {Conroy}, {Conseil}, {Craig}, {Cross}, {Cruz}, {D'Eugenio}, {Dencheva}, {Devillepoix}, {Dietrich}, {Eigenbrot}, {Erben}, {Ferreira}, {Foreman-Mackey}, {Fox}, {Freij}, {Garg}, {Geda}, {Glattly}, {Gondhalekar}, {Gordon}, {Grant}, {Greenfield}, {Groener}, {Guest}, {Gurovich}, {Handberg}, {Hart}, {Hatfield-Dodds}, {Homeier}, {Hosseinzadeh}, {Jenness}, {Jones}, {Joseph}, {Kalmbach}, {Karamehmetoglu}, {Ka{\l}uszy{\'n}ski}, {Kelley}, {Kern}, {Kerzendorf}, {Koch}, {Kulumani}, {Lee}, {Ly}, {Ma}, {MacBride}, {Maljaars}, {Muna}, {Murphy}, {Norman},
  {O'Steen}, {Oman}, {Pacifici}, {Pascual}, {Pascual-Granado}, {Patil}, {Perren}, {Pickering}, {Rastogi}, {Roulston}, {Ryan}, {Rykoff}, {Sabater}, {Sakurikar}, {Salgado}, {Sanghi}, {Saunders}, {Savchenko}, {Schwardt}, {Seifert-Eckert}, {Shih}, {Jain}, {Shukla}, {Sick}, {Simpson}, {Singanamalla}, {Singer}, {Singhal}, {Sinha}, {Sip{\H{o}}cz}, {Spitler}, {Stansby}, {Streicher}, {{\v{S}}umak}, {Swinbank}, {Taranu}, {Tewary}, {Tremblay}, {de Val-Borro}, {Van Kooten}, {Vasovi{\'c}}, {Verma}, {de Miranda Cardoso}, {Williams}, {Wilson}, {Winkel}, {Wood-Vasey}, {Xue}, {Yoachim}, {Zhang}, {Zonca}, \& {Astropy Project Contributors}}]{astropy2022}
{Astropy Collaboration}, {Price-Whelan}, A.~M., {Lim}, P.~L., {et~al.} 2022, \bibinfo{title}{{The Astropy Project: Sustaining and Growing a Community-oriented Open-source Project and the Latest Major Release (v5.0) of the Core Package},} \apj, 935, 167, \dodoi{10.3847/1538-4357/ac7c74}

\bibitem[{W.~M. {Baker} {et~al.}(2025{\natexlab{a}}){Baker}, {Valentino}, {Lagos}, {Ito}, {Jespersen}, {Gottumukkala}, {Hjorth}, {Langeroodi}, \& {Sedgewick}}]{Baker2025b}
{Baker}, W.~M., {Valentino}, F., {Lagos}, C. d.~P., {et~al.} 2025{\natexlab{a}}, \bibinfo{title}{{Exploring over 700 massive quiescent galaxies at z = 2─7: Demographics and stellar mass functions},} \aap, 702, A270, \dodoi{10.1051/0004-6361/202555829}

\bibitem[{W.~M. {Baker} {et~al.}(2025{\natexlab{b}}){Baker}, {Lim}, {D'Eugenio}, {Maiolino}, {Ji}, {Arribas}, {Bunker}, {Carniani}, {Charlot}, {de Graaff}, {Hainline}, {Looser}, {Lyu}, {Rinaldi}, {Robertson}, {Schaller}, {Schaye}, {Scholtz}, {{\"U}bler}, {Williams}, {Willmer}, {Willott}, \& {Zhu}}]{Baker2025}
{Baker}, W.~M., {Lim}, S., {D'Eugenio}, F., {et~al.} 2025{\natexlab{b}}, \bibinfo{title}{{The abundance and nature of high-redshift quiescent galaxies from JADES spectroscopy and the FLAMINGO simulations},} \mnras, 539, 557, \dodoi{10.1093/mnras/staf475}

\bibitem[{J.~A. {Baldwin} {et~al.}(1981){Baldwin}, {Phillips}, \& {Terlevich}}]{Baldwin1981}
{Baldwin}, J.~A., {Phillips}, M.~M., \& {Terlevich}, R. 1981, \bibinfo{title}{{Classification parameters for the emission-line spectra of extragalactic objects.},} \pasp, 93, 5, \dodoi{10.1086/130766}

\bibitem[{M.~L. {Balogh} {et~al.}(1999){Balogh}, {Morris}, {Yee}, {Carlberg}, \& {Ellingson}}]{Balogh1999}
{Balogh}, M.~L., {Morris}, S.~L., {Yee}, H.~K.~C., {Carlberg}, R.~G., \& {Ellingson}, E. 1999, \bibinfo{title}{{Differential Galaxy Evolution in Cluster and Field Galaxies at z\raisebox{-0.5ex}\textasciitilde0.3},} \apj, 527, 54, \dodoi{10.1086/308056}

\bibitem[{D. {Baron} {et~al.}(2022){Baron}, {Netzer}, {Lutz}, {Prochaska}, \& {Davies}}]{Baron2022}
{Baron}, D., {Netzer}, H., {Lutz}, D., {Prochaska}, J.~X., \& {Davies}, R.~I. 2022, \bibinfo{title}{{Multiphase outflows in post-starburst E+A galaxies - I. General sample properties and the prevalence of obscured starbursts},} \mnras, 509, 4457, \dodoi{10.1093/mnras/stab3232}

\bibitem[{S. {Belli} {et~al.}(2024){Belli}, {Park}, {Davies}, {Mendel}, {Johnson}, {Conroy}, {Benton}, {Bugiani}, {Emami}, {Leja}, {Li}, {Maheson}, {Mathews}, {Naidu}, {Nelson}, {Tacchella}, {Terrazas}, \& {Weinberger}}]{Belli2024}
{Belli}, S., {Park}, M., {Davies}, R.~L., {et~al.} 2024, \bibinfo{title}{{Star formation shut down by multiphase gas outflow in a galaxy at a redshift of 2.45},} \nat, 630, 54, \dodoi{10.1038/s41586-024-07412-1}

\bibitem[{S. {Belli} {et~al.}(2025){Belli}, {Bugiani}, {Park}, {Mendel}, {Davies}, {Khoram}, {Johnson}, {Leja}, {Tacchella}, {Brown}, {Conroy}, {Emami}, {Li}, {Liboni}, {Maheson}, {Mathews}, {Naidu}, {Nelson}, {Terrazas}, \& {Weinberger}}]{Belli2025}
{Belli}, S., {Bugiani}, L., {Park}, M., {et~al.} 2025, \bibinfo{title}{{The Blue Jay Survey: Deep JWST Spectroscopy for a Representative Sample of Galaxies at Cosmic Noon},} arXiv e-prints, arXiv:2510.11775, \dodoi{10.48550/arXiv.2510.11775}

\bibitem[{R. {Bezanson} {et~al.}(2022){Bezanson}, {Spilker}, {Suess}, {Setton}, {Feldmann}, {Greene}, {Kriek}, {Narayanan}, \& {Verrico}}]{Bezanson2022}
{Bezanson}, R., {Spilker}, J.~S., {Suess}, K.~A., {et~al.} 2022, \bibinfo{title}{{Now You See It, Now You Don't: Star Formation Truncation Precedes the Loss of Molecular Gas by 100 Myr in Massive Poststarburst Galaxies at z 0.6},} \apj, 925, 153, \dodoi{10.3847/1538-4357/ac3dfa}

\bibitem[{J. {Brinchmann} {et~al.}(2004){Brinchmann}, {Charlot}, {White}, {Tremonti}, {Kauffmann}, {Heckman}, \& {Brinkmann}}]{Brinchmann2004}
{Brinchmann}, J., {Charlot}, S., {White}, S.~D.~M., {et~al.} 2004, \bibinfo{title}{{The physical properties of star-forming galaxies in the low-redshift Universe},} \mnras, 351, 1151, \dodoi{10.1111/j.1365-2966.2004.07881.x}

\bibitem[{L. {Bugiani} {et~al.}(2025){Bugiani}, {Belli}, {Park}, {Davies}, {Mendel}, {Johnson}, {Khoram}, {Benton}, {Cimatti}, {Conroy}, {Emami}, {Leja}, {Li}, {Maheson}, {Mathews}, {Naidu}, {Nelson}, {Tacchella}, {Terrazas}, \& {Weinberger}}]{Bugiani2025}
{Bugiani}, L., {Belli}, S., {Park}, M., {et~al.} 2025, \bibinfo{title}{{Active Galactic Nucleus Feedback in Quiescent Galaxies at Cosmic Noon Traced by Ionized Gas Emission},} \apj, 981, 25, \dodoi{10.3847/1538-4357/adaeaf}

\bibitem[{A.~J. {Bunker} {et~al.}(2024){Bunker}, {Cameron}, {Curtis-Lake}, {Jakobsen}, {Carniani}, {Curti}, {Witstok}, {Maiolino}, {D'Eugenio}, {Looser}, {Willott}, {Bonaventura}, {Hainline}, {{\"U}bler}, {Willmer}, {Saxena}, {Smit}, {Alberts}, {Arribas}, {Baker}, {Baum}, {Bhatawdekar}, {Bowler}, {Boyett}, {Charlot}, {Chen}, {Chevallard}, {Circosta}, {DeCoursey}, {de Graaff}, {Egami}, {Eisenstein}, {Endsley}, {Ferruit}, {Giardino}, {Hausen}, {Helton}, {Hviding}, {Ji}, {Johnson}, {Jones}, {Kumari}, {Laseter}, {L{\"u}tzgendorf}, {Maseda}, {Nelson}, {Parlanti}, {Perna}, {Rauscher}, {Rawle}, {Rix}, {Rieke}, {Robertson}, {Rodr{\'\i}guez Del Pino}, {Sandles}, {Scholtz}, {Sharpe}, {Skarbinski}, {Stark}, {Sun}, {Tacchella}, {Topping}, {Villanueva}, {Wallace}, {Williams}, \& {Woodrum}}]{Bunker2024}
{Bunker}, A.~J., {Cameron}, A.~J., {Curtis-Lake}, E., {et~al.} 2024, \bibinfo{title}{{JADES NIRSpec initial data release for the Hubble Ultra Deep Field: Redshifts and line fluxes of distant galaxies from the deepest JWST Cycle 1 NIRSpec multi-object spectroscopy},} \aap, 690, A288, \dodoi{10.1051/0004-6361/202347094}

\bibitem[{N. {Byler} {et~al.}(2017){Byler}, {Dalcanton}, {Conroy}, \& {Johnson}}]{Byler2017}
{Byler}, N., {Dalcanton}, J.~J., {Conroy}, C., \& {Johnson}, B.~D. 2017, \bibinfo{title}{{Nebular Continuum and Line Emission in Stellar Population Synthesis Models},} \apj, 840, 44, \dodoi{10.3847/1538-4357/aa6c66}

\bibitem[{D. {Calzetti} {et~al.}(2000){Calzetti}, {Armus}, {Bohlin}, {Kinney}, {Koornneef}, \& {Storchi-Bergmann}}]{Calzetti2000}
{Calzetti}, D., {Armus}, L., {Bohlin}, R.~C., {et~al.} 2000, \bibinfo{title}{{The Dust Content and Opacity of Actively Star-forming Galaxies},} \apj, 533, 682, \dodoi{10.1086/308692}

\bibitem[{M. Cappellari(2017)Cappellari}]{cappellari2017}
Cappellari, M. 2017, \bibinfo{title}{Improving the full spectrum fitting method: accurate convolution with {Gauss}–{Hermite} functions,} \mnras, 466, 798, \dodoi{10.1093/mnras/stw3020}

\bibitem[{M. {Cappellari}(2023){Cappellari}}]{Cappellari2023}
{Cappellari}, M. 2023, \bibinfo{title}{{Full spectrum fitting with photometry in PPXF: stellar population versus dynamical masses, non-parametric star formation history and metallicity for 3200 LEGA-C galaxies at redshift z {\ensuremath{\approx}} 0.8},} \mnras, 526, 3273, \dodoi{10.1093/mnras/stad2597}

\bibitem[{M. Cappellari \& E. Emsellem(2004)Cappellari \& Emsellem}]{cappellari2004}
Cappellari, M., \& Emsellem, E. 2004, \bibinfo{title}{Parametric {Recovery} of {Line}‐of‐{Sight} {Velocity} {Distributions} from {Absorption}‐{Line} {Spectra} of {Galaxies} via {Penalized} {Likelihood},} \pasp, 116, 138, \dodoi{10.1086/381875}

\bibitem[{A.~C. {Carnall} {et~al.}(2018){Carnall}, {McLure}, {Dunlop}, \& {Dav{\'e}}}]{Carnall2018}
{Carnall}, A.~C., {McLure}, R.~J., {Dunlop}, J.~S., \& {Dav{\'e}}, R. 2018, \bibinfo{title}{{Inferring the star formation histories of massive quiescent galaxies with BAGPIPES: evidence for multiple quenching mechanisms},} \mnras, 480, 4379, \dodoi{10.1093/mnras/sty2169}

\bibitem[{S. {Cazzoli} {et~al.}(2016){Cazzoli}, {Arribas}, {Maiolino}, \& {Colina}}]{Cazzoli2016}
{Cazzoli}, S., {Arribas}, S., {Maiolino}, R., \& {Colina}, L. 2016, \bibinfo{title}{{Neutral gas outflows in nearby [U]LIRGs via optical NaD feature},} \aap, 590, A125, \dodoi{10.1051/0004-6361/201526788}

\bibitem[{Y.-M. {Chen} {et~al.}(2010){Chen}, {Tremonti}, {Heckman}, {Kauffmann}, {Weiner}, {Brinchmann}, \& {Wang}}]{Chen2010}
{Chen}, Y.-M., {Tremonti}, C.~A., {Heckman}, T.~M., {et~al.} 2010, \bibinfo{title}{{Absorption-line Probes of the Prevalence and Properties of Outflows in Present-day Star-forming Galaxies},} \aj, 140, 445, \dodoi{10.1088/0004-6256/140/2/445}

\bibitem[{E. {Cheung} {et~al.}(2016){Cheung}, {Bundy}, {Cappellari}, {Peirani}, {Rujopakarn}, {Westfall}, {Yan}, {Bershady}, {Greene}, {Heckman}, {Drory}, {Law}, {Masters}, {Thomas}, {Wake}, {Weijmans}, {Rubin}, {Belfiore}, {Vulcani}, {Chen}, {Zhang}, {Gelfand}, {Bizyaev}, {Roman-Lopes}, \& {Schneider}}]{Cheung2016}
{Cheung}, E., {Bundy}, K., {Cappellari}, M., {et~al.} 2016, \bibinfo{title}{{Suppressing star formation in quiescent galaxies with supermassive black hole winds},} \nat, 533, 504, \dodoi{10.1038/nature18006}

\bibitem[{J. {Choi} {et~al.}(2016){Choi}, {Dotter}, {Conroy}, {Cantiello}, {Paxton}, \& {Johnson}}]{Choi2016}
{Choi}, J., {Dotter}, A., {Conroy}, C., {et~al.} 2016, \bibinfo{title}{{Mesa Isochrones and Stellar Tracks (MIST). I. Solar-scaled Models},} \apj, 823, 102, \dodoi{10.3847/0004-637X/823/2/102}

\bibitem[{A.~L. {Coil} {et~al.}(2011){Coil}, {Weiner}, {Holz}, {Cooper}, {Yan}, \& {Aird}}]{Coil2011}
{Coil}, A.~L., {Weiner}, B.~J., {Holz}, D.~E., {et~al.} 2011, \bibinfo{title}{{Outflowing Galactic Winds in Post-starburst and Active Galactic Nucleus Host Galaxies at 0.2 < z < 0.8},} \apj, 743, 46, \dodoi{10.1088/0004-637X/743/1/46}

\bibitem[{A. {Concas} {et~al.}(2017){Concas}, {Popesso}, {Brusa}, {Mainieri}, {Erfanianfar}, \& {Morselli}}]{Concas2017}
{Concas}, A., {Popesso}, P., {Brusa}, M., {et~al.} 2017, \bibinfo{title}{{Light breeze in the local Universe},} \aap, 606, A36, \dodoi{10.1051/0004-6361/201629519}

\bibitem[{A. {Concas} {et~al.}(2019){Concas}, {Popesso}, {Brusa}, {Mainieri}, \& {Thomas}}]{Concas2019}
{Concas}, A., {Popesso}, P., {Brusa}, M., {Mainieri}, V., \& {Thomas}, D. 2019, \bibinfo{title}{{Two-face(s): ionized and neutral gas winds in the local Universe},} \aap, 622, A188, \dodoi{10.1051/0004-6361/201732152}

\bibitem[{C. {Conroy} \& J.~E. {Gunn}(2010){Conroy} \& {Gunn}}]{Conroy2010}
{Conroy}, C., \& {Gunn}, J.~E. 2010, \bibinfo{title}{{The Propagation of Uncertainties in Stellar Population Synthesis Modeling. III. Model Calibration, Comparison, and Evaluation},} \apj, 712, 833, \dodoi{10.1088/0004-637X/712/2/833}

\bibitem[{C. {Conroy} {et~al.}(2009){Conroy}, {Gunn}, \& {White}}]{Conroy2009}
{Conroy}, C., {Gunn}, J.~E., \& {White}, M. 2009, \bibinfo{title}{{The Propagation of Uncertainties in Stellar Population Synthesis Modeling. I. The Relevance of Uncertain Aspects of Stellar Evolution and the Initial Mass Function to the Derived Physical Properties of Galaxies},} \apj, 699, 486, \dodoi{10.1088/0004-637X/699/1/486}

\bibitem[{L.~L. {Cowie} {et~al.}(2017){Cowie}, {Barger}, {Hsu}, {Chen}, {Owen}, \& {Wang}}]{Cowie2017}
{Cowie}, L.~L., {Barger}, A.~J., {Hsu}, L.-Y., {et~al.} 2017, \bibinfo{title}{{A Submillimeter Perspective on the GOODS Fields (SUPER GOODS). I. An Ultradeep SCUBA-2 Survey of the GOODS-N},} \apj, 837, 139, \dodoi{10.3847/1538-4357/aa60bb}

\bibitem[{E. {Curtis-Lake} {et~al.}(2025){Curtis-Lake}, {Cameron}, {Bunker}, {Scholtz}, {Carniani}, {Parlanti}, {D'Eugenio}, {Jakobsen}, {Willmer}, {Arribas}, {Baker}, {Charlot}, {Chevallard}, {Circosta}, {Curti}, {Eisenstein}, {Hainline}, {Ji}, {Johnson}, {Jones}, {Maiolino}, {Maseda}, {P{\'e}rez-Gonz{\'a}lez}, {Rawle}, {Rieke}, {Rinaldi}, {Robertson}, {Rodr{\'\i}gez Del Pino}, {Saxena}, {Shivaei}, {Smit}, {Tacchella}, {{\"U}bler}, {Venturi}, {Williams}, {Willott}, \& {Duan}}]{Curtis-Lake2025}
{Curtis-Lake}, E., {Cameron}, A.~J., {Bunker}, A.~J., {et~al.} 2025, \bibinfo{title}{{JADES Data Release 4 Paper I: Sample Selection, Observing Strategy and Redshifts of the complete spectroscopic sample},} arXiv e-prints, arXiv:2510.01033, \dodoi{10.48550/arXiv.2510.01033}

\bibitem[{R. {Dav{\'e}} {et~al.}(2011){Dav{\'e}}, {Finlator}, \& {Oppenheimer}}]{Dave2011}
{Dav{\'e}}, R., {Finlator}, K., \& {Oppenheimer}, B.~D. 2011, \bibinfo{title}{{Galaxy evolution in cosmological simulations with outflows - II. Metallicities and gas fractions},} \mnras, 416, 1354, \dodoi{10.1111/j.1365-2966.2011.19132.x}

\bibitem[{R.~L. {Davies} {et~al.}(2024){Davies}, {Belli}, {Park}, {Mendel}, {Johnson}, {Conroy}, {Benton}, {Bugiani}, {Emami}, {Leja}, {Li}, {Maheson}, {Mathews}, {Naidu}, {Nelson}, {Tacchella}, {Terrazas}, \& {Weinberger}}]{Davies2024}
{Davies}, R.~L., {Belli}, S., {Park}, M., {et~al.} 2024, \bibinfo{title}{{JWST reveals widespread AGN-driven neutral gas outflows in massive z 2 galaxies},} \mnras, 528, 4976, \dodoi{10.1093/mnras/stae327}

\bibitem[{A. {de Graaff} {et~al.}(2025){de Graaff}, {Brammer}, {Weibel}, {Lewis}, {Maseda}, {Oesch}, {Bezanson}, {Boogaard}, {Cleri}, {Cooper}, {Gottumukkala}, {Greene}, {Hirschmann}, {Hviding}, {Katz}, {Labb{\'e}}, {Leja}, {Matthee}, {McConachie}, {Miller}, {Naidu}, {Price}, {Rix}, {Setton}, {Suess}, {Wang}, {Whitaker}, \& {Williams}}]{deGraaff2025}
{de Graaff}, A., {Brammer}, G., {Weibel}, A., {et~al.} 2025, \bibinfo{title}{{RUBIES: A complete census of the bright and red distant Universe with JWST/NIRSpec},} \aap, 697, A189, \dodoi{10.1051/0004-6361/202452186}

\bibitem[{F. {D'Eugenio} {et~al.}(2024{\natexlab{a}}){D'Eugenio}, {P{\'e}rez-Gonz{\'a}lez}, {Maiolino}, {Scholtz}, {Perna}, {Circosta}, {{\"U}bler}, {Arribas}, {B{\"o}ker}, {Bunker}, {Carniani}, {Charlot}, {Chevallard}, {Cresci}, {Curtis-Lake}, {Jones}, {Kumari}, {Lamperti}, {Looser}, {Parlanti}, {Rix}, {Robertson}, {Rodr{\'\i}guez Del Pino}, {Tacchella}, {Venturi}, \& {Willott}}]{D'Eugenio2024}
{D'Eugenio}, F., {P{\'e}rez-Gonz{\'a}lez}, P.~G., {Maiolino}, R., {et~al.} 2024{\natexlab{a}}, \bibinfo{title}{{A fast-rotator post-starburst galaxy quenched by supermassive black-hole feedback at z = 3},} Nature Astronomy, 8, 1443, \dodoi{10.1038/s41550-024-02345-1}

\bibitem[{F. {D'Eugenio} {et~al.}(2024{\natexlab{b}}){D'Eugenio}, {P{\'e}rez-Gonz{\'a}lez}, {Maiolino}, {Scholtz}, {Perna}, {Circosta}, {{\"U}bler}, {Arribas}, {B{\"o}ker}, {Bunker}, {Carniani}, {Charlot}, {Chevallard}, {Cresci}, {Curtis-Lake}, {Jones}, {Kumari}, {Lamperti}, {Looser}, {Parlanti}, {Rix}, {Robertson}, {Rodr{\'\i}guez Del Pino}, {Tacchella}, {Venturi}, \& {Willott}}]{DEugenio2024}
{D'Eugenio}, F., {P{\'e}rez-Gonz{\'a}lez}, P.~G., {Maiolino}, R., {et~al.} 2024{\natexlab{b}}, \bibinfo{title}{{A fast-rotator post-starburst galaxy quenched by supermassive black-hole feedback at z = 3},} Nature Astronomy, 8, 1443, \dodoi{10.1038/s41550-024-02345-1}

\bibitem[{A.~M. {Diamond-Stanic} {et~al.}(2009){Diamond-Stanic}, {Rieke}, \& {Rigby}}]{Diamond-Stanic2009}
{Diamond-Stanic}, A.~M., {Rieke}, G.~H., \& {Rigby}, J.~R. 2009, \bibinfo{title}{{Isotropic Luminosity Indicators in a Complete AGN Sample},} \apj, 698, 623, \dodoi{10.1088/0004-637X/698/1/623}

\bibitem[{F. {Duras} {et~al.}(2020){Duras}, {Bongiorno}, {Ricci}, {Piconcelli}, {Shankar}, {Lusso}, {Bianchi}, {Fiore}, {Maiolino}, {Marconi}, {Onori}, {Sani}, {Schneider}, {Vignali}, \& {La Franca}}]{Duras2020}
{Duras}, F., {Bongiorno}, A., {Ricci}, F., {et~al.} 2020, \bibinfo{title}{{Universal bolometric corrections for active galactic nuclei over seven luminosity decades},} \aap, 636, A73, \dodoi{10.1051/0004-6361/201936817}

\bibitem[{D.~J. {Eisenstein} {et~al.}(2023){Eisenstein}, {Johnson}, {Robertson}, {Tacchella}, {Hainline}, {Jakobsen}, {Maiolino}, {Bonaventura}, {Bunker}, {Cameron}, {Cargile}, {Curtis-Lake}, {Hausen}, {Pusk{\'a}s}, {Rieke}, {Sun}, {Willmer}, {Willott}, {Alberts}, {Arribas}, {Baker}, {Baum}, {Bhatawdekar}, {Carniani}, {Charlot}, {Chen}, {Chevallard}, {Curti}, {DeCoursey}, {D'Eugenio}, {de Graaff}, {Egami}, {Helton}, {Ji}, {Jones}, {Kumari}, {L{\"u}tzgendorf}, {Laseter}, {Looser}, {Lyu}, {Maseda}, {Nelson}, {Parlanti}, {Rauscher}, {Rawle}, {Rieke}, {Rix}, {Rujopakarn}, {Sandles}, {Saxena}, {Scholtz}, {Sharpe}, {Shivaei}, {Simmonds}, {Smit}, {Topping}, {{\"U}bler}, {Venturi}, {Williams}, {Witstok}, \& {Woodrum}}]{Eisenstein2023b}
{Eisenstein}, D.~J., {Johnson}, B.~D., {Robertson}, B., {et~al.} 2023, \bibinfo{title}{{The JADES Origins Field: A New JWST Deep Field in the JADES Second NIRCam Data Release},} arXiv e-prints, arXiv:2310.12340, \dodoi{10.48550/arXiv.2310.12340}

\bibitem[{D.~J. {Eisenstein} {et~al.}(2026){Eisenstein}, {Willott}, {Alberts}, {Arribas}, {Bonaventura}, {Bunker}, {Cameron}, {Carniani}, {Charlot}, {Curtis-Lake}, {D'Eugenio}, {Ferruit}, {Giardino}, {Hainline}, {Hausen}, {Jakobsen}, {Johnson}, {Maiolino}, {Rauscher}, {Rieke}, {Rieke}, {Rix}, {Robertson}, {Stark}, {Tacchella}, {Williams}, {Willmer}, {Baker}, {Baum}, {Bhatawdekar}, {Boyett}, {Chen}, {Chevallard}, {Circosta}, {Curti}, {Danhaive}, {DeCoursey}, {Endsley}, {de Graaff}, {Dressler}, {Egami}, {Helton}, {Hviding}, {Ji}, {Jones}, {Kumari}, {L{\"u}tzgendorf}, {Laseter}, {Looser}, {Lyu}, {Maseda}, {Nelson}, {Parlanti}, {Perna}, {Pusk{\'a}s}, {Rawle}, {Rodr{\'\i}guez Del Pino}, {Rujopakarn}, {Sandles}, {Saxena}, {Scholtz}, {Sharpe}, {Shivaei}, {Silcock}, {Simmonds}, {Skarbinski}, {Smit}, {Stone}, {Suess}, {Sun}, {Tang}, {Topping}, {{\"U}bler}, {Villanueva}, {Wallace}, {Whitler}, {Witstok}, \& {Woodrum}}]{Eisenstein2026}
{Eisenstein}, D.~J., {Willott}, C., {Alberts}, S., {et~al.} 2026, \bibinfo{title}{{Overview of the JWST Advanced Deep Extragalactic Survey (JADES)},} \apjs, 283, 6, \dodoi{10.3847/1538-4365/ae3163}

\bibitem[{J. {Falc{\'o}n-Barroso} {et~al.}(2011){Falc{\'o}n-Barroso}, {S{\'a}nchez-Bl{\'a}zquez}, {Vazdekis}, {Ricciardelli}, {Cardiel}, {Cenarro}, {Gorgas}, \& {Peletier}}]{Falcon-Barroso2011}
{Falc{\'o}n-Barroso}, J., {S{\'a}nchez-Bl{\'a}zquez}, P., {Vazdekis}, A., {et~al.} 2011, \bibinfo{title}{{An updated MILES stellar library and stellar population models},} \aap, 532, A95, \dodoi{10.1051/0004-6361/201116842}

\bibitem[{P. {Ferruit} {et~al.}(2022){Ferruit}, {Jakobsen}, {Giardino}, {Rawle}, {Alves de Oliveira}, {Arribas}, {Beck}, {Birkmann}, {B{\"o}ker}, {Bunker}, {Charlot}, {de Marchi}, {Franx}, {Henry}, {Karakla}, {Kassin}, {Kumari}, {L{\'o}pez-Caniego}, {L{\"u}tzgendorf}, {Maiolino}, {Manjavacas}, {Marston}, {Moseley}, {Muzerolle}, {Pirzkal}, {Rauscher}, {Rix}, {Sabbi}, {Sirianni}, {te Plate}, {Valenti}, {Willott}, \& {Zeidler}}]{Ferruit2022}
{Ferruit}, P., {Jakobsen}, P., {Giardino}, G., {et~al.} 2022, \bibinfo{title}{{The Near-Infrared Spectrograph (NIRSpec) on the James Webb Space Telescope. II. Multi-object spectroscopy (MOS)},} \aap, 661, A81, \dodoi{10.1051/0004-6361/202142673}

\bibitem[{A. {Fluetsch} {et~al.}(2021){Fluetsch}, {Maiolino}, {Carniani}, {Arribas}, {Belfiore}, {Bellocchi}, {Cazzoli}, {Cicone}, {Cresci}, {Fabian}, {Gallagher}, {Ishibashi}, {Mannucci}, {Marconi}, {Perna}, {Sturm}, \& {Venturi}}]{Fluetsch2021}
{Fluetsch}, A., {Maiolino}, R., {Carniani}, S., {et~al.} 2021, \bibinfo{title}{{Properties of the multiphase outflows in local (ultra)luminous infrared galaxies},} \mnras, 505, 5753, \dodoi{10.1093/mnras/stab1666}

\bibitem[{D. {Foreman-Mackey} {et~al.}(2013){Foreman-Mackey}, {Hogg}, {Lang}, \& {Goodman}}]{Foreman-Mackey2013}
{Foreman-Mackey}, D., {Hogg}, D.~W., {Lang}, D., \& {Goodman}, J. 2013, \bibinfo{title}{{emcee: The MCMC Hammer},} \pasp, 125, 306, \dodoi{10.1086/670067}

\bibitem[{N.~M. {F{\"o}rster Schreiber} {et~al.}(2019){F{\"o}rster Schreiber}, {{\"U}bler}, {Davies}, {Genzel}, {Wisnioski}, {Belli}, {Shimizu}, {Lutz}, {Fossati}, {Herrera-Camus}, {Mendel}, {Tacconi}, {Wilman}, {Beifiori}, {Brammer}, {Burkert}, {Carollo}, {Davies}, {Eisenhauer}, {Fabricius}, {Lilly}, {Momcheva}, {Naab}, {Nelson}, {Price}, {Renzini}, {Saglia}, {Sternberg}, {van Dokkum}, \& {Wuyts}}]{Forster2019}
{F{\"o}rster Schreiber}, N.~M., {{\"U}bler}, H., {Davies}, R.~L., {et~al.} 2019, \bibinfo{title}{{The KMOS$^{3D}$ Survey: Demographics and Properties of Galactic Outflows at z = 0.6-2.7},} \apj, 875, 21, \dodoi{10.3847/1538-4357/ab0ca2}

\bibitem[{M. {Franx} {et~al.}(2008){Franx}, {van Dokkum}, {F{\"o}rster Schreiber}, {Wuyts}, {Labb{\'e}}, \& {Toft}}]{Franx2008}
{Franx}, M., {van Dokkum}, P.~G., {F{\"o}rster Schreiber}, N.~M., {et~al.} 2008, \bibinfo{title}{{Structure and Star Formation in Galaxies out to z = 3: Evidence for Surface Density Dependent Evolution and Upsizing},} \apj, 688, 770, \dodoi{10.1086/592431}

\bibitem[{K.~D. {French} {et~al.}(2015){French}, {Yang}, {Zabludoff}, {Narayanan}, {Shirley}, {Walter}, {Smith}, \& {Tremonti}}]{French2015}
{French}, K.~D., {Yang}, Y., {Zabludoff}, A., {et~al.} 2015, \bibinfo{title}{{Discovery of Large Molecular Gas Reservoirs in Post-starburst Galaxies},} \apj, 801, 1, \dodoi{10.1088/0004-637X/801/1/1}

\bibitem[{J.~P. {Gardner} {et~al.}(2006){Gardner}, {Mather}, {Clampin}, {Doyon}, {Greenhouse}, {Hammel}, {Hutchings}, {Jakobsen}, {Lilly}, {Long}, {Lunine}, {McCaughrean}, {Mountain}, {Nella}, {Rieke}, {Rieke}, {Rix}, {Smith}, {Sonneborn}, {Stiavelli}, {Stockman}, {Windhorst}, \& {Wright}}]{Gardner2006}
{Gardner}, J.~P., {Mather}, J.~C., {Clampin}, M., {et~al.} 2006, \bibinfo{title}{{The James Webb Space Telescope},} \ssr, 123, 485, \dodoi{10.1007/s11214-006-8315-7}

\bibitem[{A. {Genin} {et~al.}(2025){Genin}, {Shuntov}, {Brammer}, {Allen}, {Ito}, {Magdis}, {Matharu}, {Oesch}, {Toft}, \& {Valentino}}]{Genin2025}
{Genin}, A., {Shuntov}, M., {Brammer}, G., {et~al.} 2025, \bibinfo{title}{{DAWN JWST Archive: Morphology from profile fitting of over 340 000 galaxies in major JWST fields: Morphology evolution with redshift and galaxy type},} \aap, 699, A343, \dodoi{10.1051/0004-6361/202555504}

\bibitem[{E. {Gonz{\'a}lez-Alfonso} {et~al.}(2017){Gonz{\'a}lez-Alfonso}, {Fischer}, {Spoon}, {Stewart}, {Ashby}, {Veilleux}, {Smith}, {Sturm}, {Farrah}, {Falstad}, {Mel{\'e}ndez}, {Graci{\'a}-Carpio}, {Janssen}, \& {Lebouteiller}}]{Gonzalez-Alfonso2017}
{Gonz{\'a}lez-Alfonso}, E., {Fischer}, J., {Spoon}, H.~W.~W., {et~al.} 2017, \bibinfo{title}{{Molecular Outflows in Local ULIRGs: Energetics from Multitransition OH Analysis},} \apj, 836, 11, \dodoi{10.3847/1538-4357/836/1/11}

\bibitem[{K.~E. {Heintz} {et~al.}(2025){Heintz}, {Brammer}, {Watson}, {Oesch}, {Keating}, {Hayes}, {Abdurro'uf}, {Arellano-C{\'o}rdova}, {Carnall}, {Christiansen}, {Cullen}, {Dav{\'e}}, {Dayal}, {Ferrara}, {Finlator}, {Fynbo}, {Flury}, {Gelli}, {Gillman}, {Gottumukkala}, {Gould}, {Greve}, {Hardin}, {Hsiao}, {Hutter}, {Jakobsson}, {Killi}, {Khosravaninezhad}, {Laursen}, {Lee}, {Magdis}, {Matthee}, {Naidu}, {Narayanan}, {Pollock}, {Prescott}, {Rusakov}, {Shuntov}, {Sneppen}, {Smit}, {Tanvir}, {Terp}, {Toft}, {Valentino}, {Vijayan}, {Weaver}, {Wise}, \& {Witstok}}]{heintz2025}
{Heintz}, K.~E., {Brammer}, G.~B., {Watson}, D., {et~al.} 2025, \bibinfo{title}{{The JWST-PRIMAL archival survey: A JWST/NIRSpec reference sample for the physical properties and Lyman-{\ensuremath{\alpha}} absorption and emission of {\ensuremath{\sim}}600 galaxies at z = 5.0 ‑ 13.4},} \aap, 693, A60, \dodoi{10.1051/0004-6361/202450243}

\bibitem[{P.~F. {Hopkins} {et~al.}(2008){Hopkins}, {Hernquist}, {Cox}, \& {Kere{\v{s}}}}]{Hopkins2008}
{Hopkins}, P.~F., {Hernquist}, L., {Cox}, T.~J., \& {Kere{\v{s}}}, D. 2008, \bibinfo{title}{{A Cosmological Framework for the Co-Evolution of Quasars, Supermassive Black Holes, and Elliptical Galaxies. I. Galaxy Mergers and Quasar Activity},} \apjs, 175, 356, \dodoi{10.1086/524362}

\bibitem[{P.~F. {Hopkins} {et~al.}(2014){Hopkins}, {Kere{\v{s}}}, {O{\~n}orbe}, {Faucher-Gigu{\`e}re}, {Quataert}, {Murray}, \& {Bullock}}]{Hopkins2014}
{Hopkins}, P.~F., {Kere{\v{s}}}, D., {O{\~n}orbe}, J., {et~al.} 2014, \bibinfo{title}{{Galaxies on FIRE (Feedback In Realistic Environments): stellar feedback explains cosmologically inefficient star formation},} \mnras, 445, 581, \dodoi{10.1093/mnras/stu1738}

\bibitem[{P. {Jakobsen} {et~al.}(2022){Jakobsen}, {Ferruit}, {Alves de Oliveira}, {Arribas}, {Bagnasco}, {Barho}, {Beck}, {Birkmann}, {B{\"o}ker}, {Bunker}, {Charlot}, {de Jong}, {de Marchi}, {Ehrenwinkler}, {Falcolini}, {Fels}, {Franx}, {Franz}, {Funke}, {Giardino}, {Gnata}, {Holota}, {Honnen}, {Jensen}, {Jentsch}, {Johnson}, {Jollet}, {Karl}, {Kling}, {K{\"o}hler}, {Kolm}, {Kumari}, {Lander}, {Lemke}, {L{\'o}pez-Caniego}, {L{\"u}tzgendorf}, {Maiolino}, {Manjavacas}, {Marston}, {Maschmann}, {Maurer}, {Messerschmidt}, {Moseley}, {Mosner}, {Mott}, {Muzerolle}, {Pirzkal}, {Pittet}, {Plitzke}, {Posselt}, {Rapp}, {Rauscher}, {Rawle}, {Rix}, {R{\"o}del}, {Rumler}, {Sabbi}, {Salvignol}, {Schmid}, {Sirianni}, {Smith}, {Strada}, {te Plate}, {Valenti}, {Wettemann}, {Wiehe}, {Wiesmayer}, {Willott}, {Wright}, {Zeidler}, \& {Zincke}}]{Jakobsen2022}
{Jakobsen}, P., {Ferruit}, P., {Alves de Oliveira}, C., {et~al.} 2022, \bibinfo{title}{{The Near-Infrared Spectrograph (NIRSpec) on the James Webb Space Telescope. I. Overview of the instrument and its capabilities},} \aap, 661, A80, \dodoi{10.1051/0004-6361/202142663}

\bibitem[{Z. {Ji} \& M. {Giavalisco}(2023){Ji} \& {Giavalisco}}]{Ji2023}
{Ji}, Z., \& {Giavalisco}, M. 2023, \bibinfo{title}{{Reconstructing the Assembly of Massive Galaxies. II. Galaxies Develop Massive and Dense Stellar Cores as They Evolve and Head toward Quiescence at Cosmic Noon},} \apj, 943, 54, \dodoi{10.3847/1538-4357/aca807}

\bibitem[{Z. {Ji} {et~al.}(2024{\natexlab{a}}){Ji}, {Williams}, {Tacchella}, {Suess}, {Baker}, {Alberts}, {Bunker}, {Johnson}, {Robertson}, {Sun}, {Eisenstein}, {Rieke}, {Maseda}, {Hainline}, {Hausen}, {Rieke}, {Willmer}, {Egami}, {Shivaei}, {Carniani}, {Charlot}, {Chevallard}, {Curtis-Lake}, {Looser}, {Maiolino}, {Willott}, {Chen}, {Helton}, {Lyu}, {Nelson}, {Bhatawdekar}, {Boyett}, \& {Sandles}}]{Ji2024}
{Ji}, Z., {Williams}, C.~C., {Tacchella}, S., {et~al.} 2024{\natexlab{a}}, \bibinfo{title}{{JADES + JEMS: A Detailed Look at the Buildup of Central Stellar Cores and Suppression of Star Formation in Galaxies at Redshifts 3 < z < 4.5},} \apj, 974, 135, \dodoi{10.3847/1538-4357/ad6e7f}

\bibitem[{Z. {Ji} {et~al.}(2024{\natexlab{b}}){Ji}, {Williams}, {Rieke}, {Lyu}, {Alberts}, {Sun}, {Helton}, {Rieke}, {Shivaei}, {D'Eugenio}, {Tacchella}, {Robertson}, {Zhu}, {Maiolino}, {Bunker}, {Sun}, \& {Willmer}}]{Ji2024b}
{Ji}, Z., {Williams}, C.~C., {Rieke}, G.~H., {et~al.} 2024{\natexlab{b}}, \bibinfo{title}{{Extended hot dust emission around the earliest massive quiescent galaxy},} arXiv e-prints, arXiv:2409.17233, \dodoi{10.48550/arXiv.2409.17233}

\bibitem[{Z. {Ji} {et~al.}(2026){Ji}, {Williams}, {Behroozi}, {Weibel}, {Jespersen}, {Oesch}, {Bezanson}, {Whitaker}, {Greene}, {Brammer}, {Dayal}, {Labb{\'e}}, {Manning}, {Rinaldi}, {Xiao}, \& {Zhang}}]{Ji2026}
{Ji}, Z., {Williams}, C.~C., {Behroozi}, P., {et~al.} 2026, \bibinfo{title}{{PANORAMIC: The Dawn of Massive Quiescent Galaxies I. Number Density and Cosmic Variance from 1000 arcmin$^2$ NIRCam Imaging},} arXiv e-prints, arXiv:2604.05022, \dodoi{10.48550/arXiv.2604.05022}

\bibitem[{B.~D. {Johnson} {et~al.}(2021){Johnson}, {Leja}, {Conroy}, \& {Speagle}}]{Johnson2021}
{Johnson}, B.~D., {Leja}, J., {Conroy}, C., \& {Speagle}, J.~S. 2021, \bibinfo{title}{{Stellar Population Inference with Prospector},} \apjs, 254, 22, \dodoi{10.3847/1538-4365/abef67}

\bibitem[{B.~D. {Johnson} {et~al.}(2026){Johnson}, {Robertson}, {Eisenstein}, {Tacchella}, {Pusk{\'a}s}, {Duan}, {Wu}, {Hainline}, {Rieke}, {Willott}, {Willmer}, {Trussler}, {Alberts}, {Arribas}, {Baker}, {Bunker}, {Cameron}, {Carniani}, {Carreira}, {Cargile}, {Curtis-Lake}, {Egami}, {Hausen}, {Helton}, {Ji}, {Maiolino}, {P{\'e}rez-Gonz{\'a}lez}, {Rinaldi}, {Sun}, {Sun}, {Villanueva}, {Williams}, \& {Zhu}}]{Johnson2026}
{Johnson}, B.~D., {Robertson}, B.~E., {Eisenstein}, D.~J., {et~al.} 2026, \bibinfo{title}{{JWST Advanced Deep Extragalactic Survey (JADES) Data Release 5: NIRCam Imaging in GOODS-S and GOODS-N},} arXiv e-prints, arXiv:2601.15954, \dodoi{10.48550/arXiv.2601.15954}

\bibitem[{I. {Juod{\v{z}}balis} {et~al.}(2026){Juod{\v{z}}balis}, {Maiolino}, {Baker}, {Lake}, {Scholtz}, {D'Eugenio}, {Trefoloni}, {Isobe}, {Tacchella}, {Bunker}, {Carniani}, {Charlot}, {Jones}, {Parlanti}, {Perna}, {Rinaldi}, {Robertson}, {{\"U}bler}, {Venturi}, \& {Willott}}]{Juodzbalis2026}
{Juod{\v{z}}balis}, I., {Maiolino}, R., {Baker}, W.~M., {et~al.} 2026, {JADES: comprehensive census of broad-line AGN from reionization to cosmic noon revealed by JWST}, \dodoi{10.1093/mnras/stag086}

\bibitem[{G. {Kauffmann} {et~al.}(2003){Kauffmann}, {Heckman}, {Tremonti}, {Brinchmann}, {Charlot}, {White}, {Ridgway}, {Brinkmann}, {Fukugita}, {Hall}, {Ivezi{\'c}}, {Richards}, \& {Schneider}}]{Kauffmann2003}
{Kauffmann}, G., {Heckman}, T.~M., {Tremonti}, C., {et~al.} 2003, \bibinfo{title}{{The host galaxies of active galactic nuclei},} \mnras, 346, 1055, \dodoi{10.1111/j.1365-2966.2003.07154.x}

\bibitem[{R.~C. {Kennicutt} \& N.~J. {Evans}(2012){Kennicutt} \& {Evans}}]{Kennicutt2012}
{Kennicutt}, R.~C., \& {Evans}, N.~J. 2012, \bibinfo{title}{{Star Formation in the Milky Way and Nearby Galaxies},} \araa, 50, 531, \dodoi{10.1146/annurev-astro-081811-125610}

\bibitem[{D. {Kere{\v{s}}} {et~al.}(2009){Kere{\v{s}}}, {Katz}, {Fardal}, {Dav{\'e}}, \& {Weinberg}}]{Keres2009}
{Kere{\v{s}}}, D., {Katz}, N., {Fardal}, M., {Dav{\'e}}, R., \& {Weinberg}, D.~H. 2009, \bibinfo{title}{{Galaxies in a simulated {\ensuremath{\Lambda}}CDM Universe - I. Cold mode and hot cores},} \mnras, 395, 160, \dodoi{10.1111/j.1365-2966.2009.14541.x}

\bibitem[{L.~J. {Kewley} {et~al.}(2001){Kewley}, {Dopita}, {Sutherland}, {Heisler}, \& {Trevena}}]{Kewley2001}
{Kewley}, L.~J., {Dopita}, M.~A., {Sutherland}, R.~S., {Heisler}, C.~A., \& {Trevena}, J. 2001, \bibinfo{title}{{Theoretical Modeling of Starburst Galaxies},} \apj, 556, 121, \dodoi{10.1086/321545}

\bibitem[{L.~J. {Kewley} {et~al.}(2013){Kewley}, {Maier}, {Yabe}, {Ohta}, {Akiyama}, {Dopita}, \& {Yuan}}]{Kewley2013}
{Kewley}, L.~J., {Maier}, C., {Yabe}, K., {et~al.} 2013, \bibinfo{title}{{The Cosmic BPT Diagram: Confronting Theory with Observations},} \apjl, 774, L10, \dodoi{10.1088/2041-8205/774/1/L10}

\bibitem[{A. {King} \& K. {Pounds}(2015){King} \& {Pounds}}]{King2015}
{King}, A., \& {Pounds}, K. 2015, \bibinfo{title}{{Powerful Outflows and Feedback from Active Galactic Nuclei},} \araa, 53, 115, \dodoi{10.1146/annurev-astro-082214-122316}

\bibitem[{P. {Kroupa}(2001){Kroupa}}]{Kroupa2001}
{Kroupa}, P. 2001, \bibinfo{title}{{On the variation of the initial mass function},} \mnras, 322, 231, \dodoi{10.1046/j.1365-8711.2001.04022.x}

\bibitem[{C. {Leitherer} {et~al.}(1999){Leitherer}, {Schaerer}, {Goldader}, {Delgado}, {Robert}, {Kune}, {de Mello}, {Devost}, \& {Heckman}}]{Leitherer1999}
{Leitherer}, C., {Schaerer}, D., {Goldader}, J.~D., {et~al.} 1999, \bibinfo{title}{{Starburst99: Synthesis Models for Galaxies with Active Star Formation},} \apjs, 123, 3, \dodoi{10.1086/313233}

\bibitem[{J. {Leja} {et~al.}(2017){Leja}, {Johnson}, {Conroy}, {van Dokkum}, \& {Byler}}]{Leja2017}
{Leja}, J., {Johnson}, B.~D., {Conroy}, C., {van Dokkum}, P.~G., \& {Byler}, N. 2017, \bibinfo{title}{{Deriving Physical Properties from Broadband Photometry with Prospector: Description of the Model and a Demonstration of its Accuracy Using 129 Galaxies in the Local Universe},} \apj, 837, 170, \dodoi{10.3847/1538-4357/aa5ffe}

\bibitem[{J. {Leja} {et~al.}(2022){Leja}, {Speagle}, {Ting}, {Johnson}, {Conroy}, {Whitaker}, {Nelson}, {van Dokkum}, \& {Franx}}]{Leja2022}
{Leja}, J., {Speagle}, J.~S., {Ting}, Y.-S., {et~al.} 2022, \bibinfo{title}{{A New Census of the 0.2 < z < 3.0 Universe. II. The Star-forming Sequence},} \apj, 936, 165, \dodoi{10.3847/1538-4357/ac887d}

\bibitem[{D. {Liu} {et~al.}(2019){Liu}, {Lang}, {Magnelli}, {Schinnerer}, {Leslie}, {Fudamoto}, {Bondi}, {Groves}, {Jim{\'e}nez-Andrade}, {Harrington}, {Karim}, {Oesch}, {Sargent}, {Vardoulaki}, {B{\v{a}}descu}, {Moser}, {Bertoldi}, {Battisti}, {da Cunha}, {Zavala}, {Vaccari}, {Davidzon}, {Riechers}, \& {Aravena}}]{Liu2019}
{Liu}, D., {Lang}, P., {Magnelli}, B., {et~al.} 2019, \bibinfo{title}{{Automated Mining of the ALMA Archive in the COSMOS Field (A$^{3}$COSMOS). I. Robust ALMA Continuum Photometry Catalogs and Stellar Mass and Star Formation Properties for {\ensuremath{\sim}}700 Galaxies at z = 0.5-6},} \apjs, 244, 40, \dodoi{10.3847/1538-4365/ab42da}

\bibitem[{B. {Luo} {et~al.}(2017){Luo}, {Brandt}, {Xue}, {Lehmer}, {Alexander}, {Bauer}, {Vito}, {Yang}, {Basu-Zych}, {Comastri}, {Gilli}, {Gu}, {Hornschemeier}, {Koekemoer}, {Liu}, {Mainieri}, {Paolillo}, {Ranalli}, {Rosati}, {Schneider}, {Shemmer}, {Smail}, {Sun}, {Tozzi}, {Vignali}, \& {Wang}}]{Luo2017}
{Luo}, B., {Brandt}, W.~N., {Xue}, Y.~Q., {et~al.} 2017, \bibinfo{title}{{The Chandra Deep Field-South Survey: 7 Ms Source Catalogs},} \apjs, 228, 2, \dodoi{10.3847/1538-4365/228/1/2}

\bibitem[{Y. {Luo} {et~al.}(2022){Luo}, {Rowlands}, {Alatalo}, {Sazonova}, {Abdurro'uf}, {Heckman}, {Medling}, {Deustua}, {Nyland}, {Lanz}, {Petric}, {Otter}, {Aalto}, {Dimassimo}, {French}, {Gallagher}, {Roediger}, \& {Stepanoff}}]{Luo2022}
{Luo}, Y., {Rowlands}, K., {Alatalo}, K., {et~al.} 2022, \bibinfo{title}{{A Multiwavelength View of IC 860: What Is in Action inside Quenching Galaxies},} \apj, 938, 63, \dodoi{10.3847/1538-4357/ac8b7d}

\bibitem[{C. {Lyu} {et~al.}(2026){Lyu}, {Yu}, {Wang}, {Wang}, {Jia}, {Song}, {Chen}, {Wang}, {Chen}, {Ma}, {Wang}, \& {Kong}}]{Lyu2026}
{Lyu}, C., {Yu}, H., {Wang}, E., {et~al.} 2026, \bibinfo{title}{{First Statistical Detection of Mg II-traced Cool Gas Outflows with JWST toward Cosmic Dawn},} \apjl, 1000, L3, \dodoi{10.3847/2041-8213/ae48ee}

\bibitem[{J. {Lyu} {et~al.}(2022){Lyu}, {Alberts}, {Rieke}, \& {Rujopakarn}}]{Lyu2022}
{Lyu}, J., {Alberts}, S., {Rieke}, G.~H., \& {Rujopakarn}, W. 2022, \bibinfo{title}{{AGN Selection and Demographics in GOODS-S/HUDF from X-Ray to Radio},} \apj, 941, 191, \dodoi{10.3847/1538-4357/ac9e5d}

\bibitem[{J. {Lyu} {et~al.}(2024){Lyu}, {Alberts}, {Rieke}, {Shivaei}, {P{\'e}rez-Gonz{\'a}lez}, {Sun}, {Hainline}, {Baum}, {Bonaventura}, {Bunker}, {Egami}, {Eisenstein}, {Florian}, {Ji}, {Johnson}, {Morrison}, {Rieke}, {Robertson}, {Rujopakarn}, {Tacchella}, {Scholtz}, \& {Willmer}}]{Lyu2024}
{Lyu}, J., {Alberts}, S., {Rieke}, G.~H., {et~al.} 2024, \bibinfo{title}{{Active Galactic Nuclei Selection and Demographics: A New Age with JWST/MIRI},} \apj, 966, 229, \dodoi{10.3847/1538-4357/ad3643}

\bibitem[{P. {Madau}(1995){Madau}}]{Madau1995}
{Madau}, P. 1995, \bibinfo{title}{{Radiative Transfer in a Clumpy Universe: The Colors of High-Redshift Galaxies},} \apj, 441, 18, \dodoi{10.1086/175332}

\bibitem[{P. {Madau} \& M. {Dickinson}(2014){Madau} \& {Dickinson}}]{Madau2014}
{Madau}, P., \& {Dickinson}, M. 2014, \bibinfo{title}{{Cosmic Star-Formation History},} \araa, 52, 415, \dodoi{10.1146/annurev-astro-081811-125615}

\bibitem[{A. {Man} \& S. {Belli}(2018){Man} \& {Belli}}]{Man2018}
{Man}, A., \& {Belli}, S. 2018, \bibinfo{title}{{Star formation quenching in massive galaxies},} Nature Astronomy, 2, 695, \dodoi{10.1038/s41550-018-0558-1}

\bibitem[{A. {Moghni} {et~al.}(2026){Moghni}, {Roy}, {Heckman}, {Bundy}, {Westfall}, \& {Rubin}}]{Moghni2026}
{Moghni}, A., {Roy}, N., {Heckman}, T.~M., {et~al.} 2026, \bibinfo{title}{{Galactic Rain: Cool Gas Inflows in Red Geyser Galaxies and Their Connection to Active Galactic Nuclei Activity and Interactions},} \apj, 1003, 31, \dodoi{10.3847/1538-4357/ae5f90}

\bibitem[{A. {Muzzin} {et~al.}(2013){Muzzin}, {Marchesini}, {Stefanon}, {Franx}, {McCracken}, {Milvang-Jensen}, {Dunlop}, {Fynbo}, {Brammer}, {Labb{\'e}}, \& {van Dokkum}}]{Muzzin2013}
{Muzzin}, A., {Marchesini}, D., {Stefanon}, M., {et~al.} 2013, \bibinfo{title}{{The Evolution of the Stellar Mass Functions of Star-forming and Quiescent Galaxies to z = 4 from the COSMOS/UltraVISTA Survey},} \apj, 777, 18, \dodoi{10.1088/0004-637X/777/1/18}

\bibitem[{T. {Naab} \& J.~P. {Ostriker}(2017){Naab} \& {Ostriker}}]{Naab2017}
{Naab}, T., \& {Ostriker}, J.~P. 2017, \bibinfo{title}{{Theoretical Challenges in Galaxy Formation},} \araa, 55, 59, \dodoi{10.1146/annurev-astro-081913-040019}

\bibitem[{M. {Nenkova} {et~al.}(2008{\natexlab{a}}){Nenkova}, {Sirocky}, {Ivezi{\'c}}, \& {Elitzur}}]{Nenkova2008a}
{Nenkova}, M., {Sirocky}, M.~M., {Ivezi{\'c}}, {\v{Z}}., \& {Elitzur}, M. 2008{\natexlab{a}}, \bibinfo{title}{{AGN Dusty Tori. I. Handling of Clumpy Media},} \apj, 685, 147, \dodoi{10.1086/590482}

\bibitem[{M. {Nenkova} {et~al.}(2008{\natexlab{b}}){Nenkova}, {Sirocky}, {Nikutta}, {Ivezi{\'c}}, \& {Elitzur}}]{Nenkova2008b}
{Nenkova}, M., {Sirocky}, M.~M., {Nikutta}, R., {Ivezi{\'c}}, {\v{Z}}., \& {Elitzur}, M. 2008{\natexlab{b}}, \bibinfo{title}{{AGN Dusty Tori. II. Observational Implications of Clumpiness},} \apj, 685, 160, \dodoi{10.1086/590483}

\bibitem[{H. {Netzer}(2009){Netzer}}]{Netzer2009}
{Netzer}, H. 2009, \bibinfo{title}{{Accretion and star formation rates in low-redshift type II active galactic nuclei},} \mnras, 399, 1907, \dodoi{10.1111/j.1365-2966.2009.15434.x}

\bibitem[{H. {Netzer}(2019){Netzer}}]{Netzer2019}
{Netzer}, H. 2019, \bibinfo{title}{{Bolometric correction factors for active galactic nuclei},} \mnras, 488, 5185, \dodoi{10.1093/mnras/stz2016}

\bibitem[{K.~G. {Noeske} {et~al.}(2007){Noeske}, {Weiner}, {Faber}, {Papovich}, {Koo}, {Somerville}, {Bundy}, {Conselice}, {Newman}, {Schiminovich}, {Le Floc'h}, {Coil}, {Rieke}, {Lotz}, {Primack}, {Barmby}, {Cooper}, {Davis}, {Ellis}, {Fazio}, {Guhathakurta}, {Huang}, {Kassin}, {Martin}, {Phillips}, {Rich}, {Small}, {Willmer}, \& {Wilson}}]{Noeske2007}
{Noeske}, K.~G., {Weiner}, B.~J., {Faber}, S.~M., {et~al.} 2007, \bibinfo{title}{{Star Formation in AEGIS Field Galaxies since z=1.1: The Dominance of Gradually Declining Star Formation, and the Main Sequence of Star-forming Galaxies},} \apjl, 660, L43, \dodoi{10.1086/517926}

\bibitem[{S. {Noll} {et~al.}(2009){Noll}, {Burgarella}, {Giovannoli}, {Buat}, {Marcillac}, \& {Mu{\~n}oz-Mateos}}]{Noll2009}
{Noll}, S., {Burgarella}, D., {Giovannoli}, E., {et~al.} 2009, \bibinfo{title}{{Analysis of galaxy spectral energy distributions from far-UV to far-IR with CIGALE: studying a SINGS test sample},} \aap, 507, 1793, \dodoi{10.1051/0004-6361/200912497}

\bibitem[{M. {Park} {et~al.}(2025){Park}, {Conroy}, {Johnson}, {Leja}, {Dotter}, \& {Cargile}}]{park2025}
{Park}, M., {Conroy}, C., {Johnson}, B.~D., {et~al.} 2025, \bibinfo{title}{{{\ensuremath{\alpha}}-MC: Self-consistent {\ensuremath{\alpha}}-enhanced Stellar Population Models Covering a Wide Range of Age, Metallicity, and Wavelength},} \apj, 994, 165, \dodoi{10.3847/1538-4357/ae0cba}

\bibitem[{M. {Park} {et~al.}(2024){Park}, {Belli}, {Conroy}, {Johnson}, {Davies}, {Leja}, {Tacchella}, {Mendel}, {Benton}, {Bugiani}, {Emami}, {Khoram}, {Li}, {Maheson}, {Mathews}, {Naidu}, {Nelson}, {Terrazas}, \& {Weinberger}}]{Park2024}
{Park}, M., {Belli}, S., {Conroy}, C., {et~al.} 2024, \bibinfo{title}{{Widespread Rapid Quenching at Cosmic Noon Revealed by JWST Deep Spectroscopy},} \apj, 976, 72, \dodoi{10.3847/1538-4357/ad7e15}

\bibitem[{A. {Pennell} {et~al.}(2017){Pennell}, {Runnoe}, \& {Brotherton}}]{Pennell2017}
{Pennell}, A., {Runnoe}, J.~C., \& {Brotherton}, M.~S. 2017, \bibinfo{title}{{Updating quasar bolometric luminosity corrections - III. [O III] bolometric corrections},} \mnras, 468, 1433, \dodoi{10.1093/mnras/stx556}

\bibitem[{P.~G. {P{\'e}rez-Gonz{\'a}lez} {et~al.}(2025){P{\'e}rez-Gonz{\'a}lez}, {D'Eugenio}, {Rodr{\'\i}guez del Pino}, {Perna}, {{\"U}bler}, {Maiolino}, {Arribas}, {Cresci}, {Lamperti}, {Bunker}, {Carniani}, {Charlot}, {Willott}, {B{\"o}ker}, {Parlanti}, {Scholtz}, {Venturi}, {Barro}, {Costantin}, {Mart{\'\i}n-Navarro}, {Dunlop}, \& {Magee}}]{Perez-Gonzalez2025}
{P{\'e}rez-Gonz{\'a}lez}, P.~G., {D'Eugenio}, F., {Rodr{\'\i}guez del Pino}, B., {et~al.} 2025, \bibinfo{title}{{Accelerated quenching and chemical enhancement of massive galaxies in a z {\ensuremath{\approx}} 4 gas-rich halo},} Nature Astronomy, 9, 1240, \dodoi{10.1038/s41550-025-02586-8}

\bibitem[{M. {Perna} {et~al.}(2019){Perna}, {Cresci}, {Brusa}, {Lanzuisi}, {Concas}, {Mainieri}, {Mannucci}, \& {Marconi}}]{Perna2019}
{Perna}, M., {Cresci}, G., {Brusa}, M., {et~al.} 2019, \bibinfo{title}{{Multi-phase outflows in Mkn 848 observed with SDSS-MaNGA integral field spectroscopy},} \aap, 623, A171, \dodoi{10.1051/0004-6361/201834193}

\bibitem[{D. {Poznanski} {et~al.}(2012){Poznanski}, {Prochaska}, \& {Bloom}}]{Poznanski2012}
{Poznanski}, D., {Prochaska}, J.~X., \& {Bloom}, J.~S. 2012, \bibinfo{title}{{An empirical relation between sodium absorption and dust extinction},} \mnras, 426, 1465, \dodoi{10.1111/j.1365-2966.2012.21796.x}

\bibitem[{J.~X. {Prochaska} {et~al.}(2011){Prochaska}, {Kasen}, \& {Rubin}}]{Prochaska2011}
{Prochaska}, J.~X., {Kasen}, D., \& {Rubin}, K. 2011, \bibinfo{title}{{Simple Models of Metal-line Absorption and Emission from Cool Gas Outflows},} \apj, 734, 24, \dodoi{10.1088/0004-637X/734/1/24}

\bibitem[{G.~H. {Rieke} {et~al.}(2024){Rieke}, {Alberts}, {Shivaei}, {Lyu}, {Willmer}, {P{\'e}rez-Gonz{\'a}lez}, \& {Williams}}]{Rieke2024}
{Rieke}, G.~H., {Alberts}, S., {Shivaei}, I., {et~al.} 2024, \bibinfo{title}{{SMILES: A Prototype JWST Multiband Mid-infrared Survey},} \apj, 975, 83, \dodoi{10.3847/1538-4357/ad6cd2}

\bibitem[{B.~E. {Robertson} {et~al.}(2026){Robertson}, {Johnson}, {Tacchella}, {Eisenstein}, {Hainline}, {Alberts}, {Arribas}, {Baker}, {Bunker}, {Cameron}, {Carniani}, {Carreira}, {Chevallard}, {Circosta}, {Curtis-Lake}, {Danhaive}, {Duan}, {Egami}, {Hausen}, {Helton}, {Ji}, {Maiolino}, {P{\'e}rez-Gonz{\'a}lez}, {Pusk{\'a}s}, {Rieke}, {Rinaldi}, {Sun}, {Sun}, {{\"U}bler}, {Trussler}, {Villanueva}, {Whitler}, {Williams}, {Willmer}, {Willott}, {Wu}, \& {Zhu}}]{Robertson2026}
{Robertson}, B.~E., {Johnson}, B.~D., {Tacchella}, S., {et~al.} 2026, \bibinfo{title}{{JWST Advanced Deep Extragalactic Survey (JADES) Data Release 5: Photometric Catalog},} arXiv e-prints, arXiv:2601.15956, \dodoi{10.48550/arXiv.2601.15956}

\bibitem[{N. {Roy} {et~al.}(2018){Roy}, {Bundy}, {Cheung}, {Rujopakarn}, {Cappellari}, {Belfiore}, {Yan}, {Heckman}, {Bershady}, {Greene}, {Westfall}, {Drory}, {Rubin}, {Law}, {Zhang}, {Gelfand}, {Bizyaev}, {Wake}, {Masters}, {Thomas}, {Li}, \& {Riffel}}]{Roy2018}
{Roy}, N., {Bundy}, K., {Cheung}, E., {et~al.} 2018, \bibinfo{title}{{Detecting Radio AGN Signatures in Red Geysers},} \apj, 869, 117, \dodoi{10.3847/1538-4357/aaee72}

\bibitem[{D.~S. {Rupke} {et~al.}(2005{\natexlab{a}}){Rupke}, {Veilleux}, \& {Sanders}}]{Rupke2005b}
{Rupke}, D.~S., {Veilleux}, S., \& {Sanders}, D.~B. 2005{\natexlab{a}}, \bibinfo{title}{{Outflows in Infrared-Luminous Starbursts at z < 0.5. II. Analysis and Discussion},} \apjs, 160, 115, \dodoi{10.1086/432889}

\bibitem[{D.~S. {Rupke} {et~al.}(2005{\natexlab{b}}){Rupke}, {Veilleux}, \& {Sanders}}]{Rupke2005a}
{Rupke}, D.~S., {Veilleux}, S., \& {Sanders}, D.~B. 2005{\natexlab{b}}, \bibinfo{title}{{Outflows in Infrared-Luminous Starbursts at z < 0.5. I. Sample, Na I D Spectra, and Profile Fitting},} \apjs, 160, 87, \dodoi{10.1086/432886}

\bibitem[{D.~S. {Rupke} {et~al.}(2005{\natexlab{c}}){Rupke}, {Veilleux}, \& {Sanders}}]{Rupke2005c}
{Rupke}, D.~S., {Veilleux}, S., \& {Sanders}, D.~B. 2005{\natexlab{c}}, \bibinfo{title}{{Outflows in Active Galactic Nucleus/Starburst-Composite Ultraluminous Infrared Galaxies1,},} \apj, 632, 751, \dodoi{10.1086/444451}

\bibitem[{D.~S.~N. {Rupke}(2018){Rupke}}]{Rupke2018}
{Rupke}, D. S.~N. 2018, \bibinfo{title}{{A Review of Recent Observations of Galactic Winds Driven by Star Formation},} Galaxies, 6, 138, \dodoi{10.3390/galaxies6040138}

\bibitem[{E.~E. {Salpeter}(1955){Salpeter}}]{Salpeter1955}
{Salpeter}, E.~E. 1955, \bibinfo{title}{{The Luminosity Function and Stellar Evolution.},} \apj, 121, 161, \dodoi{10.1086/145971}

\bibitem[{P. {S{\'a}nchez-Bl{\'a}zquez} {et~al.}(2006){S{\'a}nchez-Bl{\'a}zquez}, {Peletier}, {Jim{\'e}nez-Vicente}, {Cardiel}, {Cenarro}, {Falc{\'o}n-Barroso}, {Gorgas}, {Selam}, \& {Vazdekis}}]{Sanchez-Blazquez2006}
{S{\'a}nchez-Bl{\'a}zquez}, P., {Peletier}, R.~F., {Jim{\'e}nez-Vicente}, J., {et~al.} 2006, \bibinfo{title}{{Medium-resolution Isaac Newton Telescope library of empirical spectra},} \mnras, 371, 703, \dodoi{10.1111/j.1365-2966.2006.10699.x}

\bibitem[{M. {Sarzi} {et~al.}(2016){Sarzi}, {Kaviraj}, {Nedelchev}, {Tiffany}, {Shabala}, {Deller}, \& {Middelberg}}]{Sarzi2016}
{Sarzi}, M., {Kaviraj}, S., {Nedelchev}, B., {et~al.} 2016, \bibinfo{title}{{Cold-gas outflows in typical low-redshift galaxies are driven by star formation, not AGN},} \mnras, 456, L25, \dodoi{10.1093/mnrasl/slv165}

\bibitem[{K. {Schawinski} {et~al.}(2015){Schawinski}, {Koss}, {Berney}, \& {Sartori}}]{Schawinski2015}
{Schawinski}, K., {Koss}, M., {Berney}, S., \& {Sartori}, L.~F. 2015, \bibinfo{title}{{Active galactic nuclei flicker: an observational estimate of the duration of black hole growth phases of {\ensuremath{\sim}}{}10$^{5}$ yr},} \mnras, 451, 2517, \dodoi{10.1093/mnras/stv1136}

\bibitem[{J. {Scholtz} {et~al.}(2025){Scholtz}, {Carniani}, {Parlanti}, {D'Eugenio}, {Curtis-Lake}, {Jakobsen}, {Bunker}, {Cameron}, {Arribas}, {Baker}, {Charlot}, {Chevellard}, {Circosta}, {Curti}, {Duan}, {Eisenstein}, {Hainline}, {Ji}, {Johnson}, {Jones}, {Kumari}, {Maiolino}, {Maseda}, {Perna}, {P{\'e}rez-Gonz{\'a}lez}, {Rawle}, {Rieke}, {Rinaldi}, {Robertson}, {Saxena}, {Shivaei}, {Silcock}, {Sun}, {Rodr{\'\i}guez Del Pino}, {Tacchella}, {{\"U}bler}, {Venturi}, {Williams}, {Willmer}, {Willott}, \& {Witstok}}]{Scholtz2025}
{Scholtz}, J., {Carniani}, S., {Parlanti}, E., {et~al.} 2025, \bibinfo{title}{{JADES Data Release 4 -- Paper II: Data reduction, analysis and emission-line fluxes of the complete spectroscopic sample},} arXiv e-prints, arXiv:2510.01034, \dodoi{10.48550/arXiv.2510.01034}

\bibitem[{J. {Scholtz} {et~al.}(2026){Scholtz}, {D'Eugenio}, {Maiolino}, {P{\'e}rez-Gonz{\'a}lez}, {Circosta}, {Tacchella}, {Williams}, {Alberts}, {Arribas}, {Baker}, {Bertola}, {Bunker}, {Carniani}, {Charlot}, {Cresci}, {Jones}, {Kumari}, {Lamperti}, {Looser}, {Pino}, {Robertson}, {Parlanti}, {Perna}, {{\"U}bler}, {Venturi}, \& {Witstok}}]{Scholtz2026}
{Scholtz}, J., {D'Eugenio}, F., {Maiolino}, R., {et~al.} 2026, \bibinfo{title}{{Measurement of the gas consumption history of a massive quiescent galaxy},} Nature Astronomy, \dodoi{10.1038/s41550-025-02751-z}

\bibitem[{A.~E. {Shapley} {et~al.}(2025){Shapley}, {Sanders}, {Topping}, {Reddy}, {Berg}, {Bouwens}, {Brammer}, {Carnall}, {Cullen}, {Dav{\'e}}, {Dunlop}, {Ellis}, {F{\"o}rster Schreiber}, {Furlanetto}, {Glazebrook}, {Illingworth}, {Jones}, {Kriek}, {McLeod}, {McLure}, {Narayanan}, {Oesch}, {Pahl}, {Pettini}, {Schaerer}, {Stark}, {Steidel}, {Tang}, {Clarke}, {Donnan}, \& {Kehoe}}]{Shapley2025}
{Shapley}, A.~E., {Sanders}, R.~L., {Topping}, M.~W., {et~al.} 2025, \bibinfo{title}{{The AURORA Survey: A New Era of Emission-line Diagrams with JWST/NIRSpec},} \apj, 980, 242, \dodoi{10.3847/1538-4357/adad68}

\bibitem[{J.~C. {Siegel} {et~al.}(2025){Siegel}, {Setton}, {Greene}, {Suess}, {Whitaker}, {Bezanson}, {Leja}, {Furtak}, {Cutler}, {de Graaff}, {Feldmann}, {Khullar}, {Labbe}, {Marchesini}, {Miller}, {Nanayakkara}, {Pan}, {Price}, {Treiber}, {van Dokkum}, {Wang}, \& {Weaver}}]{Siegel2025}
{Siegel}, J.~C., {Setton}, D.~J., {Greene}, J.~E., {et~al.} 2025, \bibinfo{title}{{UNCOVER: Significant Reddening in Cosmic Noon Quiescent Galaxies},} \apj, 985, 125, \dodoi{10.3847/1538-4357/adc7b7}

\bibitem[{J. {Silk} \& M.~J. {Rees}(1998){Silk} \& {Rees}}]{Silk1998}
{Silk}, J., \& {Rees}, M.~J. 1998, \bibinfo{title}{{Quasars and galaxy formation},} \aap, 331, L1, \dodoi{10.48550/arXiv.astro-ph/9801013}

\bibitem[{R.~S. {Somerville} {et~al.}(2015){Somerville}, {Popping}, \& {Trager}}]{Somerville2015}
{Somerville}, R.~S., {Popping}, G., \& {Trager}, S.~C. 2015, \bibinfo{title}{{Star formation in semi-analytic galaxy formation models with multiphase gas},} \mnras, 453, 4337, \dodoi{10.1093/mnras/stv1877}

\bibitem[{J.~S. {Speagle} {et~al.}(2014){Speagle}, {Steinhardt}, {Capak}, \& {Silverman}}]{Speagle2014}
{Speagle}, J.~S., {Steinhardt}, C.~L., {Capak}, P.~L., \& {Silverman}, J.~D. 2014, \bibinfo{title}{{A Highly Consistent Framework for the Evolution of the Star-Forming ``Main Sequence'' from z \raisebox{-0.5ex}\textasciitilde 0-6},} \apjs, 214, 15, \dodoi{10.1088/0067-0049/214/2/15}

\bibitem[{J. {Spilker} {et~al.}(2018){Spilker}, {Bezanson}, {Bari{\v{s}}i{\'c}}, {Bell}, {Lagos}, {Maseda}, {Muzzin}, {Pacifici}, {Sobral}, {Straatman}, {van der Wel}, {van Dokkum}, {Weiner}, {Whitaker}, {Williams}, \& {Wu}}]{Spilker2018}
{Spilker}, J., {Bezanson}, R., {Bari{\v{s}}i{\'c}}, I., {et~al.} 2018, \bibinfo{title}{{Molecular Gas Contents and Scaling Relations for Massive, Passive Galaxies at Intermediate Redshifts from the LEGA-C Survey},} \apj, 860, 103, \dodoi{10.3847/1538-4357/aac438}

\bibitem[{I. {Strateva} {et~al.}(2001){Strateva}, {Ivezi{\'c}}, {Knapp}, {Narayanan}, {Strauss}, {Gunn}, {Lupton}, {Schlegel}, {Bahcall}, {Brinkmann}, {Brunner}, {Budav{\'a}ri}, {Csabai}, {Castander}, {Doi}, {Fukugita}, {Gy{\H{o}}ry}, {Hamabe}, {Hennessy}, {Ichikawa}, {Kunszt}, {Lamb}, {McKay}, {Okamura}, {Racusin}, {Sekiguchi}, {Schneider}, {Shimasaku}, \& {York}}]{Strateva2001}
{Strateva}, I., {Ivezi{\'c}}, {\v{Z}}., {Knapp}, G.~R., {et~al.} 2001, \bibinfo{title}{{Color Separation of Galaxy Types in the Sloan Digital Sky Survey Imaging Data},} \aj, 122, 1861, \dodoi{10.1086/323301}

\bibitem[{Y. {Sun} {et~al.}(2024){Sun}, {Lee}, {Zabludoff}, {French}, {Helton}, {Kerrison}, {Tremonti}, \& {Yang}}]{Sun2024}
{Sun}, Y., {Lee}, G.-H., {Zabludoff}, A.~I., {et~al.} 2024, \bibinfo{title}{{Evolution of gas flows along the starburst to post-starburst to quiescent galaxy sequence},} \mnras, 528, 5783, \dodoi{10.1093/mnras/stae366}

\bibitem[{Y. {Sun} {et~al.}(2025){Sun}, {Lyu}, {Rieke}, {Ji}, {Sun}, {Zhu}, {Bunker}, {Cargile}, {Circosta}, {D'Eugenio}, {Egami}, {Hainline}, {Helton}, {Rinaldi}, {Robertson}, {Scholtz}, {Shivaei}, {Stone}, {Tacchella}, {Williams}, {Willmer}, \& {Willott}}]{Sun2025}
{Sun}, Y., {Lyu}, J., {Rieke}, G.~H., {et~al.} 2025, \bibinfo{title}{{No Evidence for a Significant Evolution of M$_{{\textbullet}}${\textendash}M. Relation in Massive Galaxies up to z {\ensuremath{\sim}} 4},} \apj, 978, 98, \dodoi{10.3847/1538-4357/ad973b}

\bibitem[{Y. {Sun} {et~al.}(2026){Sun}, {Ji}, {Rieke}, {D'Eugenio}, {Zhu}, {Sun}, {Lin}, {Bunker}, {Lyu}, {Rinaldi}, \& {Willmer}}]{Sun2026}
{Sun}, Y., {Ji}, Z., {Rieke}, G.~H., {et~al.} 2026, \bibinfo{title}{{Extreme Neutral Outflow in a Non-active Galactic Nucleus Quiescent Galaxy at z {\ensuremath{\sim}} 1.3},} \apj, 997, 140, \dodoi{10.3847/1538-4357/ae2c83}

\bibitem[{S. {Tacchella} {et~al.}(2016){Tacchella}, {Dekel}, {Carollo}, {Ceverino}, {DeGraf}, {Lapiner}, {Mandelker}, \& {Primack Joel}}]{Tacchella2016}
{Tacchella}, S., {Dekel}, A., {Carollo}, C.~M., {et~al.} 2016, \bibinfo{title}{{The confinement of star-forming galaxies into a main sequence through episodes of gas compaction, depletion and replenishment},} \mnras, 457, 2790, \dodoi{10.1093/mnras/stw131}

\bibitem[{S. {Tacchella} {et~al.}(2022){Tacchella}, {Conroy}, {Faber}, {Johnson}, {Leja}, {Barro}, {Cunningham}, {Deason}, {Guhathakurta}, {Guo}, {Hernquist}, {Koo}, {McKinnon}, {Rockosi}, {Speagle}, {van Dokkum}, \& {Yesuf}}]{Tacchella2022}
{Tacchella}, S., {Conroy}, C., {Faber}, S.~M., {et~al.} 2022, \bibinfo{title}{{Fast, Slow, Early, Late: Quenching Massive Galaxies at z {\ensuremath{\sim}} 0.8},} \apj, 926, 134, \dodoi{10.3847/1538-4357/ac449b}

\bibitem[{E. {Taylor} {et~al.}(2024){Taylor}, {Maltby}, {Almaini}, {Merrifield}, {Wild}, {Rowlands}, \& {Harrold}}]{Taylor2024}
{Taylor}, E., {Maltby}, D., {Almaini}, O., {et~al.} 2024, \bibinfo{title}{{High-velocity outflows persist up to 1 Gyr after a starburst in recently quenched galaxies at z > 1},} \mnras, 535, 1684, \dodoi{10.1093/mnras/stae2463}

\bibitem[{E. {Taylor} {et~al.}(2026){Taylor}, {Carnall}, {Maltby}, {Almaini}, {Leung}, {Stevenson}, {Negri}, {Cullen}, {Wild}, {McLure}, {Shapley}, {Arellano-C{\'o}rdova}, {Begley}, {Bondestam}, {de Lisle}, {Donnan}, {Dunlop}, {Ellis}, {Hewitt}, {Koekemoer}, {Frey Liu}, {McLeod}, {Rowlands}, {Sanders}, {Scholte}, {Skarbinski}, \& {Stanton}}]{Taylor2026}
{Taylor}, E., {Carnall}, A.~C., {Maltby}, D., {et~al.} 2026, \bibinfo{title}{{The JWST EXCELS survey: Outflows in 1.5 < z < 5 quiescent galaxies are likely relics from episodic AGN activity},} arXiv e-prints, arXiv:2601.02269, \dodoi{10.48550/arXiv.2601.02269}

\bibitem[{C.~A. {Tremonti} {et~al.}(2007){Tremonti}, {Moustakas}, \& {Diamond-Stanic}}]{Tremonti2007}
{Tremonti}, C.~A., {Moustakas}, J., \& {Diamond-Stanic}, A.~M. 2007, \bibinfo{title}{{The Discovery of 1000 km s$^{-1}$ Outflows in Massive Poststarburst Galaxies at z=0.6},} \apjl, 663, L77, \dodoi{10.1086/520083}

\bibitem[{F. {Valentino} {et~al.}(2025){Valentino}, {Heintz}, {Brammer}, {Ito}, {Kokorev}, {Whitaker}, {Gallazzi}, {de Graaff}, {Weibel}, {Frye}, {Kamieneski}, {Jin}, {Ceverino}, {Faisst}, {Farcy}, {Fujimoto}, {Gillman}, {Gottumukkala}, {Hamadouche}, {Harrington}, {Hirschmann}, {Jespersen}, {Kakimoto}, {Kubo}, {Lagos}, {Lee}, {Magdis}, {Man}, {Onodera}, {Rizzo}, {Shimakawa}, {Setton}, {Tanaka}, {Toft}, {Wu}, \& {Zhu}}]{Valentino2025}
{Valentino}, F., {Heintz}, K.~E., {Brammer}, G., {et~al.} 2025, \bibinfo{title}{{Gas outflows in two recently quenched galaxies at z = 4 and 7},} \aap, 699, A358, \dodoi{10.1051/0004-6361/202553908}

\bibitem[{S. {Veilleux} {et~al.}(2005){Veilleux}, {Cecil}, \& {Bland-Hawthorn}}]{Veilleux2005}
{Veilleux}, S., {Cecil}, G., \& {Bland-Hawthorn}, J. 2005, \bibinfo{title}{{Galactic Winds},} \araa, 43, 769, \dodoi{10.1146/annurev.astro.43.072103.150610}

\bibitem[{S. {Veilleux} {et~al.}(2020){Veilleux}, {Maiolino}, {Bolatto}, \& {Aalto}}]{Veilleux2020}
{Veilleux}, S., {Maiolino}, R., {Bolatto}, A.~D., \& {Aalto}, S. 2020, \bibinfo{title}{{Cool outflows in galaxies and their implications},} \aapr, 28, 2, \dodoi{10.1007/s00159-019-0121-9}

\bibitem[{S. {Veilleux} \& D.~E. {Osterbrock}(1987){Veilleux} \& {Osterbrock}}]{Veilleux1987}
{Veilleux}, S., \& {Osterbrock}, D.~E. 1987, \bibinfo{title}{{Spectral Classification of Emission-Line Galaxies},} in NASA Conference Publication, Vol. 2466, NASA Conference Publication, ed. C.~J. {Lonsdale Persson}, 737--740

\bibitem[{P. {Virtanen} {et~al.}(2020){Virtanen}, {Gommers}, {Oliphant}, {Haberland}, {Reddy}, {Cournapeau}, {Burovski}, {Peterson}, {Weckesser}, {Bright}, {van der Walt}, {Brett}, {Wilson}, {Millman}, {Mayorov}, {Nelson}, {Jones}, {Kern}, {Larson}, {Carey}, {Polat}, {Feng}, {Moore}, {VanderPlas}, {Laxalde}, {Perktold}, {Cimrman}, {Henriksen}, {Quintero}, {Harris}, {Archibald}, {Ribeiro}, {Pedregosa}, {van Mulbregt}, \& {SciPy 1. 0 Contributors}}]{Virtanen2020}
{Virtanen}, P., {Gommers}, R., {Oliphant}, T.~E., {et~al.} 2020, \bibinfo{title}{{SciPy 1.0: fundamental algorithms for scientific computing in Python},} Nature Methods, 17, 261, \dodoi{10.1038/s41592-019-0686-2}

\bibitem[{C.~C. {Williams} {et~al.}(2021){Williams}, {Spilker}, {Whitaker}, {Dav{\'e}}, {Woodrum}, {Brammer}, {Bezanson}, {Narayanan}, \& {Weiner}}]{Williams2021}
{Williams}, C.~C., {Spilker}, J.~S., {Whitaker}, K.~E., {et~al.} 2021, \bibinfo{title}{{ALMA Measures Rapidly Depleted Molecular Gas Reservoirs in Massive Quiescent Galaxies at z {\ensuremath{\sim}} 1.5},} \apj, 908, 54, \dodoi{10.3847/1538-4357/abcbf6}

\bibitem[{J.-H. {Woo} {et~al.}(2016){Woo}, {Bae}, {Son}, \& {Karouzos}}]{Woo2016}
{Woo}, J.-H., {Bae}, H.-J., {Son}, D., \& {Karouzos}, M. 2016, \bibinfo{title}{{The Prevalence of Gas Outflows in Type 2 AGNs},} \apj, 817, 108, \dodoi{10.3847/0004-637X/817/2/108}

\bibitem[{G. {Worthey} \& D.~L. {Ottaviani}(1997){Worthey} \& {Ottaviani}}]{Worthey1997}
{Worthey}, G., \& {Ottaviani}, D.~L. 1997, \bibinfo{title}{{H{\ensuremath{\gamma}} and H{\ensuremath{\delta}} Absorption Features in Stars and Stellar Populations},} \apjs, 111, 377, \dodoi{10.1086/313021}

\bibitem[{P.-F. {Wu}(2025){Wu}}]{Wu2025}
{Wu}, P.-F. 2025, \bibinfo{title}{{Ejective Feedback as a Quenching Mechanism in the First 1.5 Billion Years of the Universe: Detection of Neutral Gas Outflow in a z = 4 Recently Quenched Galaxy},} \apj, 978, 131, \dodoi{10.3847/1538-4357/ad98ef}

\bibitem[{Y.~Q. {Xue} {et~al.}(2016){Xue}, {Luo}, {Brandt}, {Alexander}, {Bauer}, {Lehmer}, \& {Yang}}]{Xue2016}
{Xue}, Y.~Q., {Luo}, B., {Brandt}, W.~N., {et~al.} 2016, \bibinfo{title}{{The 2 Ms Chandra Deep Field-North Survey and the 250 ks Extended Chandra Deep Field-South Survey: Improved Point-source Catalogs},} \apjs, 224, 15, \dodoi{10.3847/0067-0049/224/2/15}

\bibitem[{P. {Zhu} {et~al.}(2026){Zhu}, {Ito}, {Valentino}, {Hamadouche}, {Scarpe}, {Whitaker}, {Kakimoto}, {Baker}, {Gallazzi}, {Gillman}, {Gottumukkala}, {Jespersen}, {Lee}, {Man}, {Magdis}, {Onodera}, {Shimakawa}, {Vijayan}, \& {Wu}}]{ZhuP2026}
{Zhu}, P., {Ito}, K., {Valentino}, F., {et~al.} 2026, \bibinfo{title}{{There and back again? Neutral outflows in z\raisebox{-0.5ex}\textasciitilde3.5 quiescent galaxies},} arXiv e-prints, arXiv:2602.17767, \dodoi{10.48550/arXiv.2602.17767}

\bibitem[{Y. {Zhu} {et~al.}(2025){Zhu}, {Rieke}, {Ji}, {Simmonds}, {Sun}, {Sun}, {Alberts}, {Bhatawdekar}, {Bunker}, {Cargile}, {Carniani}, {de Graaff}, {Hainline}, {Helton}, {Jones}, {Lyu}, {Rieke}, {Rinaldi}, {Robertson}, {Scholtz}, {{\"U}bler}, {Williams}, \& {Willmer}}]{Zhu2025}
{Zhu}, Y., {Rieke}, M.~J., {Ji}, Z., {et~al.} 2025, \bibinfo{title}{{A Systematic Search for Galaxies with Extended Emission Lines and Potential Outflows in JADES Medium-band Images},} \apj, 986, 162, \dodoi{10.3847/1538-4357/add2f4}

\bibitem[{Y. {Zhu} {et~al.}(2026){Zhu}, {Bonaventura}, {Sun}, {Rieke}, {Alberts}, {Lyu}, {Shivaei}, {Morrison}, {Ji}, {Egami}, {Helton}, {Rieke}, {Rinaldi}, {Sun}, \& {Willmer}}]{Zhu2026}
{Zhu}, Y., {Bonaventura}, N., {Sun}, Y., {et~al.} 2026, \bibinfo{title}{{SMILES Data Release. II. Probing Galaxy Evolution during Cosmic Noon and Beyond with NIRSpec Medium-resolution Spectra},} \apj, 997, 301, \dodoi{10.3847/1538-4357/ae29b4}

\bibitem[{K. {Zubovas} {et~al.}(2022){Zubovas}, {Bialopetravi{\v{c}}ius}, \& {Kazlauskait{\.{e}}}}]{Zubovas2022}
{Zubovas}, K., {Bialopetravi{\v{c}}ius}, J., \& {Kazlauskait{\.{e}}}, M. 2022, \bibinfo{title}{{Determining active galactic nucleus luminosity histories using present-day outflow properties: a neural network-based approach},} \mnras, 515, 1705, \dodoi{10.1093/mnras/stac1887}

\end{thebibliography}
\bibliographystyle{aasjournalv7}



\end{document}